\documentclass[journal=jpclcd,manuscript=letter]{achemso}

\usepackage{amsfonts}
\usepackage{bm}
\usepackage{setspace}
\usepackage{graphicx}
\usepackage{color}
\usepackage{multirow}
\usepackage{verbatim}
\usepackage{float}
\usepackage{array}
\usepackage{booktabs}
\usepackage{float}
\usepackage{cleveref}

\usepackage{amsmath}
\usepackage{amssymb}
\usepackage{graphicx}
\usepackage[version=3]{mhchem}
\usepackage[T1]{fontenc}
\usepackage{subfigure}

\title{Approximating Quasiparticle and Excitation Energies from Ground State
Generalized Kohn-Sham Calculations}

\author{Yuncai Mei}
\affiliation[Department of Chemistry, Duke University, Durham, NC 27708]
{Department of Chemistry, Duke University, Durham, NC 27708}
\author{Chen Li}
\affiliation[Department of Chemistry, Duke University, Durham, NC 27708]
{Department of Chemistry, Duke University, Durham, NC 27708}
\author{Neil Qiang Su}
\affiliation[Department of Chemistry, Duke University, Durham, NC 27708]
{Department of Chemistry, Duke University, Durham, NC 27708}
\author{Weitao Yang}
\email{weitao.yang@duke.edu}
\affiliation[Department of Chemistry, Duke University, Durham, NC 27708]
{Department of Chemistry, Duke University, Durham, NC 27708}
\alsoaffiliation[Key Laboratory of Theoretical Chemistry of Environment, School
of Chemistry and Environment, South China Normal University, Guangzhou
510006, China]
{Key Laboratory of Theoretical Chemistry of Environment, School
of Chemistry and Environment, South China Normal University, Guangzhou
510006, China}

\date{\today}

\begin{document}

\date{\today}

\begin{tocentry}
    \includegraphics[width=\linewidth]{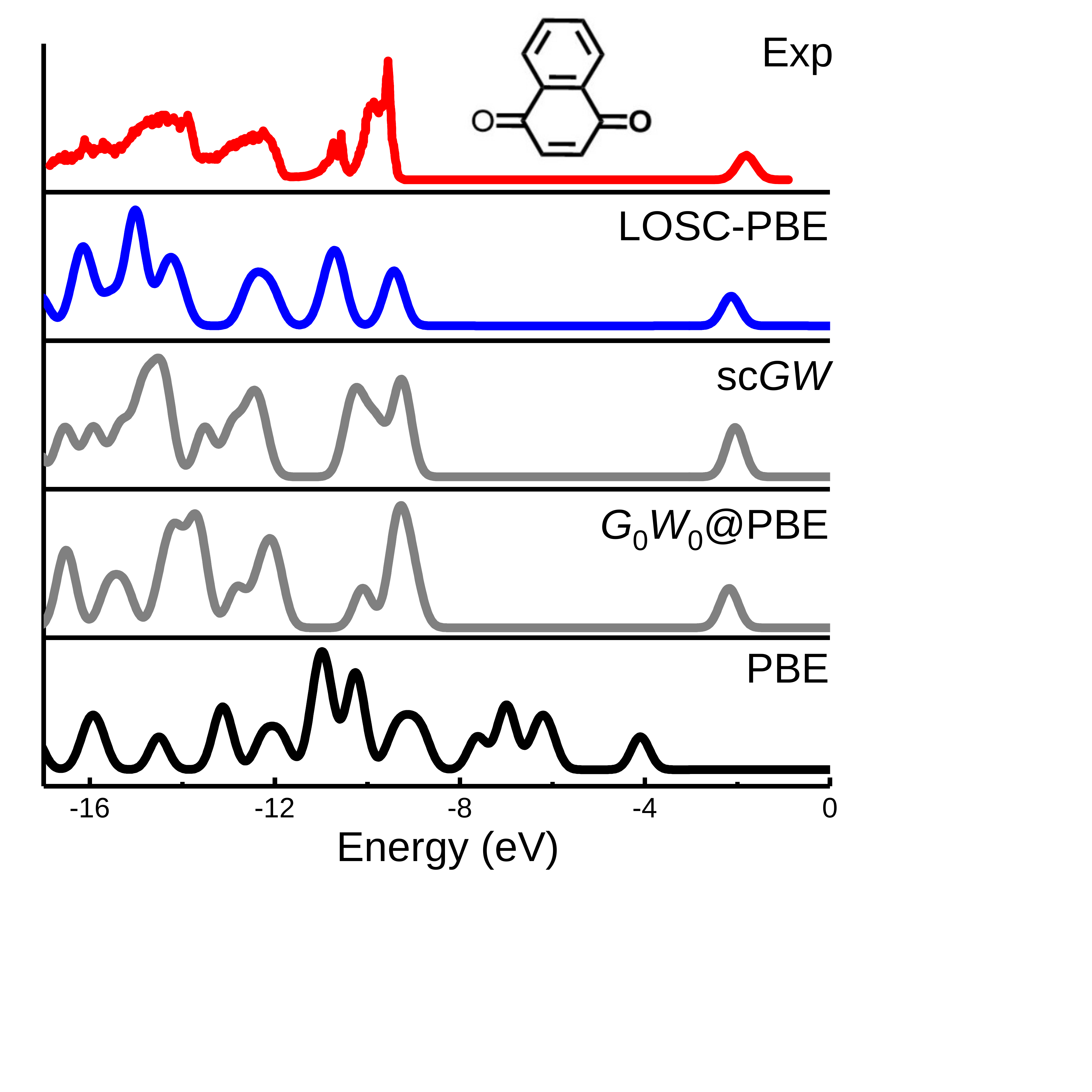}
\end{tocentry}

\begin{abstract}
Quasiparticle energies and fundamental band gaps in particular are
critical properties of molecules and materials. It was rigorously
established that the generalized Kohn-Sham HOMO and LUMO orbital energies
are the chemical potentials of electron removal and addition and thus
good approximations to band edges and fundamental gaps from a density functional approximation (DFA)
with minimal delocalization error. For other quasiparticle energies,
their connection to the generalized Kohn-Sham orbital energies has
not been established but remains highly interesting. We provide the
comparison of experimental quasiparticle energies for many finite
systems with calculations from the GW Green's function and localized
orbitals scaling correction (LOSC), a recently developed correction
to semilocal DFAs, which has minimal delocalization error. Extensive
results with over forty systems clearly show that LOSC orbital energies
achieve slightly better accuracy than the GW calculations with little
dependence on the semilocal DFA, supporting the use of LOSC DFA orbital
energies to predict quasiparticle energies. This also leads to the
calculations of excitation energies of the $N$-electron systems from
the ground state DFA calculations of the $\left(N-1\right)$-electron
systems. Results show good performance with accuracy similar to TDDFT
and the delta SCF approach for valence excitations with commonly used
DFAs with or without LOSC. For Rydberg states, good accuracy was obtained
only with the use of LOSC DFA. This work highlights the pathway to
quasiparticle and excitation energies from ground density functional
calculations. 
\end{abstract}

\maketitle
Quasiparticles are a powerful concept in electronic structure
theory of many-electron systems. In particular, accurate prediction
of quasiparticle energies is essential for interpreting the electronic
excitation spectra of molecules and materials, such as photoemission
and optical experiments. Formally, quasiparticle energies can be exactly
formulated in many-body perturbation theory \cite{Fetter2003quan,Dickhoff2008,Martin2016}.
In practice, the GW approximation \cite{Hedin1965PR,Aryasetiawan1998,Onida2002RMP,Holm1998}
is most widely used for bulk simulations. Unfortunately, GW calculations
are still expensive computationally. Therefore, a low-cost alternative
to GW approximation that offers good accuracy for the prediction of
quasiparticle energies is critical to the calculations of large-scale
systems, and for efficient high throughput study of materials.

Kohn-Sham (KS) density functional theory (DFT) \cite{hohenberg1964inhomogeneous,kohn1965self,Parr-Yang1989},
due to its good balance between accuracy and computational tractability,
is among the most popular and versatile methods available for many-electron
problems. In addition to the total electron energy, the physical interpretation
of the KS eigenvalues has also attracted great interest. It has been
known for decades that among the KS eigenvalues obtained from the
exact functional, the highest occupied molecular orbital (HOMO) energy,
$\varepsilon_{{\rm HOMO}}$, is negative vertical ionization potential
(VIP), $-I$ \cite{Perdew821691,Perdew831884,Parr-Yang1989,Levy842745,Almbladh842322,Almbladh853231,Perdew9716021,Casida994694}.
In 2008, it was rigorously proven \cite{Cohen08115123,Yang12d} that
within the generalized KS (GKS) theory, which includes KS theory as
a special case, the HOMO/LUMO energy is the chemical potential, $\left(\frac{\partial E}{\partial N}\right)_{v}$,
for electron removal/addition from the DFAs for any DFA that is a
differentiable functional of the non-interacting one-electron density
matrix in case of GKS or the density in case of KS, and consequently
approximation to $-I/-A$ following the Perdew-Parr-Levy-Balduz (PPLB)
condition \cite{Perdew821691,Zhang00346,Yang005172,Perdew0740501}.
Accurate approximation of $-I/-A$ can thus be expected from the HOMO/LUMO
energy of DFAs with minimum delocalization error \cite{Mori-Sanchez08146401}. Therefore,
the fundamental gap defined as $I-A$ can be exactly obtained from
the chemical potential difference, that is, the GKS HOMO-LUMO gap.

In addition to HOMO and LUMO, the physical meaning of other GKS eigenvalues
also has great theoretical significance and application value. Of
particular interest is the connection between the GKS spectrum and
the quasiparticle spectrum. Unfortunately, no clear connection has
been established, although there have been many attempts to approximately
attach some meanings to the occupied orbital energies within the KS
theory. It has been argued that the orbital energies
below $\varepsilon_{{\rm HOMO}}$ can be interpreted as other approximate
principle (sometimes called relaxed) VIPs, i.e., the ionized system
being in an excited state \cite{Chong021760,Gritsenko031937}. Recently,
it has been argued that the correct occupied KS orbital energies should
correspond to the exact principle VIPs using the linear response time-dependent
density functional theory (LR-TDDFT) under the adiabatic approximation
\cite{Bartlett1754,Duminda17034102,Bartlett091}. However, it has
been shown that the adiabatic approximation within TDDFT is not generally
valid \cite{Fuks2014pccp}.

Even though no theorem has been rigorously established to link the
remaining GKS orbital energies to quasiparticle energies, it is still beneficial
for practical applications to construct a good density functional
approximation (DFA) that can accurately predict quasiparticle energies
from orbital energies. For commonly used DFAs, such as local density
approximations (LDAs), generalized gradient approximations (GGAs)
and hybrid GGAs, their HOMO and LUMO energies are the corresponding
chemical potentials but have large systematic error in predicting
$-I/-A$ \cite{Cohen08115123,Mori-Sanchez08146401}. In particular,
the HOMO energy is significantly overestimated, which leads to underestimation
of $I$; while the LUMO energy is severely underestimated, so that
$A$ is overestimated. Hence, the fundamental gap is significantly
underestimated by HOMO-LUMO gap of common DFAs. From the fractional
charge perspective, this failure has been attributed \cite{Cohen08115123,Mori-Sanchez08146401}
to the violation of the PPLB condition \cite{Perdew821691,Zhang00346,Yang005172,Perdew0740501}
which requires the total energy, as a function of electron number,
to be piecewise straight lines interpolating between adjacent integer
points. And the convex deviation suffered by commonly used DFAs was
identified as the delocalization error inherent in approximate functionals
\cite{Cohen12289,Mori-Sanchez08146401,Cohen2008a}. Other occupied
and unoccupied orbitals follow the same trend as HOMO and LUMO, respectively.
Typically, energies of occupied orbitals (including HOMO) have been
seriously overestimated when serving as approximations to electron
removal energies, so that they cannot qualitatively reproduce experimental
photoemission spectrum. It is thus reasonable to believe that other
orbitals should suffer similarly from the delocalization error of
approximate functionals.

Following the perspective of fractional charges, there have been many
attempts focusing on removing delocalization error in approximate
functionals. MCY3 \cite{Cohen07191109} was the first DFA constructed
to restore the PPLB condition; long-range corrected (LC) functionals
\cite{Gill961005,Leininger97151,Iikura013540,Tsuneda10174101} and
doubly hybrid functionals \cite{Grimme06034108,Zhang094963,Su149201}
show some promise on reproducing linear fractional charge behavior;
tuned range-separated hybrid functionals \cite{Baer1085,Stein10266802}
impose extra constraints on orbital energies from total energy difference
by optimizing the range-separation parameter for each system. All
these functionals show significantly improvement on the calculations
of HOMO and LUMO energies for small molecules. Extension to large
and bulk systems lead to various issues. To achieve systematic elimination
of the delocalization error associated with commonly used DFAs, recently
developed localized orbital scaling correction (LOSC) functional \cite{Li185203}
introduces a set of auxiliary localized orbitals (LOs), or orbitalets,
and imposes PPLB condition on each of the LOs. As a result, LOSC can
achieve size-consistent corrections to both the total energy and orbital
energies.

To demonstrate that orbital energies $\bm{\varepsilon}\left(N\right)$
of LOSC can give accurate approximation to quasiparticle/quasihole
energies $\bm{\omega}^{+/-}\left(N\right)$ for an $N$-electron system,
for the description of electron addition/removal, i.e. 

\begin{align}
	\varepsilon_{m}(N)&\approx\omega_{m}^{+}(N)=E_{m}(N+1)-E_{0}(N), \nonumber \\
	\varepsilon_{n}(N)&\approx\omega_{n}^{-}(N)=E_{0}(N)-E_{n}(N-1),\label{eq:ip_ea}
\end{align}
we have already applied LOSC to generate accurate LUMO and HOMO energies
for a broad range of atoms and molecules \cite{Li185203}. In Eq.
\ref{eq:ip_ea}, $\varepsilon_{m}(N)$/$\varepsilon_{n}(N)$ is a
virtual/occupied GKS orbital energy for the $N$-electron system. The performance of LOSC for HOMO/LUMO
and other GKS orbital energies will be examined extensively in present
work.

Furthermore, Eq. \ref{eq:ip_ea} allows the calculation of excitation
energies $\Delta E_{m}(N)$ at the cost of a ground-state DFT calculation
via the particle part of the quasiparticle spectrum of the ($N-1)$
system, i.e. 
\begin{align}
\Delta E_{m}(N) & =E_{m}(N)-E_{0}(N)\nonumber \\
 & =[E_{m}(N)-E_{0}(N-1)]-[E_{0}(N)-E_{0}(N-1)]\nonumber \\
 & =\omega_{m}^{+}(N-1)-\omega_{min}^{+}(N-1)\nonumber \\
 & \approx\varepsilon_{m}(N-1)-\varepsilon_{LUMO}(N-1),\label{eq:e_n-1}
\end{align}
where $E_{m}(N)$ corresponds to the $m$th excitation of the $N$-electron
system, and $E_{0}(N-1)$ is the ground-state energy of $(N-1)$-electron
system. $E_{0}(N)-E_{0}(N-1)$ is -A of the $(N-1)$ system and can
be obtained from $\omega_{min}^{+}(N-1)$, the minimum of particle
part of the quasiparticle spectrum, and approximated as $\varepsilon_{LUMO}(N-1)$,
the LUMO energy of the DFA calculation for the $(N-1)$ system. The
excitation energy $\Delta E_{m}(N)$ can thus be obtained as the virtual
orbital energy difference $\varepsilon_{m}(N-1)-\varepsilon_{LUMO}(N-1)$
from a ground-state self-consistent field (SCF) calculation on $(N-1)$-electron
system. Similarly, excitation energies can also be calculated via
the hole part of the quasiparticle spectrum of the ($N+1)$ system,
i.e.

\begin{align}
\Delta E_{n}(N) & =E_{n}(N)-E_{0}(N)\nonumber \\
 & =[E_{0}(N+1)-E_{0}(N)]-[E_{0}(N+1)-E_{n}(N)]\nonumber \\
 & =\omega_{max}^{-}(N+1)-\omega_{n}^{-}(N+1)\nonumber \\
 & \approx\varepsilon_{HOMO}(N+1)-\varepsilon_{n}(N+1),\label{eq:e_n+1}
\end{align}
where $E_{0}(N+1)-E_{0}(N)$ is $-I$ of the $(N+1)$ system and can
be obtained from $\omega_{max}^{-}(N+1)$, the maximum of the hole
part in the quasiparticle spectrum, and approximated as $\varepsilon_{HOMO}(N+1)$,
the HOMO energy of the DFA calculation for the $(N+1)$ system. The
excitation energies can thus be obtained as occupied orbital energy
differences $\varepsilon_{HOMO}(N+1)-\varepsilon_{n}(N+1)$ from a ground-state SCF calculation on $(N+1)$-electron
system. 

Many theoretical approaches have been developed to calculate excitation
energies. High-level methods, including equation-of-motion coupled
cluster (EOM-CC) \cite{emrich1981extension,sekino1984linear,geertsen1989equation},
linear-response coupled cluster (LR-CC) \cite{monkhorst1977calculation,monkhorst1983j,koch1995excitation,christiansen2000spin},
multireference configuration interaction (MRCI) \cite{helgaker2014molecular,buenker1968ci},
complete active space configuration interaction (CASCI) \cite{potts2001improved,abrams2004natural,slavivcek2010ab},
CASPT2 \cite{andersson1990second,andersson1992second} and others,
can produce accurate results, but significantly limited in system
size and complexity. Other computationally efficient methods, such
as configuration interaction singles (CIS) \cite{bene1971self,dreuw2005single},
time dependent DFT (TDDFT) \cite{runge1984density} and $\Delta$SCF
\cite{ziegler1977calculation} have been well-known to describe excitation
energies with success, meanwhile they have important weakness. Particularly,
CIS can overestimate excitation energy by 2 eV \cite{dreuw2005single}.
TDDFT \cite{laurent2013td,dreuw2005single} and $\Delta$SCF method
\cite{tozer2000determination,liu2004density,ceresoli2004trapping,cheng2008rydberg,gavnholt2008delta,besley2009self,kowalczyk_assessment_2011,becke2016vertical}
typically yield results with good accuracy, but TDDFT faces challenges
to describe double \cite{tozer2000determination,maitra2004double,levine2006conical},
Rydberg \cite{tozer1998improving,casida1998molecular,casida2000asymptotic,tozer2003importance}
and charge transfer excitations \cite{tozer1999does,dreuw2003long,dreuw2004failure}.
 In contrast, Eqs. \ref{eq:e_n-1} and \ref{eq:e_n+1} provide the
simplest way to calculate excitation energies, with which various
excitation energies can be obtained after the corresponding ground-state
SCF calculation. Obviously, the accuracy of excitation energies from
Eqs. \ref{eq:e_n-1} and \ref{eq:e_n+1} depends on the quality of
DFA orbital energies, as approximation to the quasiparticle energies.

Next, we will show the test results of approximating quasiparticle
energies (Eq. \ref{eq:ip_ea}) and excitation energies (Eqs. \ref{eq:e_n-1}/\ref{eq:e_n+1})
by different DFAs and LOSC-DFAs. For the test of quasiparticle energies,
40 molecules were selected from Blase's \cite{blase2011first} and
Marom's \cite{potentialselectron} test set to calculate photoemission
spectrum, HOMO and LUMO energies. Polyacene (n=1-6) and other three
small molecules are used to study the valence orbital energies as
approximation to the corresponding quasiparticle energies. For the
test of excitation energies, 16 molecules are obtained from Ref \citenum{yangexcitation2014}
as a molecular set to test the low-lying excitation energies. Four
atoms (Li, Be, Mg, and Na) are selected as an atomic set to test their
excitation energies up to Rydberg states. The QM4D package \cite{qm4d}
was used to perform the DFT calculations. Several conventional functionals,
such as local density approximation (LDA) \cite{slater1974self,vosko1980accurate},
PBE \cite{perdew1996generalized}, BLYP \cite{becke1988density,lee1988development}
and B3LYP,\cite{becke1988density,becke1993becke,lee1988development}
and LOSC-DFAs were tested. For LOSC calculations, the post-SCF procedure
was applied. More details of computations and test results can be
found in SI.

\begin{table}[h]
\centering \caption{Mean absolute errors (MAEs, in eV) of orbital energies compared with
experimental quasi-particle energies. Experimental reference were
obtained from Ref. \citenum{potentialselectron,blase2011first}.}
\label{tab:eig} %
\begin{tabular}{@{}lccc@{}}
\toprule 
 & HOMO \textsuperscript{\emph{a}}  & LUMO \textsuperscript{\emph{a}}  & Valence \textsuperscript{\emph{b}} \tabularnewline
\midrule 
sc$GW$ \textsuperscript{\emph{c}}  & 0.47  & 0.34  & - \tabularnewline
$G_{0}W_{0}$@PBE \textsuperscript{\emph{c}} & 0.51  & 0.37  & - \tabularnewline
LOSC\_LDA  & 0.34  & 0.48  & 0.69 (0.53) \tabularnewline
LOSC\_PBE  & 0.37  & 0.33  & 0.60 (0.35) \tabularnewline
LOSC\_B3LYP  & 0.26  & 0.29  & 0.43 (0.36) \tabularnewline
LDA  & 2.58  & 2.43  & 3.06 (2.33) \tabularnewline
PBE  & 2.81  & 2.16  & 3.23 (2.55) \tabularnewline
B3LYP  & 2.00  & 1.57  & 2.24 (1.79) \tabularnewline
$\Delta$-DFA \textsuperscript{\emph{d}}  & 0.43  & 0.26  & 0.70 (0.73) \tabularnewline
$\Delta$-LOSC-DFA \textsuperscript{\emph{d}} & 0.34  & 0.38  & 0.41 (0.26) \tabularnewline
\bottomrule
\end{tabular}

\raggedright \textsuperscript{\emph{a}} MAE of HOMO and LUMO energies
are based on the results of 40 test molecules.\\
 \textsuperscript{\emph{b}} MAE of valence orbital energies (HOMO
and below HOMO) are based on 51 states of polyacene (n = 1 - 6) and
other three small molecules. MAE of polyacene set were listed in the
parenthesis. \\
 \textsuperscript{\emph{c}} $GW$ results were taken from Ref \citenum{potentialselectron}.\\
 \textsuperscript{\emph{d}} PBE functional was used in HOMO and LUMO
calculation. BLYP functional was used in valence orbital results.\\
 
\end{table}

First, HOMO and LUMO energies of different DFAs and LOSC-DFAs are
compared. Tab. \ref{tab:eig} summarizes mean absolute errors (MAEs)
of orbital energies in comparison with experimental quasiparticle
energies, where self-consistent GW (scGW) \cite{Aryasetiawan1998,Holm1998}
and G0W0 \cite{Hedin1965PR,Aryasetiawan1998} results are also included
for comparison. Previousely, it has been shown that LOSC can size-consistently
improve HOMO and LUMO energies on systems range from small sized molecules
to polymers \cite{Li185203}. Here, we further calculated a set of
40 organic molecules, where the molecular size is much larger than
that of the G2-97 set tested before. Due to the serious delocalization
error \cite{Cohen08115123,Mori-Sanchez08146401}, LDA and PBE show
systematic underestimation of VIPs and overestimation of VEAs, with
MAEs larger than 2.0 eV; hybrid functional B3LYP performs slightly
better with a 20\% reduction in error, but the results still qualitatively
deviate from the experiment. LOSC-DFAs significantly improve both
HOMO and LUMO energies, with MAEs much smaller than their parent DFAs.
In particular, MAEs of LOSC-B3LYP are smaller than 0.3 eV. It is also
interesting to compare LOSC with the well-recognized scGW and G0W0
methods. We find that LOSC can achieve better accuracy than scGW and
G0W0 methods for HOMO and LUMO energy calculations. Our results also
show that starting from the same reference DFA (PBE), LOSC (MAE of
HOMO 0.37 eV and of LUMO 0.33 eV) outperforms the G0W0 (MAE of HOMO
0.51 eV and of LUMO 0.37 eV). It is well-known that the G0W0 calculation
is significantly influenced by the reference DFAs. In contrast, LOSC
can provide similar accuracy based on different parent DFAs, including
hybrid functionals.

\begin{figure}[h]
\includegraphics[width=1\columnwidth]{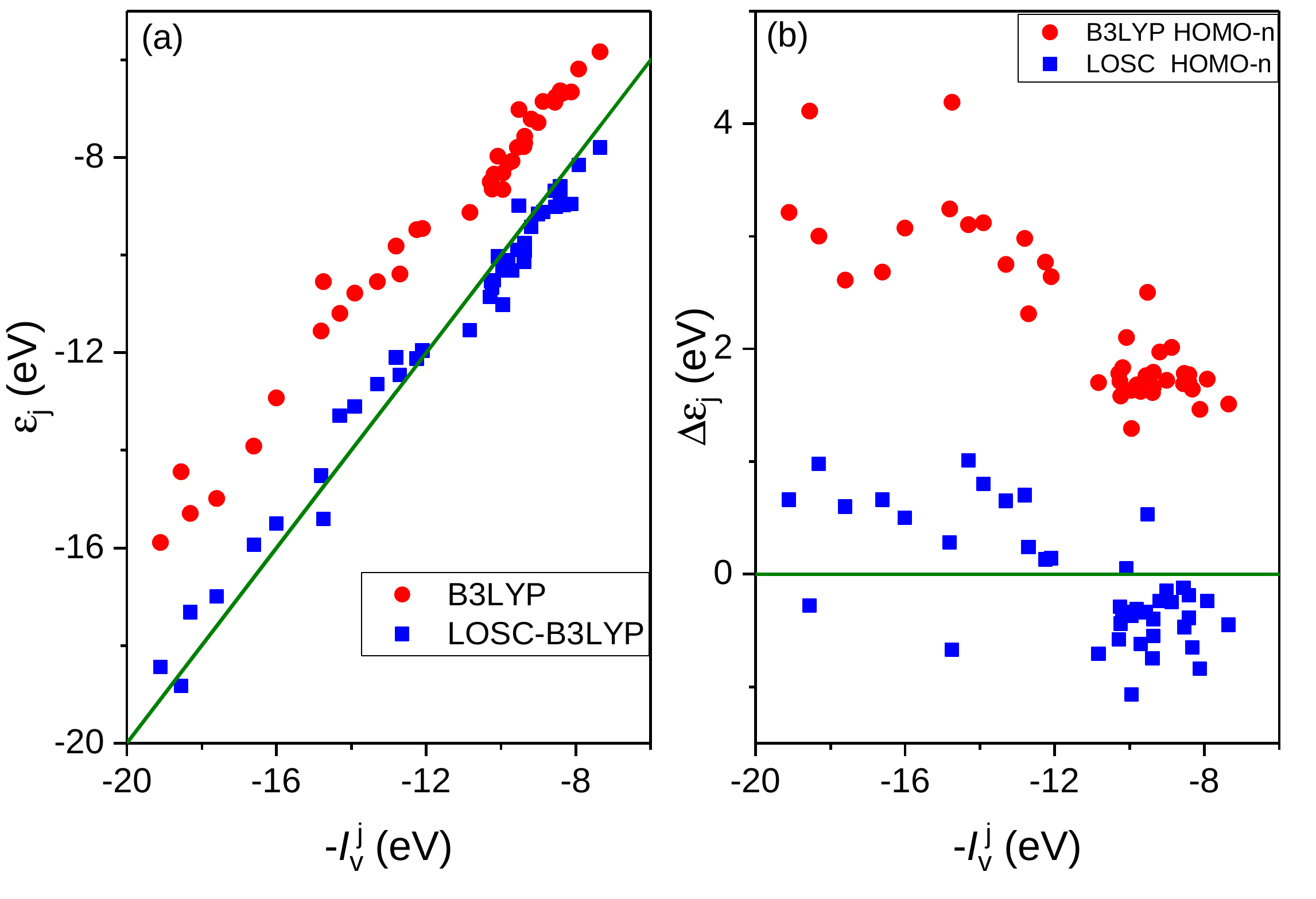} \caption{Calculated $\varepsilon_{j}$ of B3LYP and LOSC-B3LYP in comparison
with the experimental $-I_{v}^{j}$. (a) Orbital energies $\varepsilon_{j}$
for 43 states below HOMO are included. The solid line indicates $\varepsilon_{j}=-I_{v}^{j}$.
(b) The errors of calculated orbital energies with respect to the
experimental negative VIPs, $\Delta\epsilon_{j}=\epsilon_{j}+I_{v}^{j}$,
are recorded.  \label{fig1}}
\end{figure}

Besides HOMO and LUMO, Tab. \ref{tab:eig} also summarizes the results
of valence orbital energies from DFAs and LOSC-DFAs. Similarly, the
serious deviation from the experiment by commonly used DFAs can be
largely reduced by LOSC. This can be clearly seen from Fig. \ref{fig1}:
valence orbital energies of B3LYP significantly overestimate quasiparticle
energies; with LOSC, the systematic error is eliminated. By observing
Fig. \ref{fig1}(b), we find that the overestimation of quasiparticle
energies by B3LYP becomes more serious for states with lower energies,
which is corrected in LOSC-B3LYP.

\begin{figure*}[h]
\centering \subfigure[]{\includegraphics[width=0.48\linewidth]{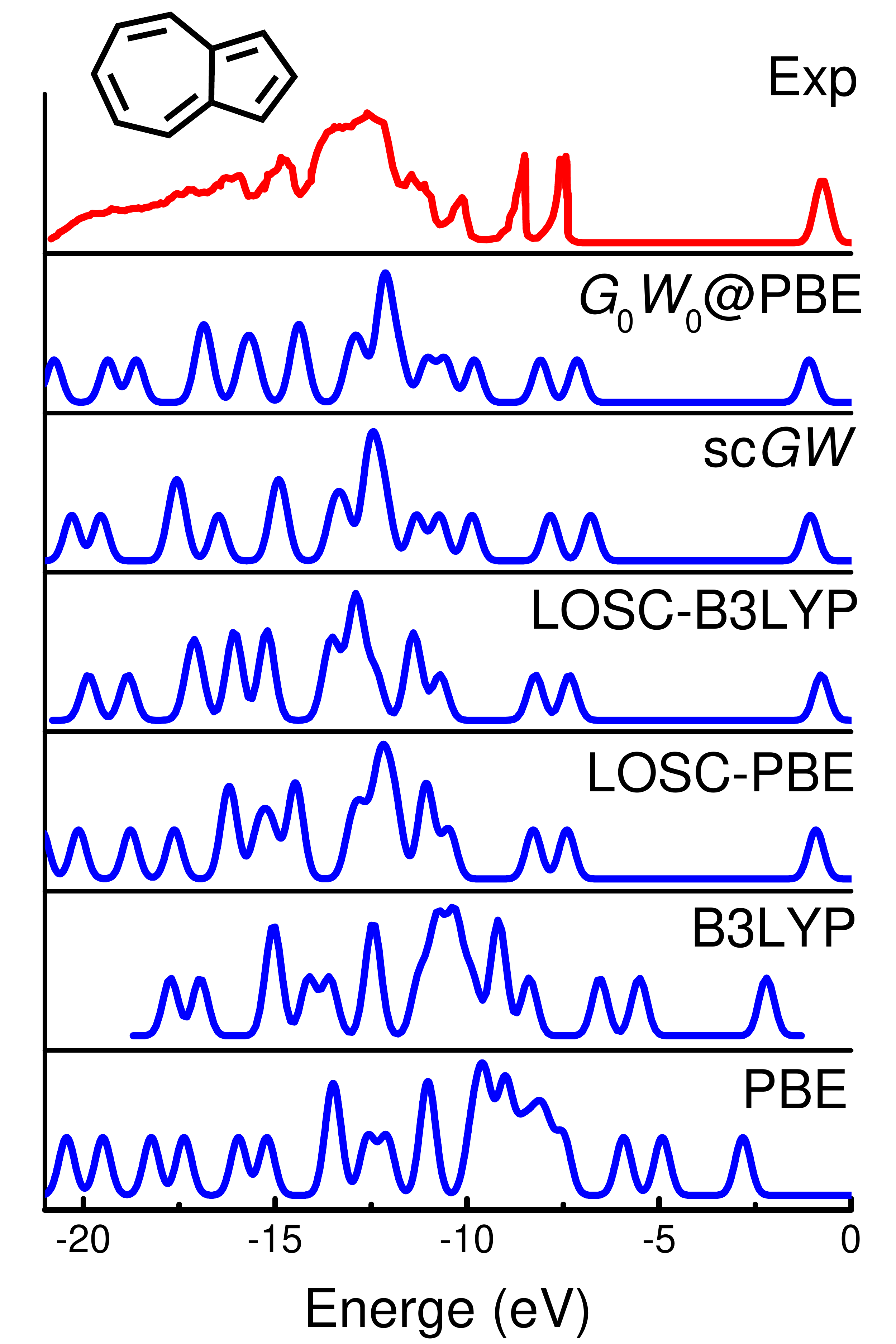}}
\subfigure[]{\includegraphics[width=0.48\linewidth]{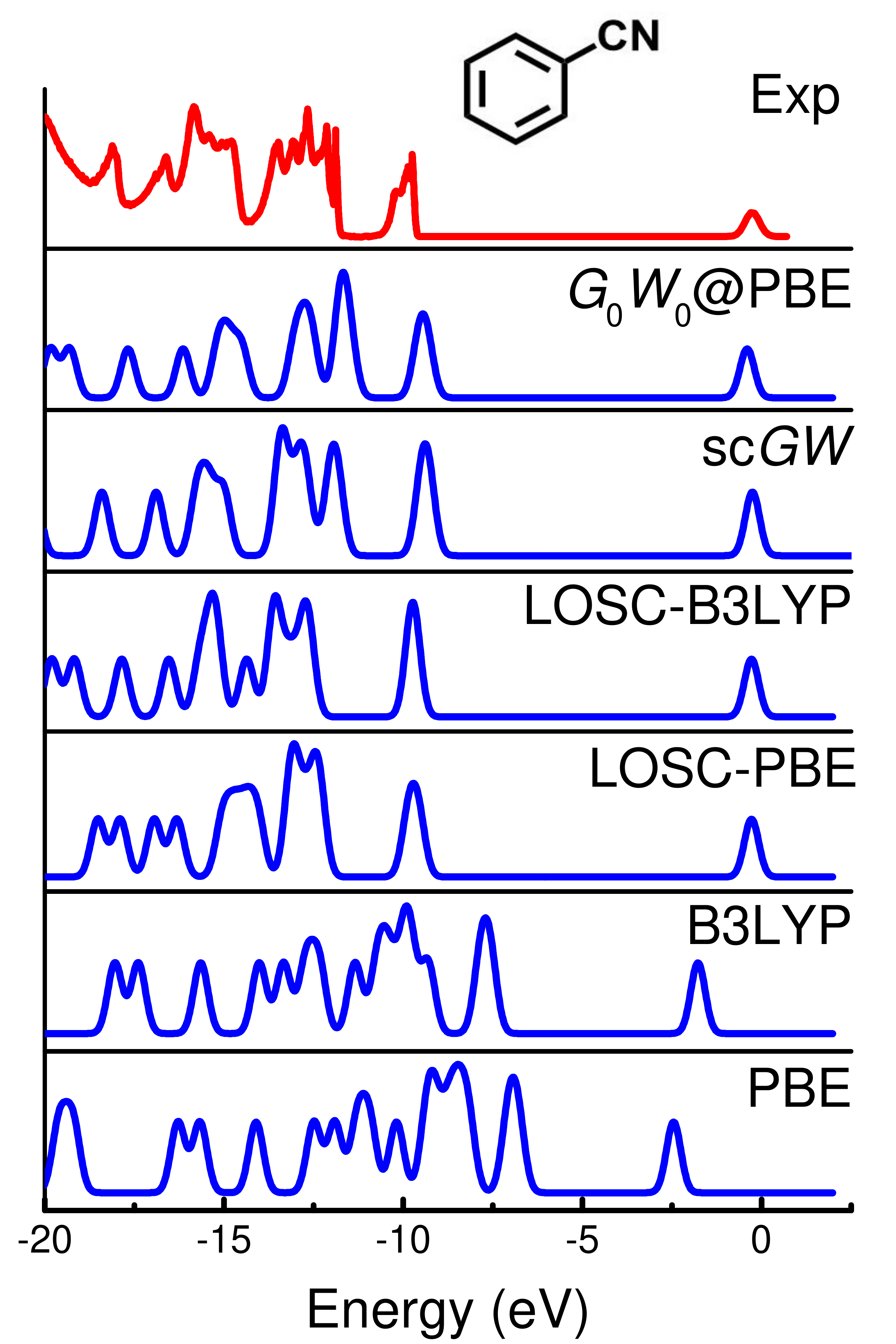}}
\caption{Photoemission spectrum of (a) azulene and (b) benzonitrile. All the
calculated spectrum are broadened by Gaussian expansion with 0.2 eV.
sc$GW$ and $G_{0}W_{0}$ results are obtained from Ref \citenum{potentialselectron}.
Experimental results are obtained from (a) Ref \citenum{dougherty1980photoelectron}
and (b) Ref \citenum{kimura1981handbook}.}
\label{fig:PES} 
\end{figure*}

To further confirm that LOSC is a reliable method for the calculation
of quasiparticle energies, GKS spectra of forty systems were plotted
and compared to the experimental photoemission spectra, along with GW results when available. 
Fig. \ref{fig:PES}
only shows the results of azulene and benzonitrile; tests on other
molecules give similar results, which can be found in SI. As can be
seen, commonly used DFAs exemplified by PBE and B3LYP give too narrow
HOMO-LUMO gaps, with the occupied levels being significantly overestimated
and LUMO energy being underestimated. LOSC greatly corrects the results.
Furthermore, spectra by LOSC-DFAs are consistent with the experimental
photoemission spectra, with the principle peaks appearing at the same
positions. Overall, LOSC shows little dependence on parent DFAs, and
can reach an accuracy that is comparable to that of GW methods in
predicting quasiparticle spectra. Note that the computational cost
of LOSC only amounts to a small portion of the parent functional,
thus it is computationally much more efficient than G0W0. Therefore,
LOSC-DFAs are a promising low-cost alternative to GW approximation for accurate
prediction of quasiparticle energies.

Accurate prediction of quasiparticle energies by LOSC-DFAs thus allows
the calculation of excitation energies from ground state DFT calculation
through Eqs. \ref{eq:e_n-1} and \ref{eq:e_n+1}. Because anionic
systems ($(N+1)$-electron systems) are generally difficult to
converge to right states, here we only discuss the excitation energies
calculated from $(N-1)$-electron systems; some results from $(N+1)$-electron
systems  can be found in SI. For the excitations of HOMO to orbitals
above HOMO, starting from the doublet ground state of $(N-1)$-electron
systems (assuming one more $\alpha$-spin electron than $\beta$-spin
electrons), there are two orbital energies of different spins for
each orbital above HOMO. Apparently, $\alpha$-spin orbital energies
should be used for triplet excitations. For singlet excitations, a
spin purification process similar to Refs. \citenum{ziegler1977calculation}
and \citenum{kowalczykassessment2011} is used here, and the excitation
energies are calculated by 
\begin{equation}
\Delta E_{m}^{{\rm singlet}}\left(N\right)\approx[2\varepsilon_{m}^{\beta}\left(N-1\right)-\varepsilon_{m}^{\alpha}\left(N-1\right)]-\varepsilon_{{\rm HOMO}}^{\beta}\left(N-1\right).\label{eq:e_singlet}
\end{equation}

\begin{table}[htpb]
\centering \caption{Mean absolute errors (MAEs, in eV) and mean sign errors (MSEs, in
eV) of 48 low-lying excitation energies obtained from HF, DFT, TDDFT
and $\Delta$SCF-B3LYP calculation on 16 molecules. Notation T1 refers
to triplet HOMO to LUMO excitation, and T2 refers to triplet HOMO
to LUMO+1 excitation. The analogy notation for S1 and S2 which stand
for singlet excitations. Reference data were obtained from Ref \citenum{schreiber2008benchmarks}.}
\label{tab:error} %
\begin{tabular}{lcccccccccc}
\toprule 
\multirow{2}{*}{Method}  & \multicolumn{2}{c}{T1} & \multicolumn{2}{c}{T2} & \multicolumn{2}{c}{S1} & \multicolumn{2}{c}{S2} & \multicolumn{2}{c}{Total}\tabularnewline
 & MAE  & MSE  & MAE  & MSE  & MAE  & MSE  & MAE  & MSE  & MAE  & MSE \tabularnewline
\midrule 
HF  & 1.08  & -0.88  & 2.04  & -1.23  & 1.12  & -0.59  & 1.49  & 0.81  & 1.35  & -0.83 \tabularnewline
BLYP  & 0.19  & -0.14  & 0.63  & -0.10  & 0.68  & -0.65  & 0.65  & -0.24  & 0.53  & -0.22 \tabularnewline
B3LYP  & 0.17  & -0.13  & 0.43  & 0.01  & 0.45  & -0.33  & 0.67  & -0.58  & 0.42  & -0.01 \tabularnewline
LDA  & 0.24  & -0.02  & 0.65  & 0.04  & 0.73  & -0.68  & 0.70  & -0.27  & 0.58  & -0.16 \tabularnewline
LOSC-BLYP  & 0.49  & -0.28  & 0.46  & -0.37  & 0.84  & -0.84  & 0.62  & 0.10  & 0.63  & -0.44 \tabularnewline
LOSC-B3LYP  & 0.30  & -0.23  & 0.28  & -0.14  & 0.60  & -0.51  & 0.69  & -0.29  & 0.49  & -0.19 \tabularnewline
LOSC-LDA  & 0.48  & -0.18  & 0.52  & -0.27  & 0.88  & -0.88  & 0.71  & 0.11  & 0.67  & -0.42 \tabularnewline
TD-B3LYP  & 0.45  & -0.45  & 0.39  & -0.39  & 0.38  & -0.35  & 0.28  & 0.27  & 0.38  & -0.37 \tabularnewline
$\Delta$-SCF & 0.20  & -0.16  & 0.33  & -0.24  & 0.56  & -0.56  & 0.18  & 0.04  & 0.35  & -0.31 \tabularnewline
\bottomrule
\end{tabular}
\end{table}

The results of 48 low-lying excitation energies obtained from different
DFAs and LOSC-DFAs are summarized in Tab. \ref{tab:error}, where
triplet and singlet excitations are categorized and presented. The
results from Hartree Fock (HF), TDDFT and $\Delta$SCF-DFT with B3LYP
functional are also listed for comparison. As expected, LOSC-DFAs
can provides good prediction for excitation energies due to their
excellent performance on quasiparticle energies. Especially, the total
MAE and MSE of LOSC-B3LYP are 0.49 eV and -0.19 eV, which are comparable
to TDDFT (MAE of 0.38 eV and MSE of -0.37 eV) and $\Delta$SCF-DFT
(MAE of 0.35 eV and MSE of -0.31 eV, based on the same reference DFA
(B3LYP). For conventional DFAs, it is surprising to find that they
have very good performance on predicting low-lying excitation energies,
even though they perform poorly in quasiparticle energy calculations.
These good results should be attributed to the fact that unoccupied
(or occupied) orbitals that are energetically close suffer from a
similar amount of systematic delocalization error, making the error
cancellation when calculating excitation energies from the difference
of orbital energies. This can be seen clearly by comparing their performance
on the T1 (HOMO-LUMO excitation) and T2 (HOMO-(LUMO+1) excitation).
Conventional DFAs tested here perform very well on T1 excitation (MAEs
are around 0.2 eV), but their performance on T2 excitation is much
worse (MAEs can be larger than 0.6 eV). In contrast, LOSC-DFAs are
consistent in their performance for these two types of excitations.
Thus, it can be inferred that for a DFT method to achieve good accuracy
for the prediction of excitation energies of low- to high-lying states,
it is necessary to provide consistently reliable quasiparticle energies
for all different states involved.

\begin{table}[h]
\centering \caption{Mean absolute errors (MAEs, in eV) and mean sign errors (MSEs, in
eV) with respect to experimental reference of excitation energies
of 4 atoms from low-lying states to Rydberg states. 12 excitations
were included for each atom. Experimental values were obtained from
Ref \citenum{NISTASD}.}
\label{tab:Rydberg} %
\begin{tabular}{@{}lccccccc@{}}
\toprule 
 &  & LDA  & BLYP  & B3LYP  & LOSC-LDA  & LOSC-BLYP  & LOSC-B3LYP \tabularnewline
\midrule 
Be singlet\textsuperscript{\emph{a}}  & MAE  & 2.37  & 1.15  & 1.85  & 0.24  & 0.54  & 0.35 \tabularnewline
 & MSE  & 2.37  & -1.15  & 1.85  & 0.07  & -0.29  & -0.06 \tabularnewline
Be triplet\textsuperscript{\emph{a}}  & MAE  & 2.30  & 1.91  & 1.79  & 0.28  & 0.60  & 0.30 \tabularnewline
 & MSE  & 2.11  & 1.68  & 1.79  & -0.04  & -0.60  & -0.29 \tabularnewline
Mg singlet\textsuperscript{\emph{b}}  & MAE  & 2.37  & 2.07  & 1.69  & 0.55  & 0.26  & 0.21 \tabularnewline
 & MSE  & 2.37  & 2.07  & 1.69  & 0.55  & 0.16  & 0.21 \tabularnewline
Mg triplet\textsuperscript{\emph{b}}  & MAE  & 2.13  & 1.82  & 1.54  & 0.40  & 0.15  & 0.14 \tabularnewline
 & MSE  & 2.12  & 1.80  & 1.52  & 0.34  & -0.11  & 0.06 \tabularnewline
\midrule 
Li doublet\textsuperscript{\emph{a}}  & MAE  & 0.97  & 1.77  & 1.40  & 0.91  & 0.17  & 0.16 \tabularnewline
 & MSE  & 0.97  & 1.77  & 1.40  & -0.89  & 0.04  & -0.03 \tabularnewline
Na doublet\textsuperscript{\emph{b}}  & MAE  & 1.52  & 2.16  & 1.69  & 0.25  & 0.57  & 0.42 \tabularnewline
 & MSE  & 1.52  & 2.16  & 1.69  & -0.11  & 0.57  & 0.42 \tabularnewline
\bottomrule
\end{tabular}

\raggedright \textsuperscript{\emph{a}} The excitation states are
calculated up to atomic orbital 6p.\\
 \textsuperscript{\emph{b}} The excitation states are calculated
up to atomic orbital 7p. 
\end{table}

To further confirm the above inference, four atoms (Li, Be, Mg, and
Na) are selected as an atomic set to test their excitation energies
up to Rydberg states. Table \ref{tab:Rydberg} summarizes the MAEs
from different DFAs and LOSC-DFAs applied to this atomic test set,
more detailed results can be found in SI. As can be seen, conventional
DFAs show large MAEs for all the four atoms. By observing Tabs. S7
to S12 in SI, it is easy to find that the higher the excited states,
the greater the deviation between the results obtained by DFAs and
the experimental values. This is because conventional DFAs show larger
errors for quasiparticle energies at higher states, thus the difference
of orbital energies cannot completely offset the systematic delocalization errors
of orbitals that are energetically far apart. In contrast, LOSC-DFAs
perform similarly for different excited states with very high accuracy,
which should be attributed to the good performance of LOSC on quasiparticle
energies of different states.

In conclusion, we have carried out a comprehensive test on calculations
of quasiparticle energies and excitation energies with the LOSC functional
and DFAs. Through a large number of comparisons with experimental
results and GW results, we demonstrated that LOSC-DFAs shows little
dependence on parent DFAs, and can reach an accuracy that is better
or comparable to that of GW methods in predicting quasiparticle spectra.
This also leads to the calculations of excitation energies of the
$N$-electron systems from ground state calculations of the
$(N-1)$-electron systems. Commonly used DFAs show good performance
for valence excitations, but not accurate for higher energy and Rydberg
states; in contrast, LOSC-DFAs can provide consistently accurate results
for excitation energies from low-lying to Rydberg states. This work
highlights the pathway to quasiparticle and excitation energies from
ground density functional calculations.

Note. When preparing the manuscript for submission, we became aware
of Ref. \citenum{Haiduke2018jcp}, which also calculated excitation
energies from orbital energy differences of the $(N-1)$-electron
systems. Different functionals from our tests and only valence excitations
were reported.

\begin{acknowledgement}
Support from the National Institutes of Health (Grant No. R01 GM061870-13)
(WY), the Center for Computational Design of Functional Layered Materials
(Award DE-SC0012575), an Energy Frontier Research Center funded by
the US Department of Energy, Office of Science, Basic Energy Sciences.
(YM, NQS), and the National Science Foundation (CHE-1362927) (CL)
is appreciated. The authors were grateful to Dr. Noa Marom for providing
detailed data from Ref. \citenum{potentialselectron}.
\end{acknowledgement}

\begin{suppinfo}
\begin{itemize}
\item SI.pdf: More details of computations and test results. 
\end{itemize}
\end{suppinfo}

\bibliography{ref}
 
\end{document}


\section{Description of LOSC calculation}
If not specified, the LOSC calculation in this paper is carried out as a post self-consistent (post-SCF) correction to the
parent functional results. In particularly, we first perform conventional DFT calculations
with the parent functional to obtain the canonical orbitals $\{\varphi_i\}$ and the orbital energies $\{\epsilon_i\}$.
Then with the restrained Boys localization (see Ref \citenum{li2017localized} for details), we obtain the localized orbitals $\{\phi_i\}$.
Finally the energy correction and orbital energy corrections are given by
\begin{equation}
    \Delta E^{\rm LOSC} = \sum_{ij} \frac{1}{2}\kappa_{ij}\lambda_{ij}(\delta_{ij}-\lambda_{ij}),
\end{equation}
and
\begin{equation}
   \Delta \epsilon_i = \sum_j \kappa_{jj} \left( \frac{1}{2}-\lambda_{jj} \right) |U_{ji}|^2
   - \sum_{j\neq l} \kappa_{jl} \lambda_{jl} U_{ji} U_{li}^\ast.  \label{delta-ep}
\end{equation}
One can also perform self-consistent field (SCF) calculation for the LOSC, however, it has been demonstrated in
Ref \citenum{li2017localized} that for small and compact molecules, the SCF only slightly differs from the post-SCF results,
especially for the orbital energy calculations. So in this paper, we stick with the post-SCF calculations, which is
computationally more efficient without much sacrifice in the accuracy of the results.
The double integrals in the curvature formula have been evaluated using the resolution of identity (or density fitting) technique
\cite{dunlap1979some,vahtras1993integral}.

\section{Photoemission spectrum}
Figure S1 - S40 show the photoemission spectrum (PES) of 40 test molecules. Most the test molecules
were from Blase's \cite{blase2011first} and Marom's \cite{potentialselectron} test set.
In addition, polyacene (n= 1 -6) and two big systems (\ce{C60} and \ce{C70}) were studied as well for interest.
Experimental PES were reproduced from literature as reference, if they were applicable. Experimental
electron affinity (see clarification of Table \ref{tab:LUMO_eig} for the data source) was broadened with Gaussian
expansion with 0.2 eV to plot a peak in the experimental spectrum. Quasi-particle energies from
sc$GW$ and $G_0W_0$@PBE were obtained from Ref \citenum{potentialselectron} for Marom's test set and used to
plot PES for comparison. For orbital energies from DFT, conventional functional (B3LYP and PBE) and LOSC functional
(LOSC-B3LYP and LOSC-PBE) were applied for calculation. cc-pVTZ were used as basis set, if not specified.
To obtain PES from $GW$ and DFT, the orbital energies were used and broadened with Gaussian expansion
by 0.2 eV for all the test cases.

\begin{figure}
    \includegraphics[width=0.7\textwidth]{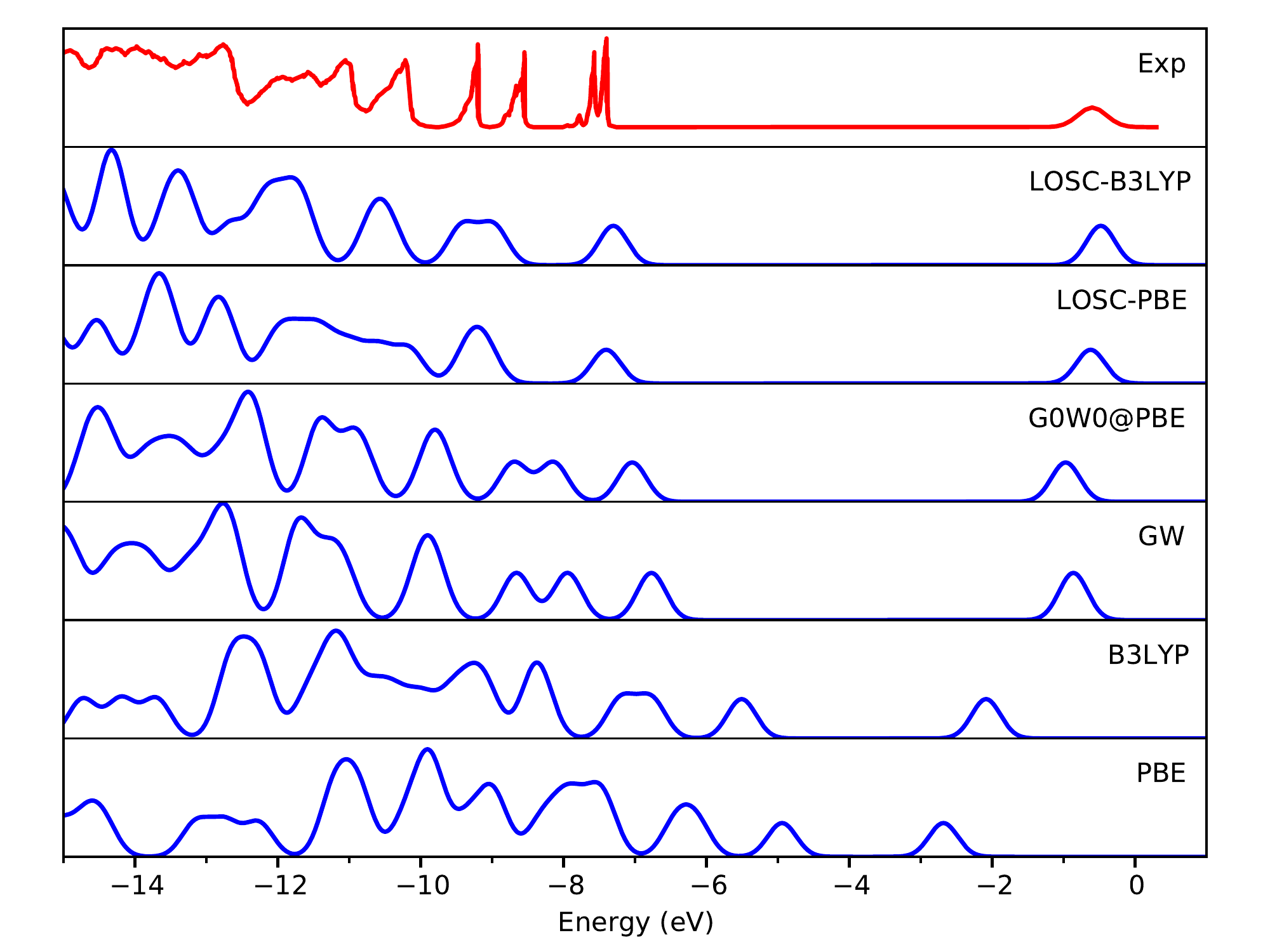}
    \caption{Photoemission spectrum of anthracene. Experimental spectrum was
             obtained from Ref \citenum{schmidt1977photoelectron}}
\end{figure}

\begin{figure}
    \includegraphics[width=0.7\textwidth]{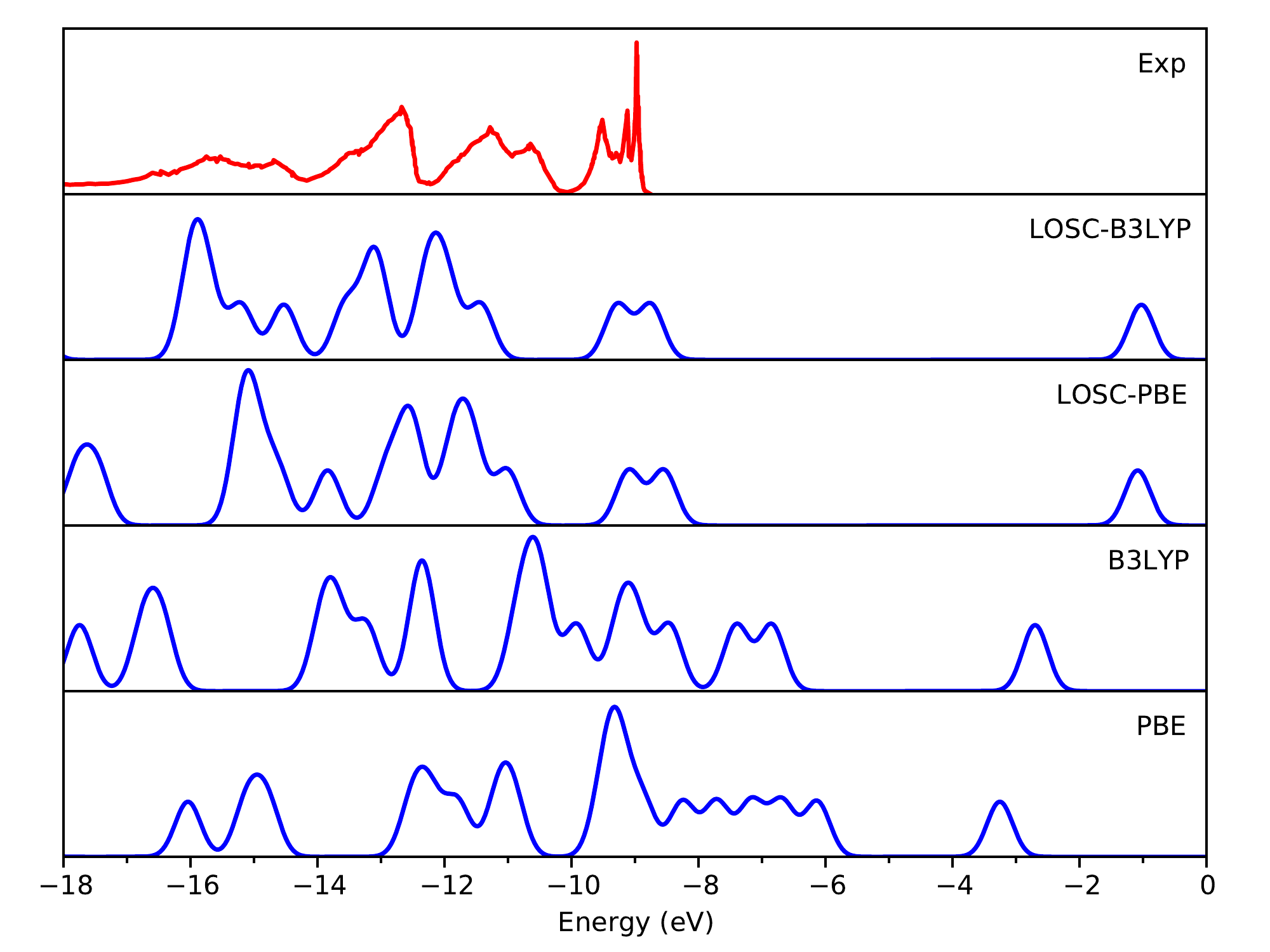}
    \caption{Photoemission spectrum of benzothiadiazole. Experimental spectrum was
             obtained from Ref \citenum{clark1973photoelectron}}
\end{figure}

\begin{figure}
    \includegraphics[width=0.7\textwidth]{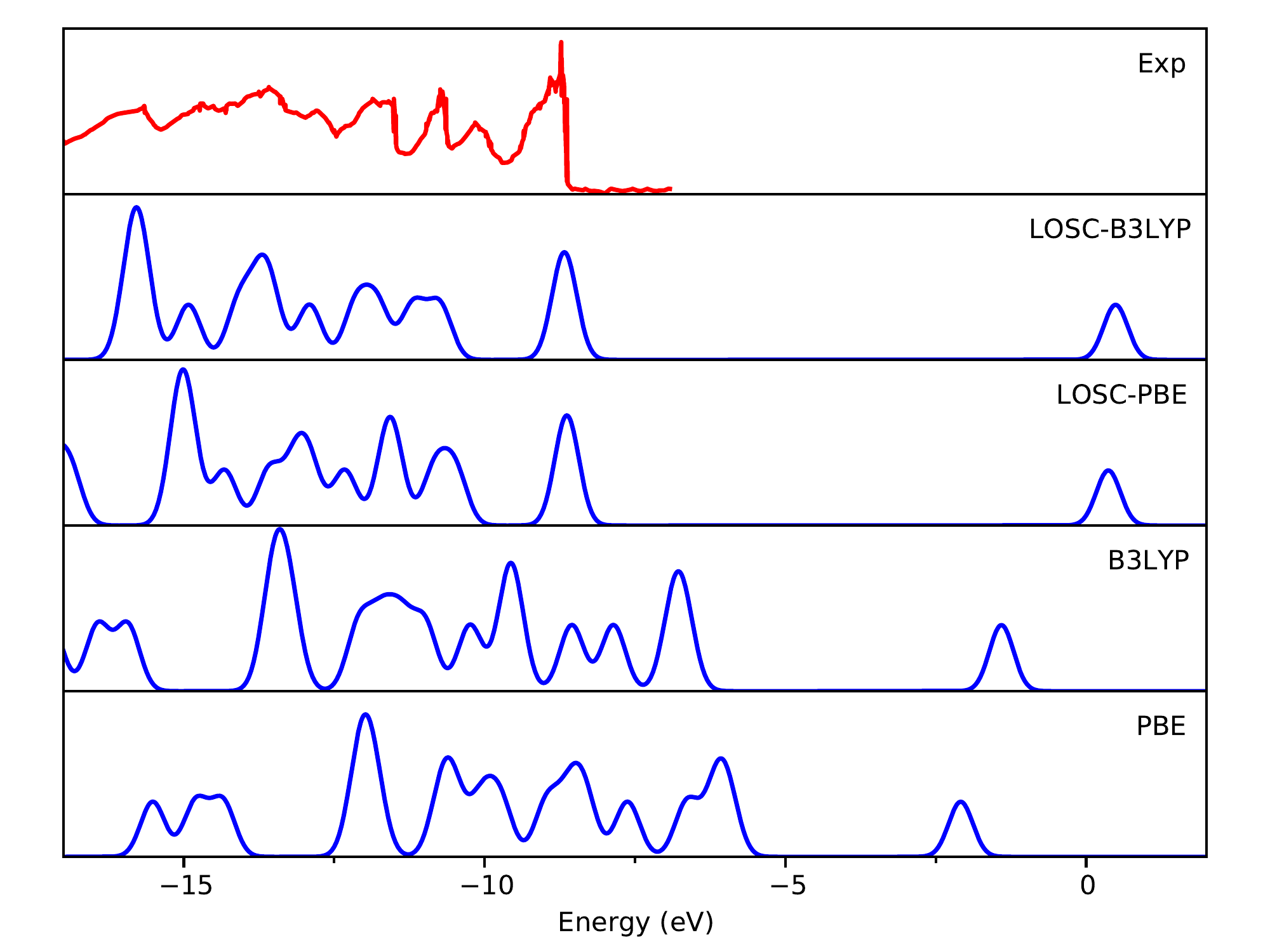}
    \caption{Photoemission spectrum of benzothiazole.
             Experimental spectrum was obtained from Ref \citenum{eland1969photoelectron}}
\end{figure}

\begin{figure}
    \includegraphics[width=0.7\textwidth]{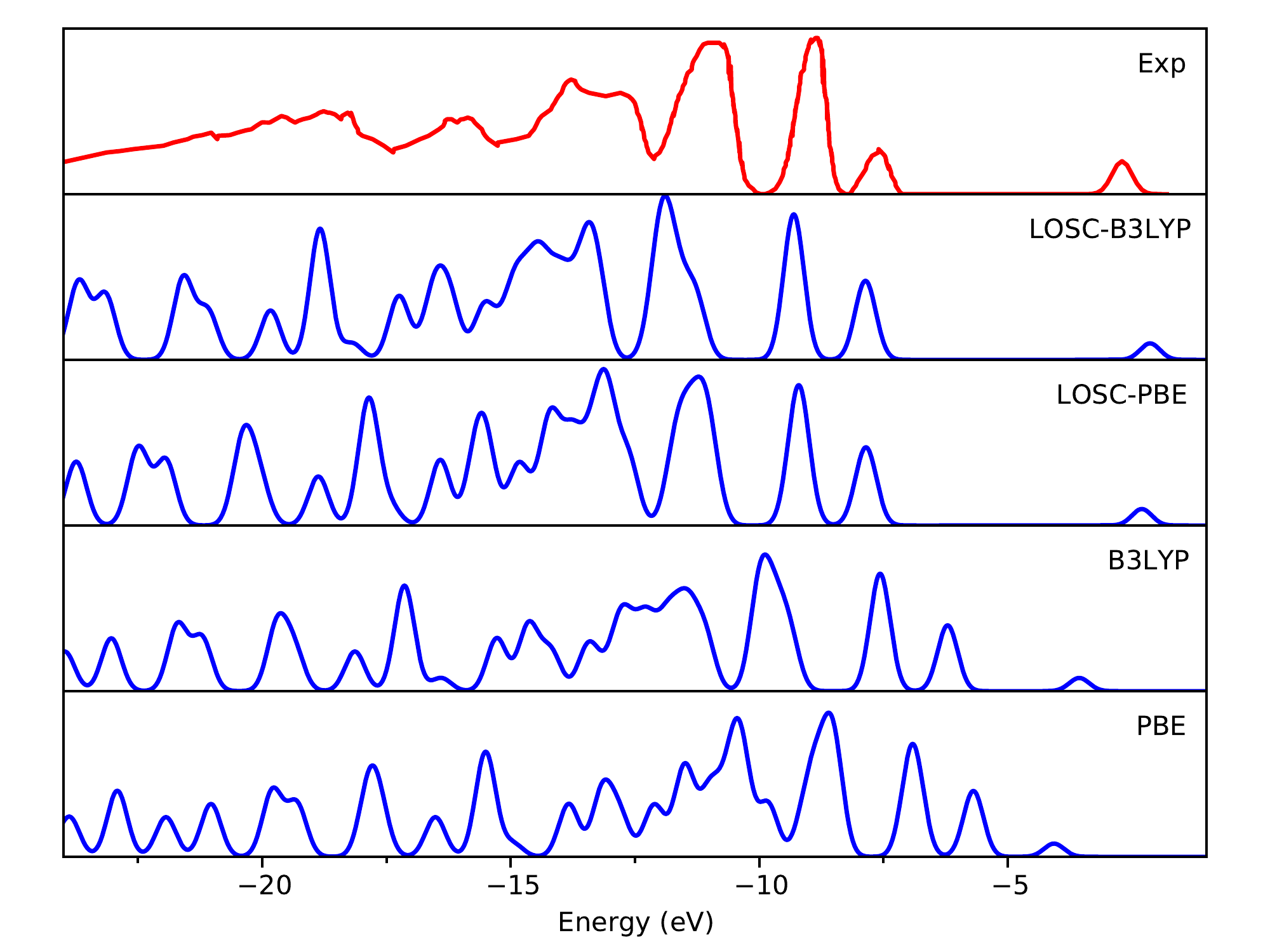}
    \caption{Photoemission spectrum of \ce{C60}.
             Experimental spectrum was obtained from Ref \citenum{liebsch1996photoelectron}}
\end{figure}

\begin{figure}
    \includegraphics[width=0.7\textwidth]{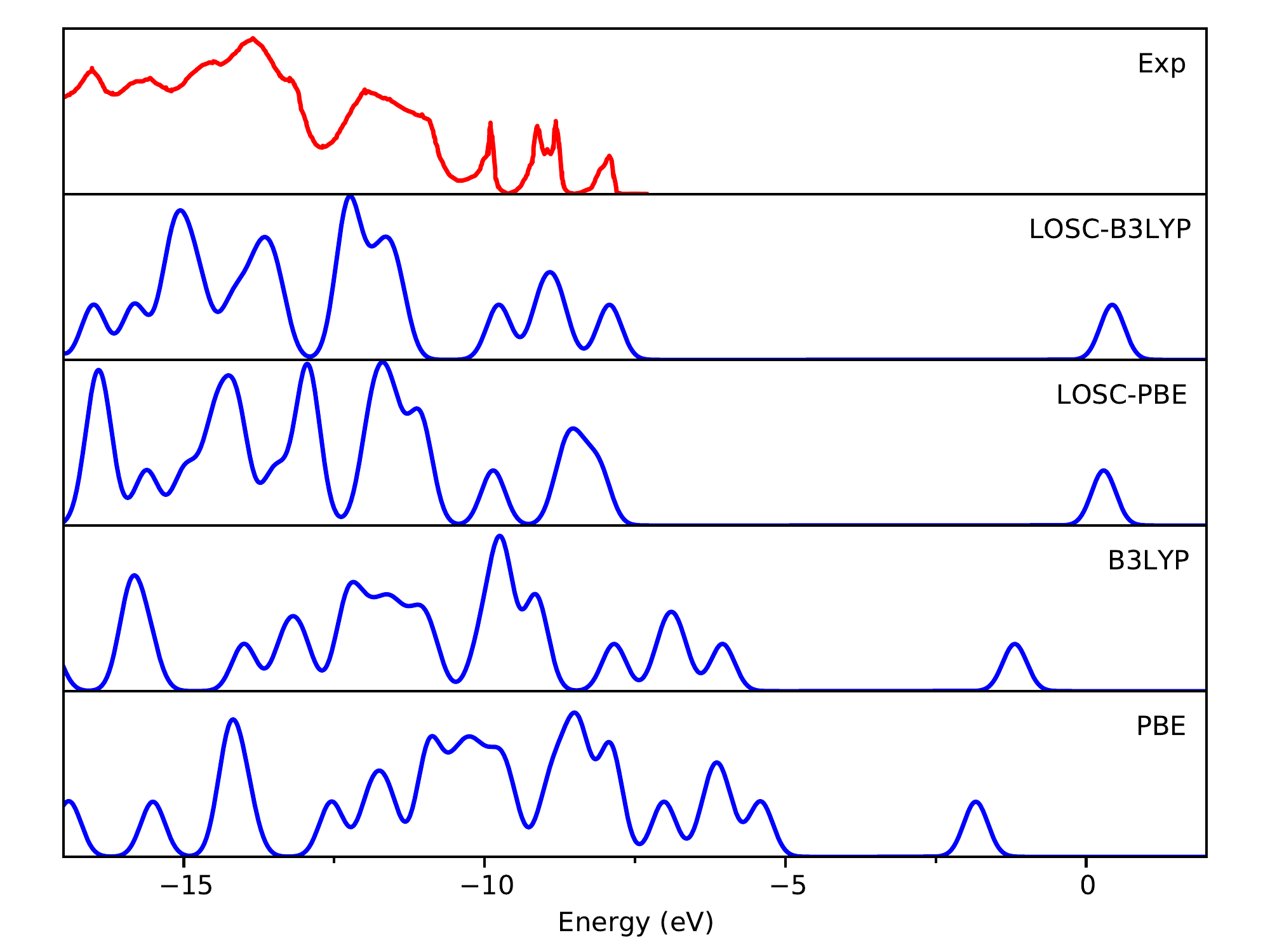}
    \caption{Photoemission spectrum of fluorene.
             Experimental spectrum was obtained from Ref \citenum{ruscic1978photoelectron}}
\end{figure}

\begin{figure}
    \includegraphics[width=0.7\textwidth]{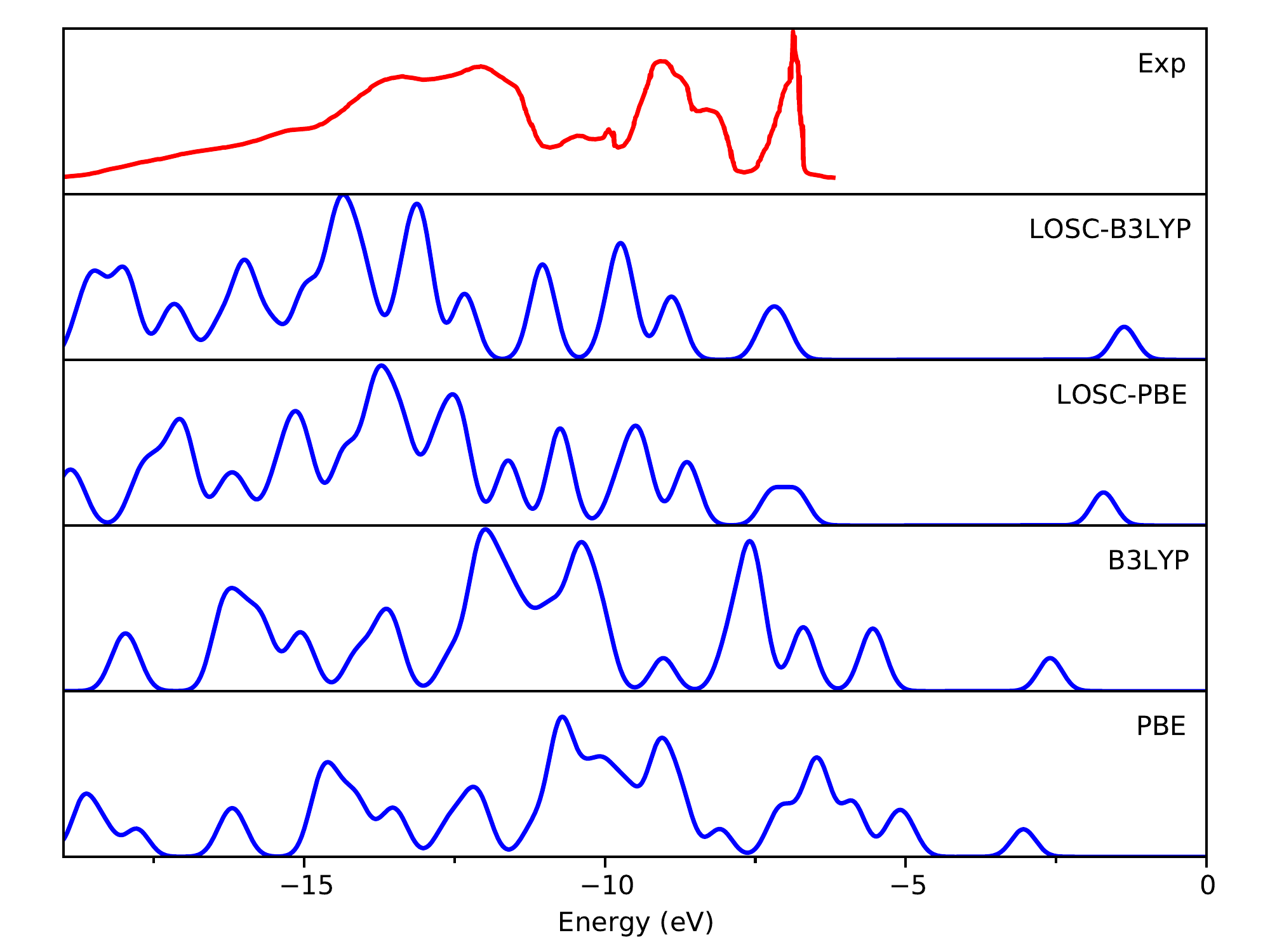}
    \caption{Photoemission spectrum of \ce{H2P}.
             Experimental spectrum was obtained from Ref \citenum{dupuis1980very}}
\end{figure}

\begin{figure}
    \includegraphics[width=0.7\textwidth]{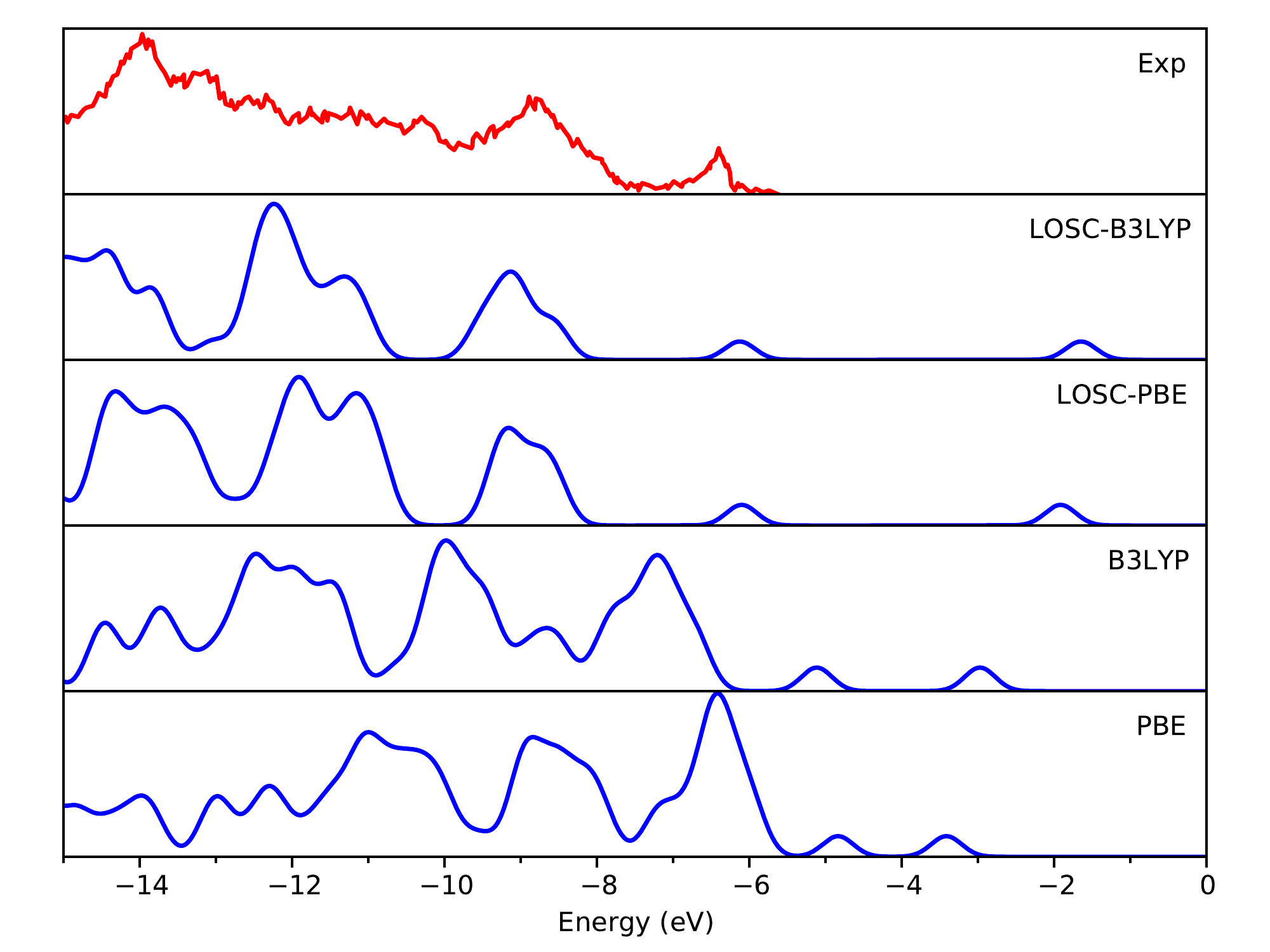}
    \caption{Photoemission spectrum of \ce{H2PC}.
             Experimental spectrum was obtained from Ref \citenum{berkowitz1979photoelectron}}
\end{figure}

\begin{figure}
    \includegraphics[width=0.7\textwidth]{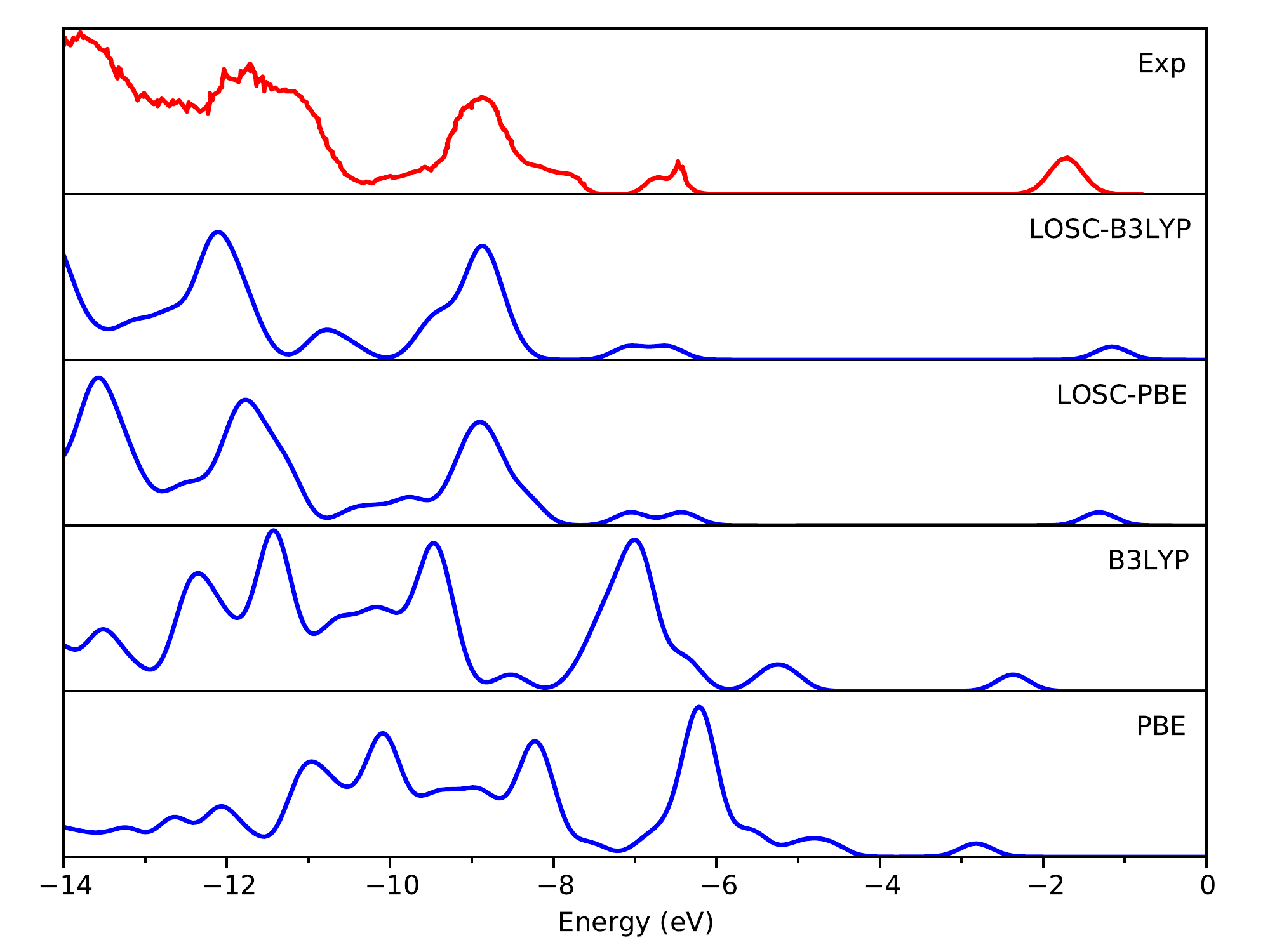}
    \caption{Photoemission spectrum of \ce{H2TPP}.
             Experimental spectrum was obtained from Ref \citenum{gruhn1999reevaluation}}
\end{figure}

\begin{figure}
    \includegraphics[width=0.7\textwidth]{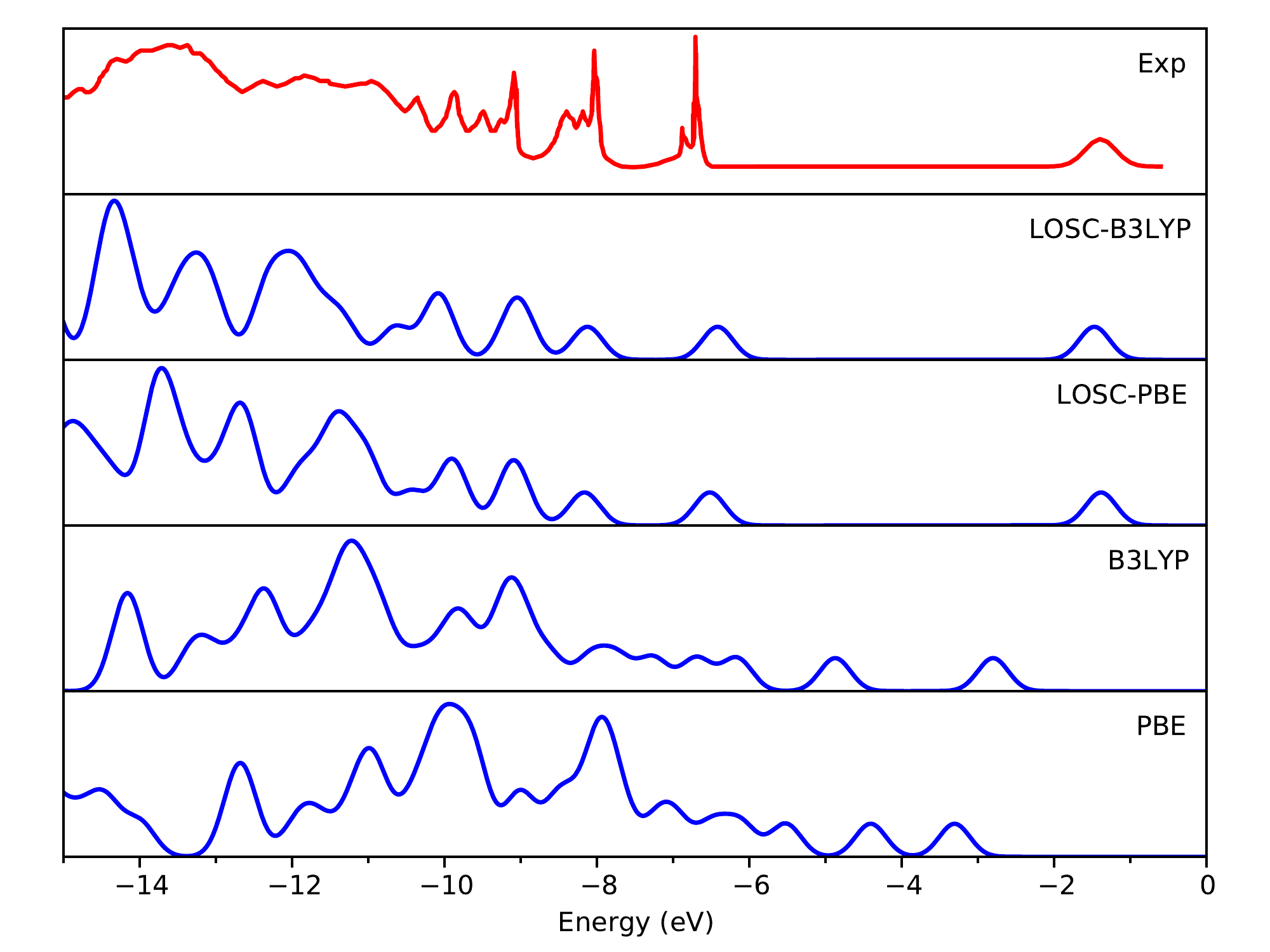}
    \caption{Photoemission spectrum of pentacene.
             Experimental spectrum was obtained from Ref \citenum{boschi1972photoelectron}}
\end{figure}

\begin{figure}
    \includegraphics[width=0.7\textwidth]{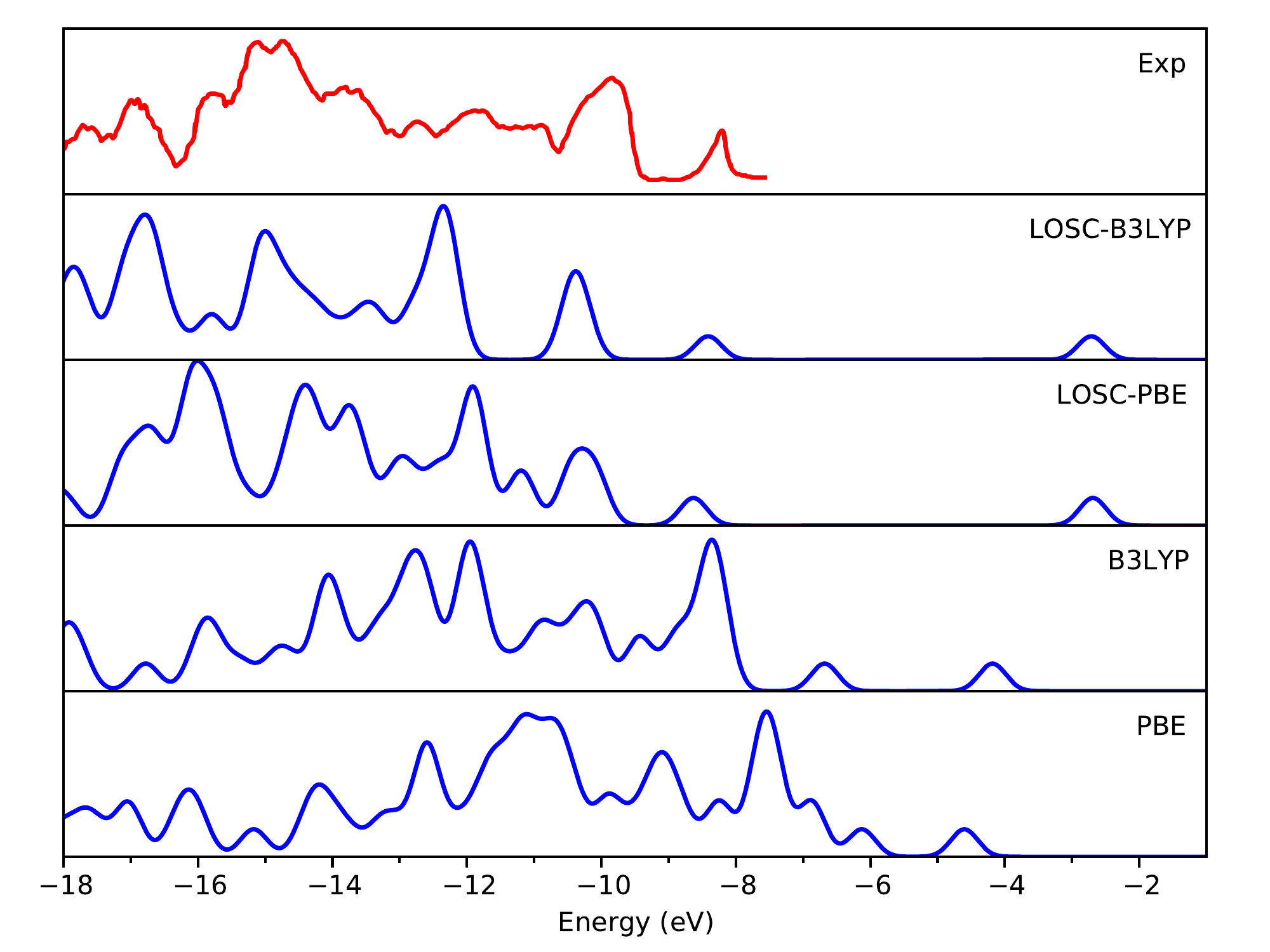}
    \caption{Photoemission spectrum of PTCDA.
             Experimental spectrum was obtained from Ref \citenum{dori2006valence}}
\end{figure}

\begin{figure}
    \includegraphics[width=0.7\textwidth]{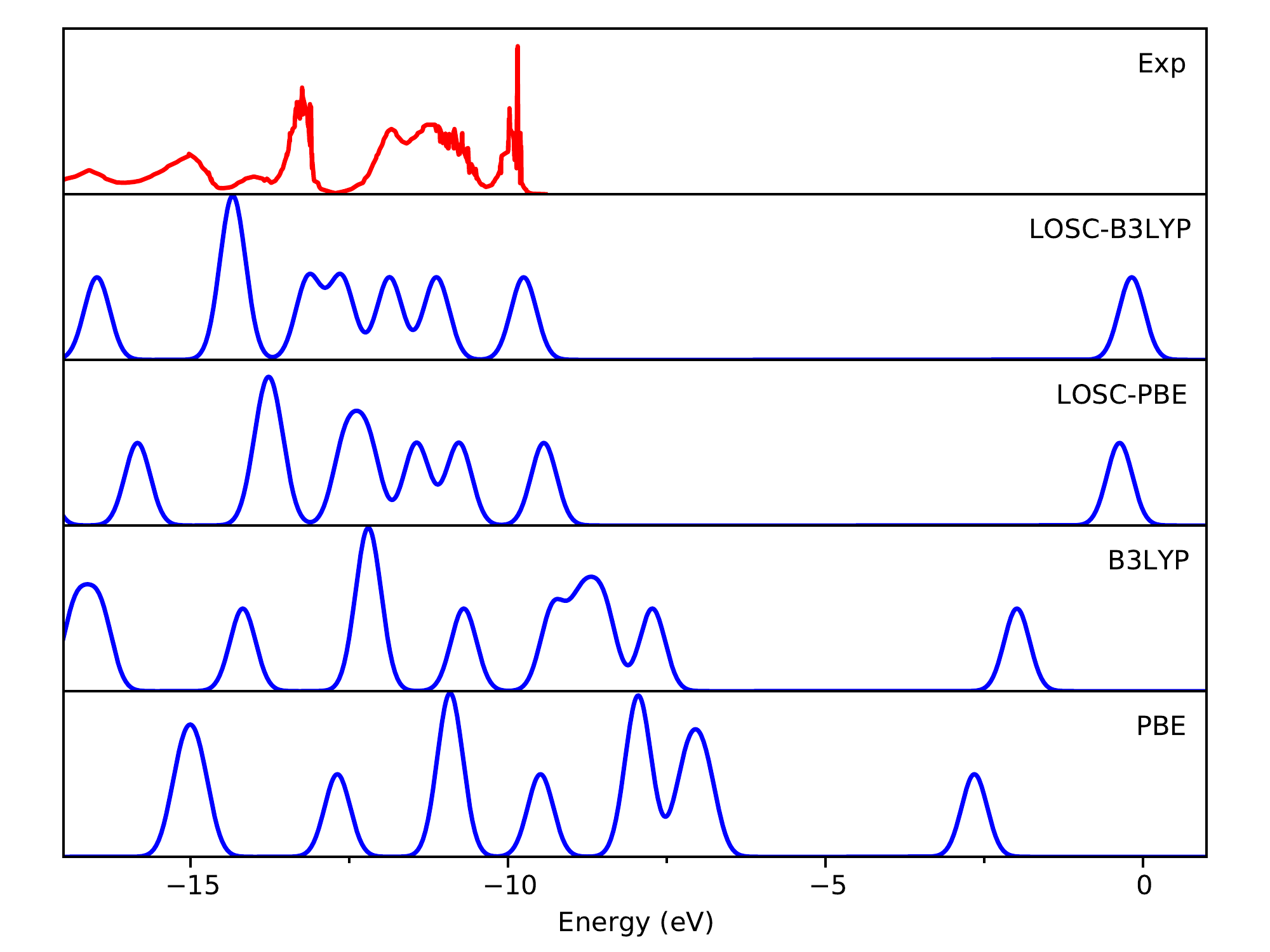}
    \caption{Photoemission spectrum of thiadiazole.
             Experimental spectrum was obtained from Ref \citenum{palmer1977electronic}}
\end{figure}

\begin{figure}
    \includegraphics[width=0.7\textwidth]{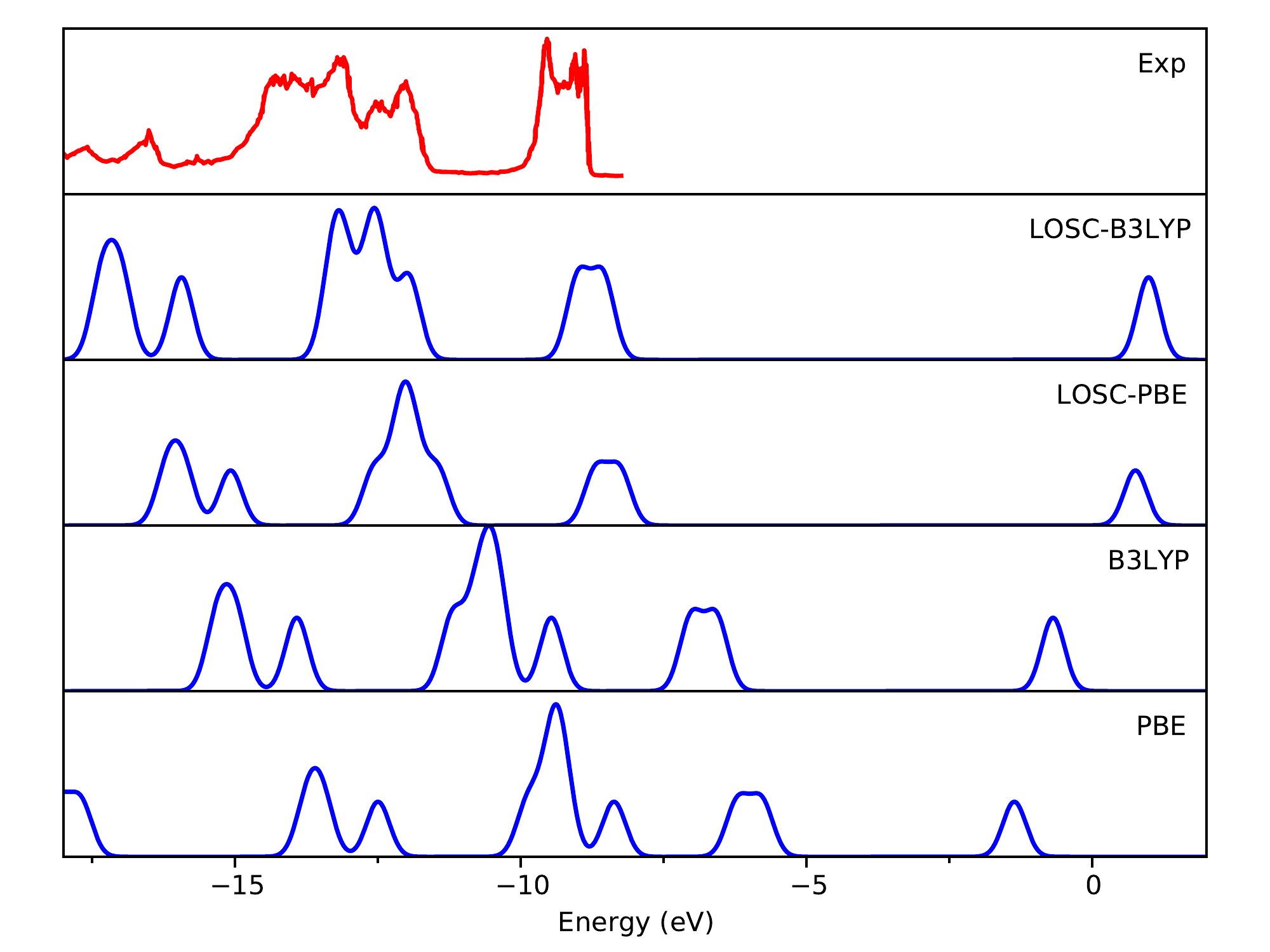}
    \caption{Photoemission spectrum of thiphene.
             Experimental spectrum was obtained from Ref \citenum{schafer1973reversal}}
\end{figure}

\begin{figure}
    \includegraphics[width=0.7\textwidth]{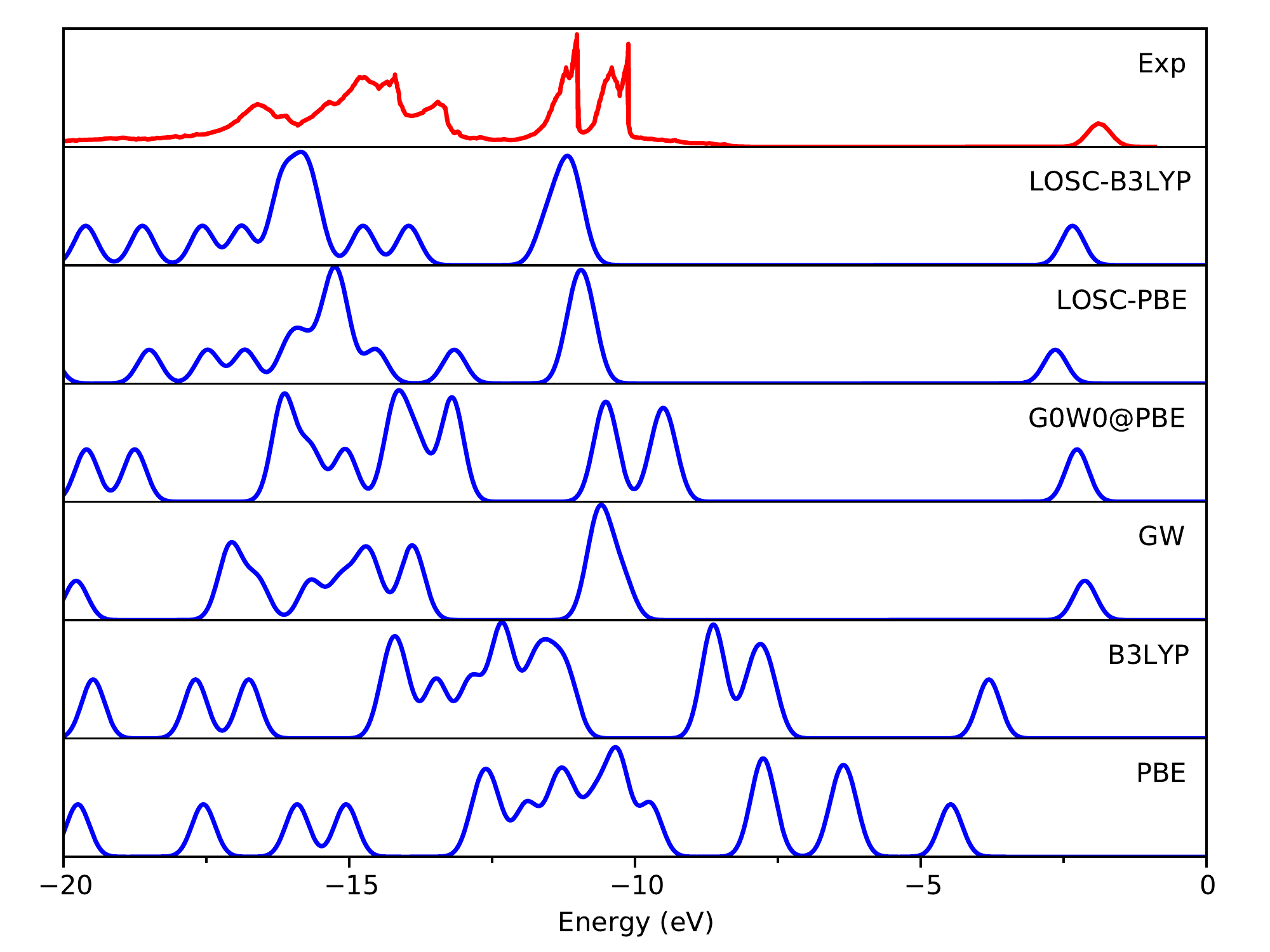}
    \caption{Photoemission spectrum of benzoquinone.
             Experimental spectrum was obtained from Ref \citenum{brundle1972perfluoro}}
\end{figure}

\begin{figure}
    \includegraphics[width=0.7\textwidth]{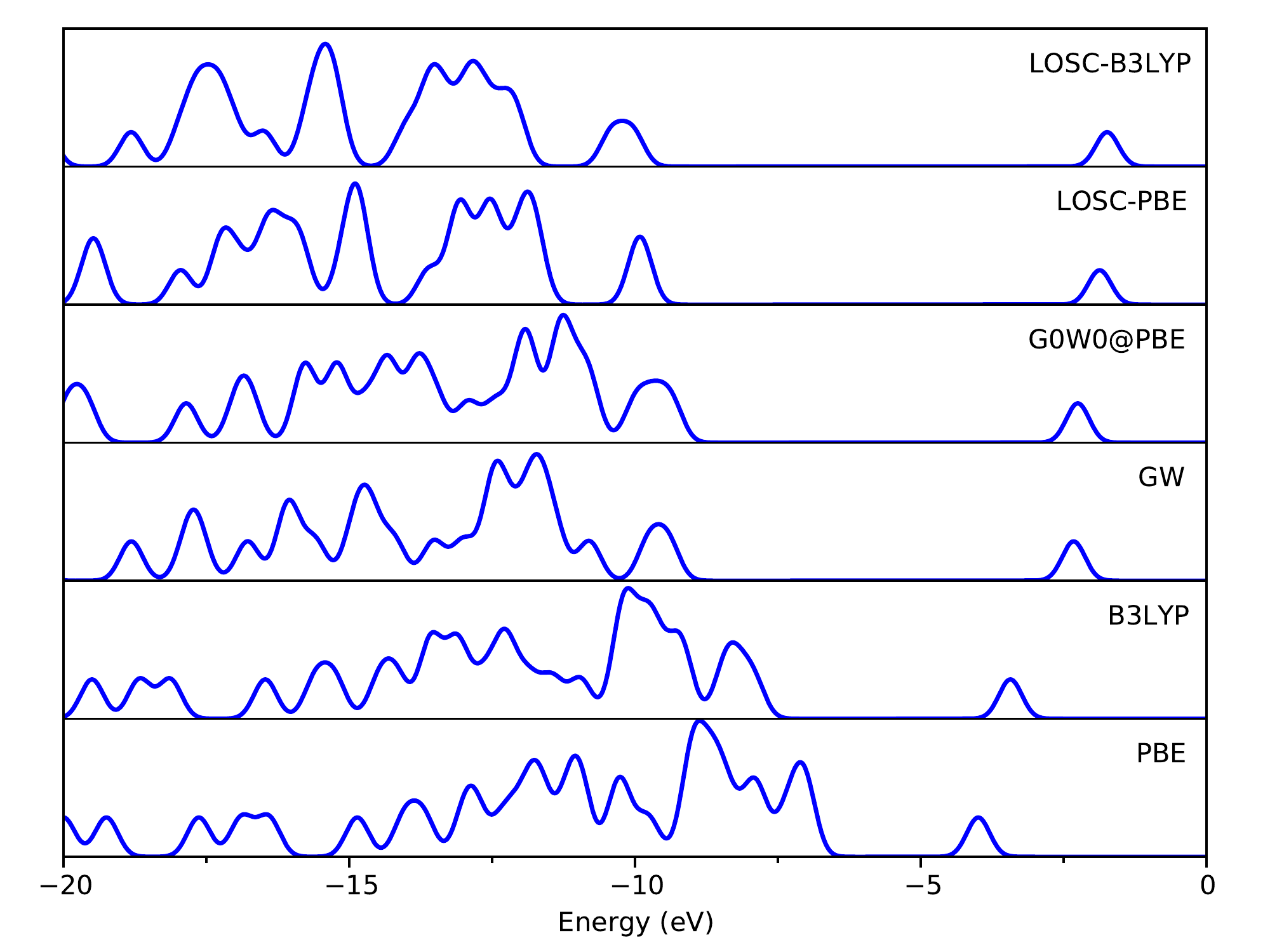}
    \caption{Photoemission spectrum of \ce{Cl4}-isobenzofuranedione.}
\end{figure}

\begin{figure}
    \includegraphics[width=0.7\textwidth]{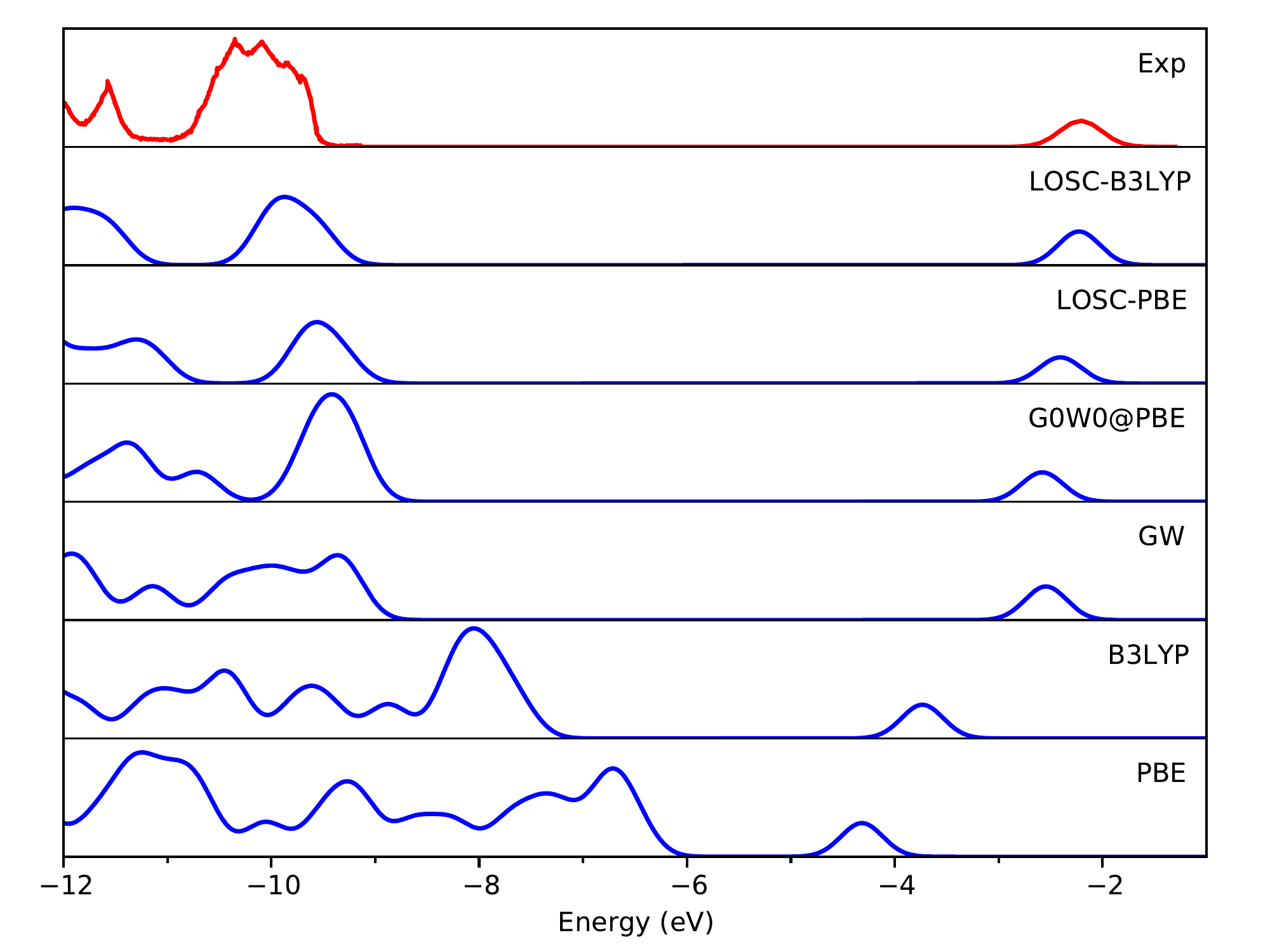}
    \caption{Photoemission spectrum of dichlone.
             Experimental spectrum was obtained from Ref \citenum{kimura1981handbook}}
\end{figure}

\begin{figure}
    \includegraphics[width=0.7\textwidth]{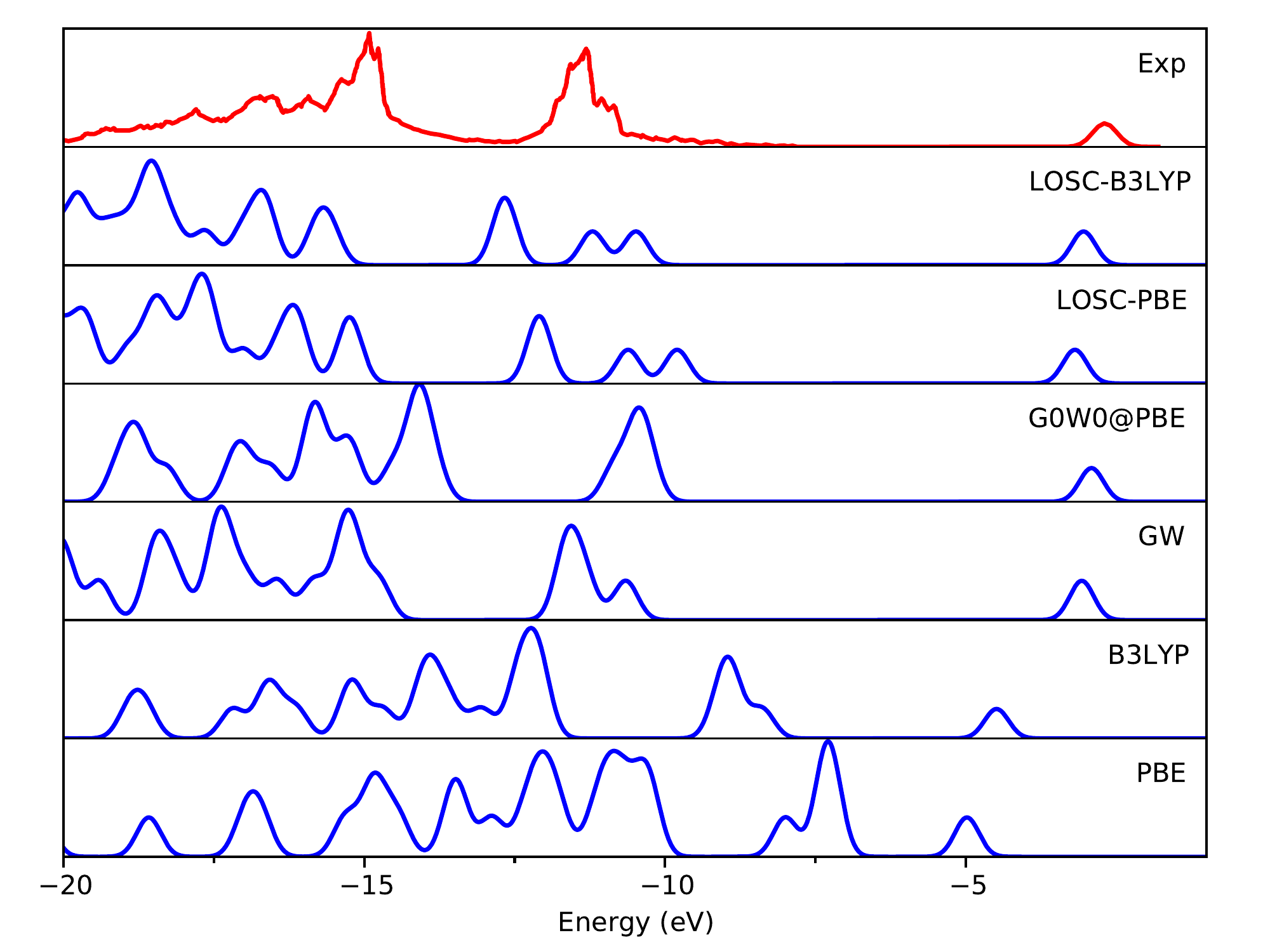}
    \caption{Photoemission spectrum of \ce{F4}-benzoquinone.
             Experimental spectrum was obtained from Ref \citenum{brundle1972perfluoro}}
\end{figure}

\begin{figure}
    \includegraphics[width=0.7\textwidth]{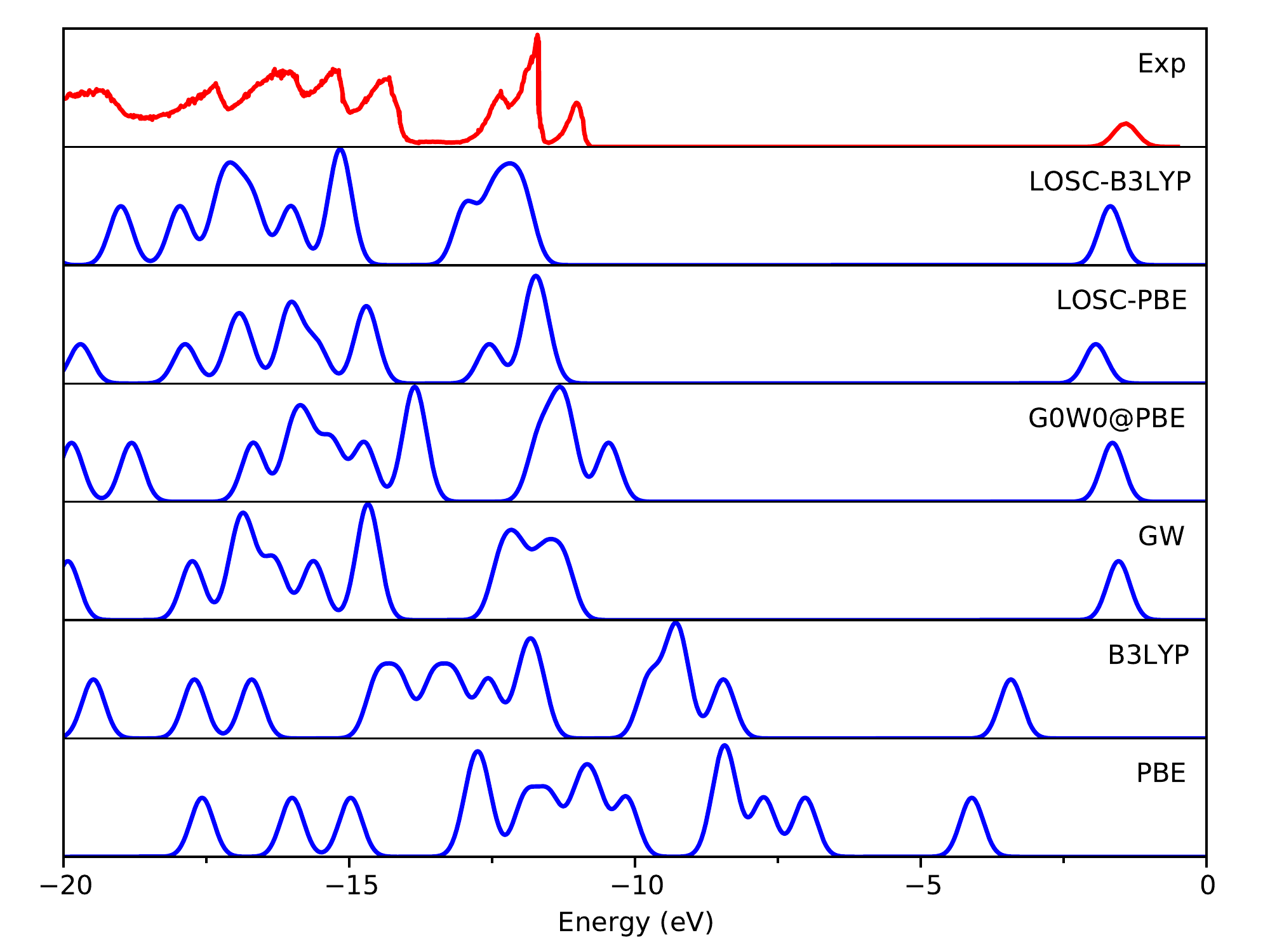}
    \caption{Photoemission spectrum of maleic anhydride.
             Experimental spectrum was obtained from Ref \citenum{potentialselectron}}
\end{figure}

\begin{figure}
    \includegraphics[width=0.7\textwidth]{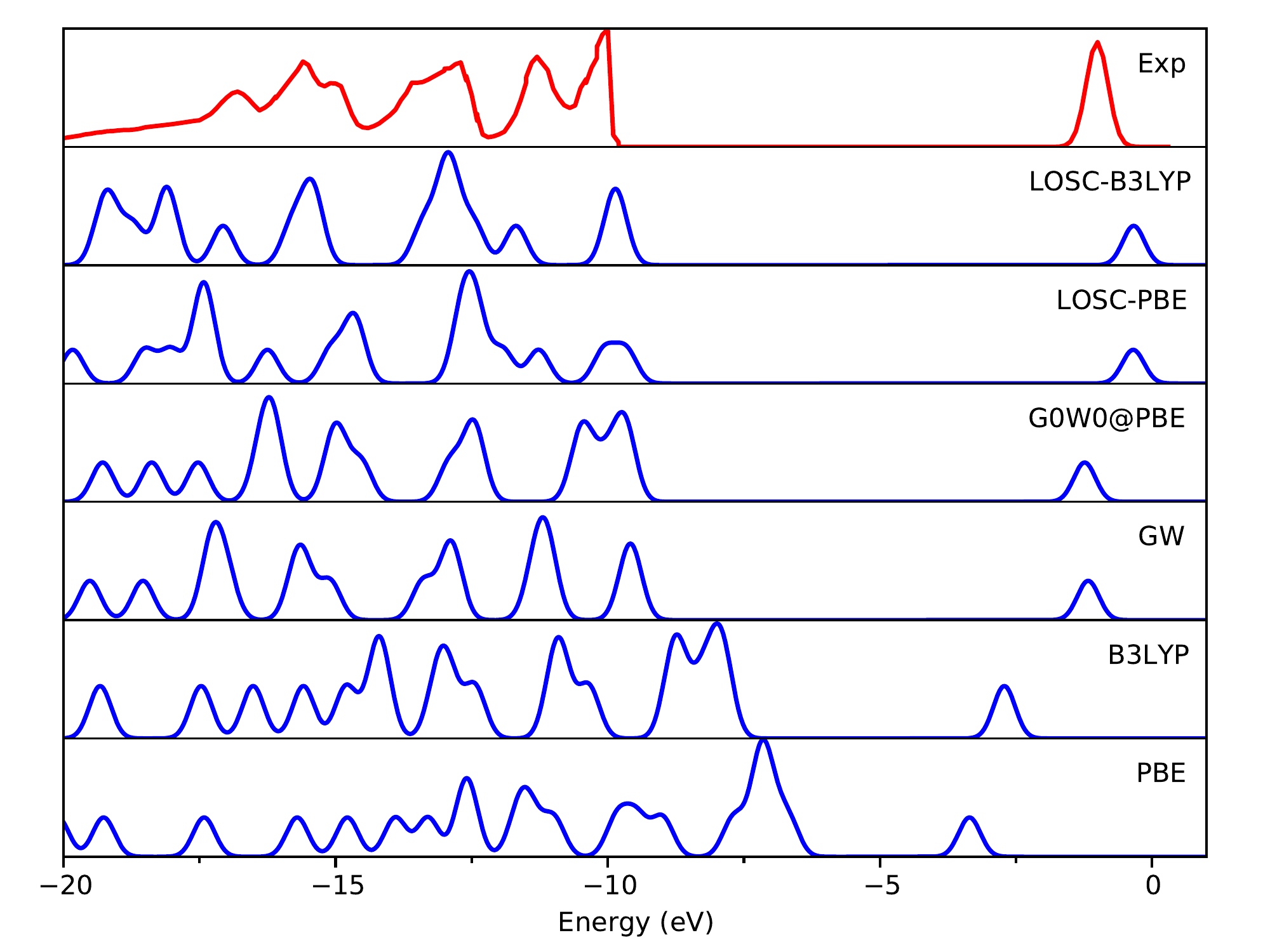}
    \caption{Photoemission spectrum of nitrobenzene.
             Experimental spectrum was obtained from Ref \citenum{rabalais1972photoelectron}}
\end{figure}

\begin{figure}
    \includegraphics[width=0.7\textwidth]{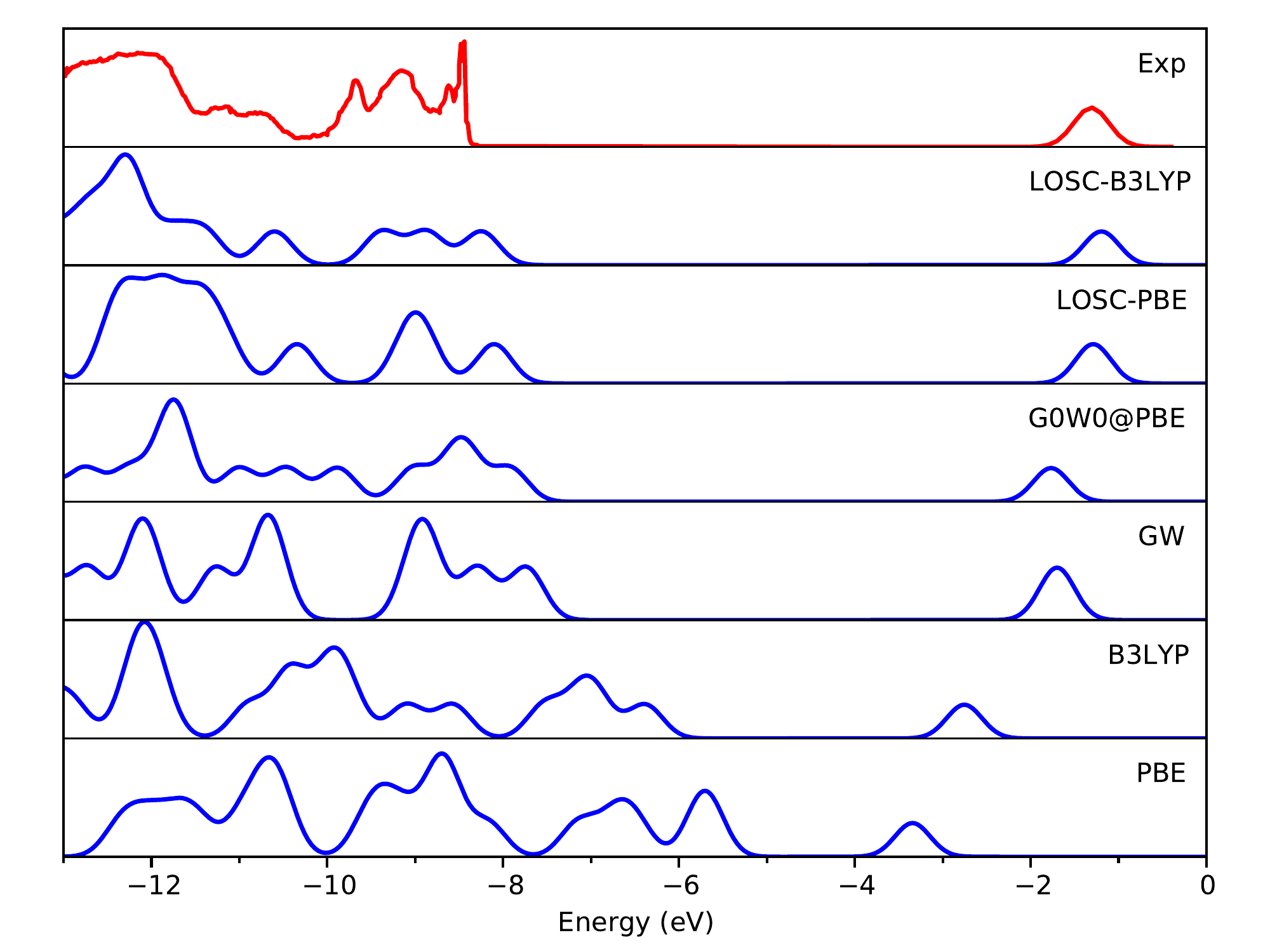}
    \caption{Photoemission spectrum of phenazine.
             Experimental spectrum was obtained from Ref \citenum{maier1975ionisation}}
\end{figure}

\begin{figure}
    \includegraphics[width=0.7\textwidth]{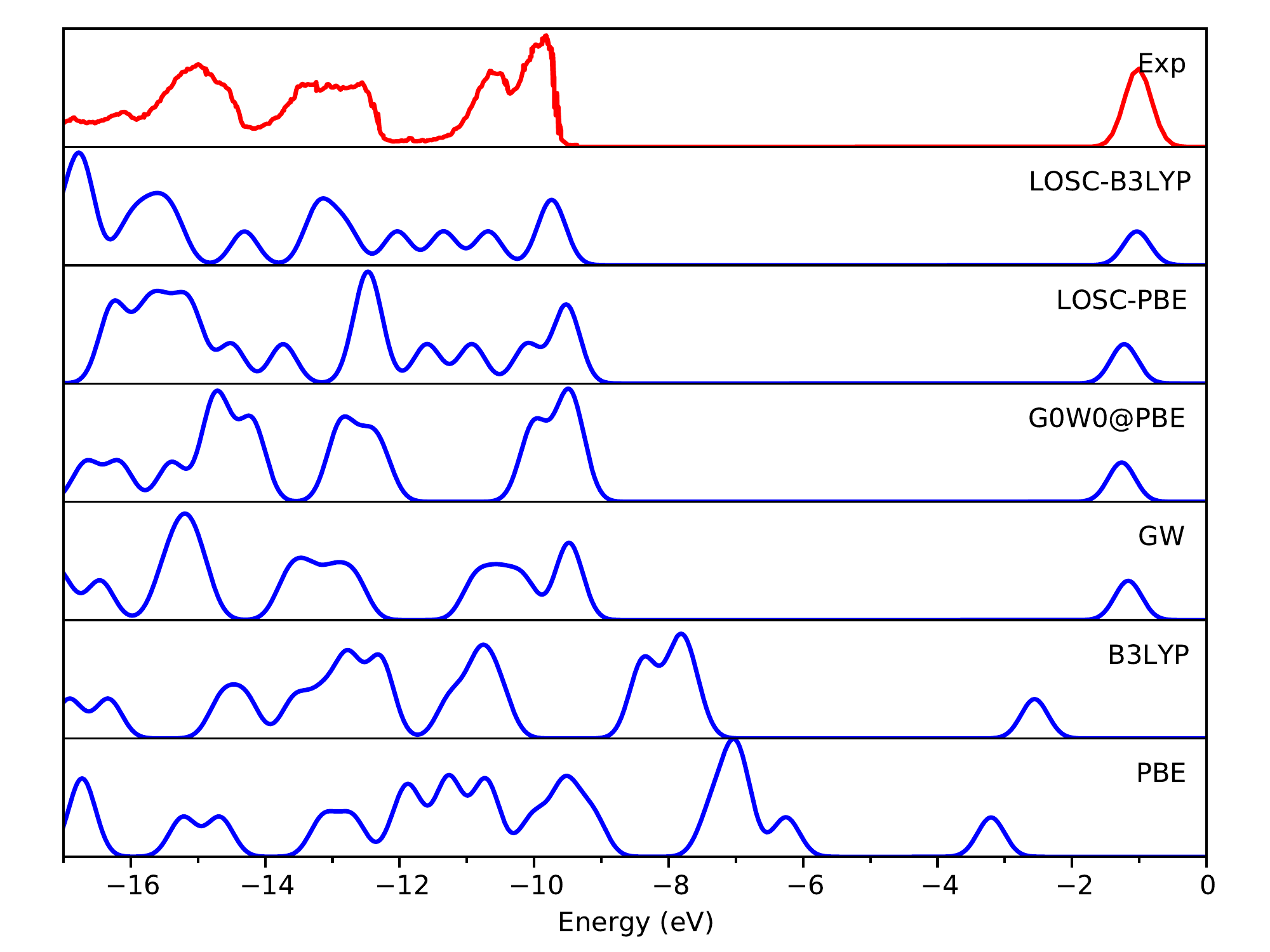}
    \caption{Photoemission spectrum of phthalimide.
             Experimental spectrum was obtained from Ref \citenum{galasso1977photoelectron}}
\end{figure}

\begin{figure}
    \includegraphics[width=0.7\textwidth]{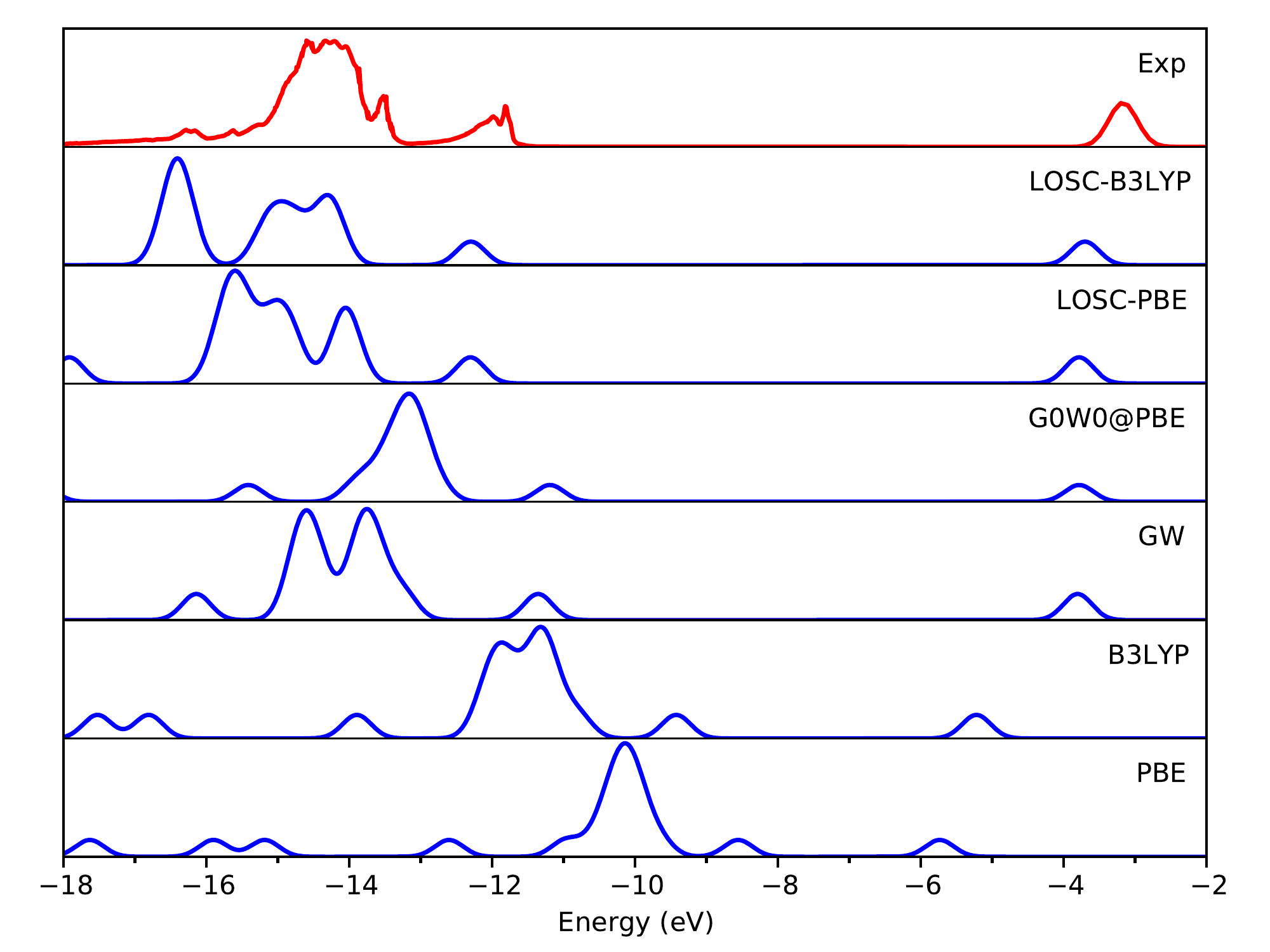}
    \caption{Photoemission spectrum of TCNE.
             Experimental spectrum was obtained from Ref \citenum{ikemoto1974photoelectron}}
\end{figure}

\begin{figure}
    \includegraphics[width=0.7\textwidth]{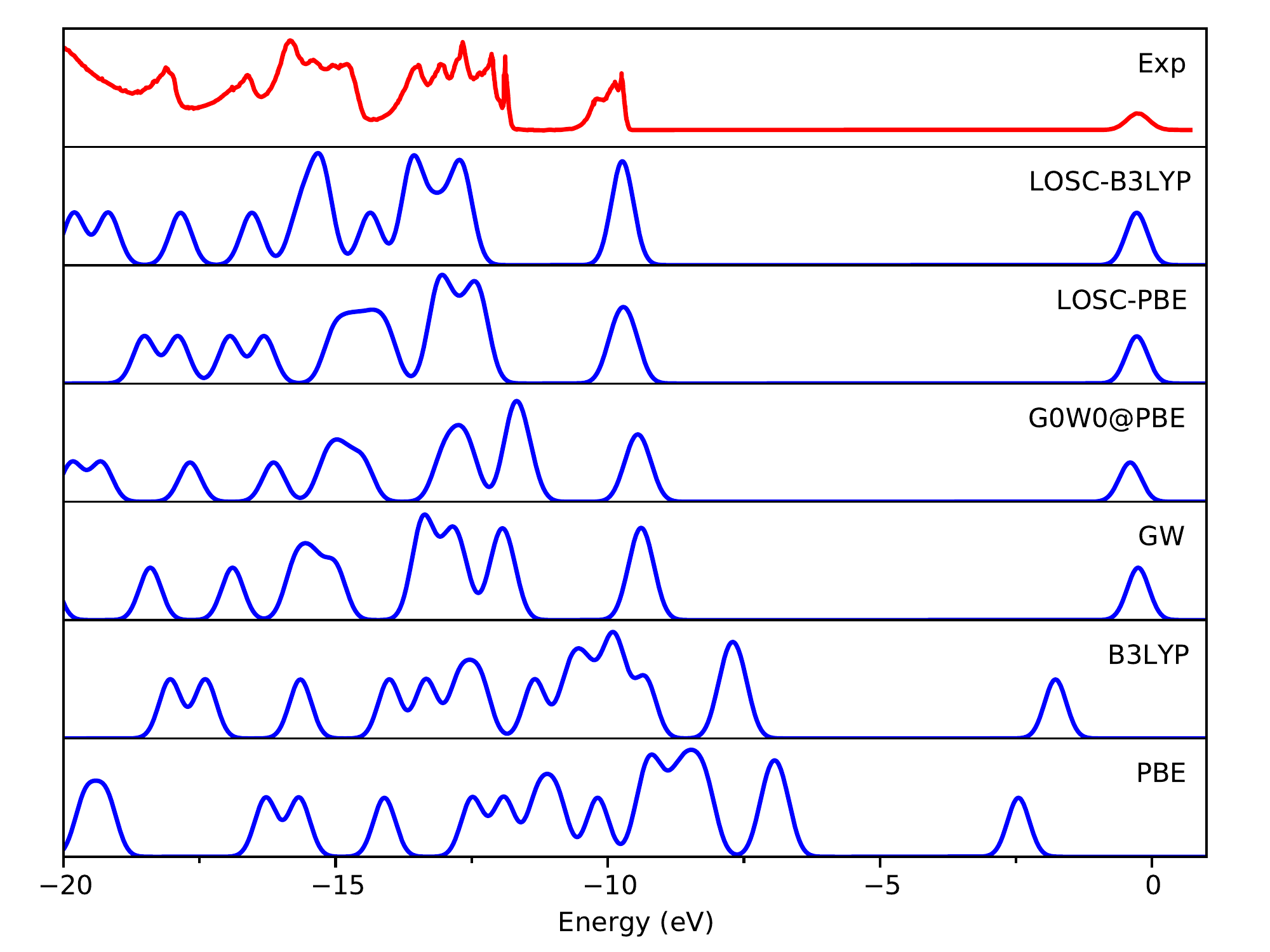}
    \caption{Photoemission spectrum of benzonitrile.
             Experimental spectrum was obtained from Ref \citenum{kimura1981handbook}}
\end{figure}

\begin{figure}
    \includegraphics[width=0.7\textwidth]{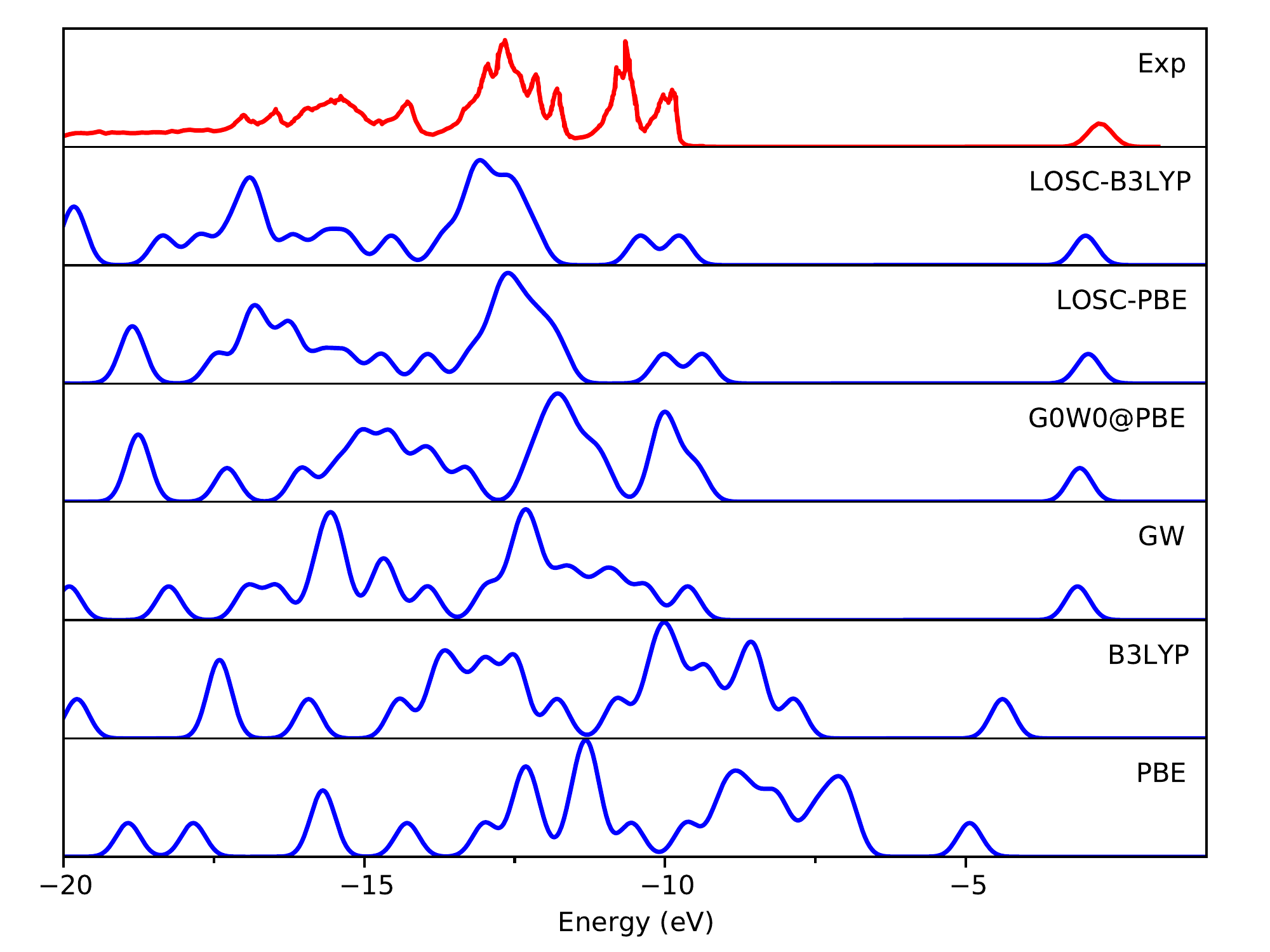}
    \caption{Photoemission spectrum of \ce{Cl4}-benzoquinone.
             Experimental spectrum was obtained from Ref \citenum{dougherty1977photoelectron}}
\end{figure}

\begin{figure}
    \includegraphics[width=0.7\textwidth]{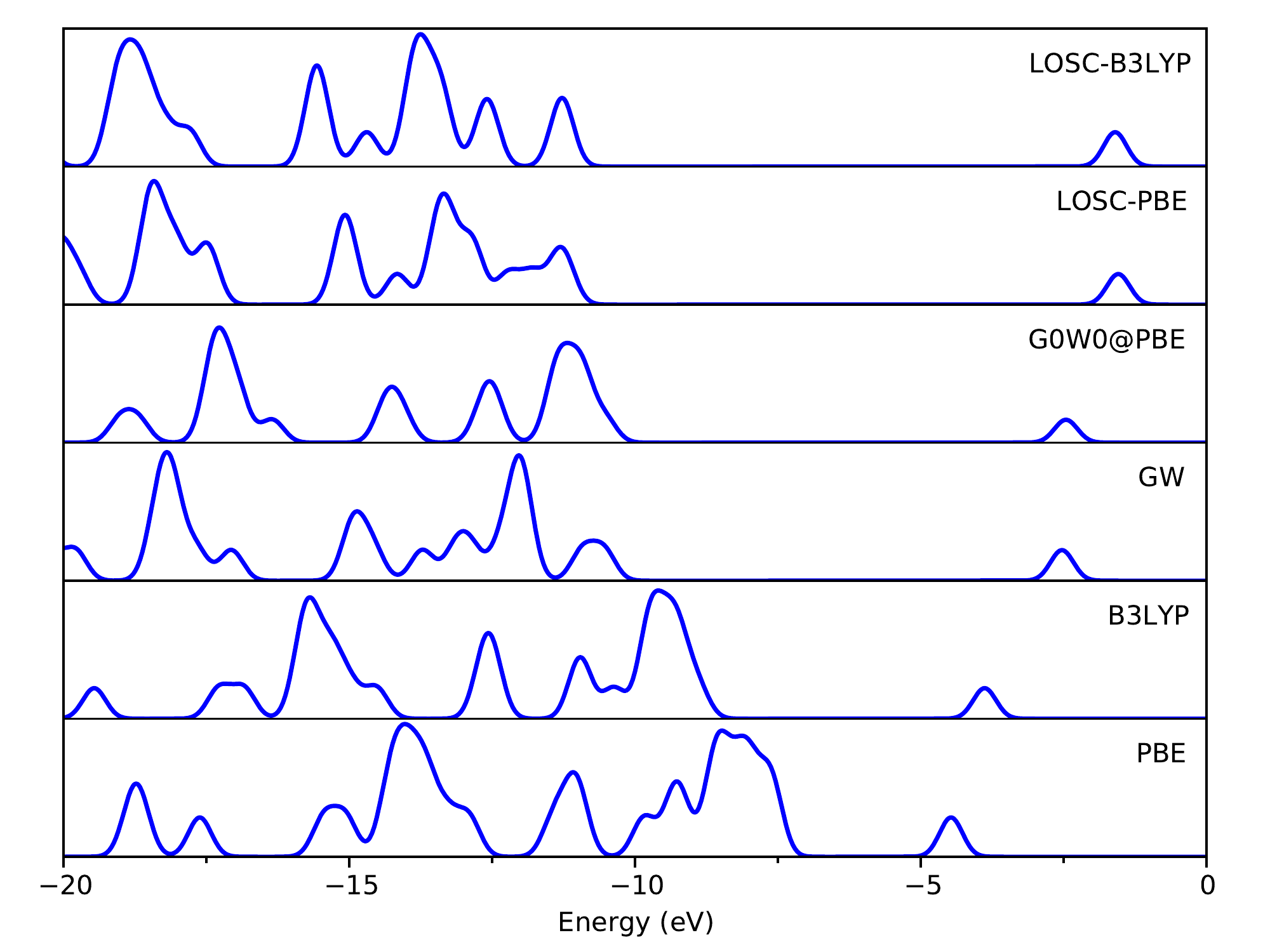}
    \caption{Photoemission spectrum of dinitrobenzonitrile.}
\end{figure}

\begin{figure}
    \includegraphics[width=0.7\textwidth]{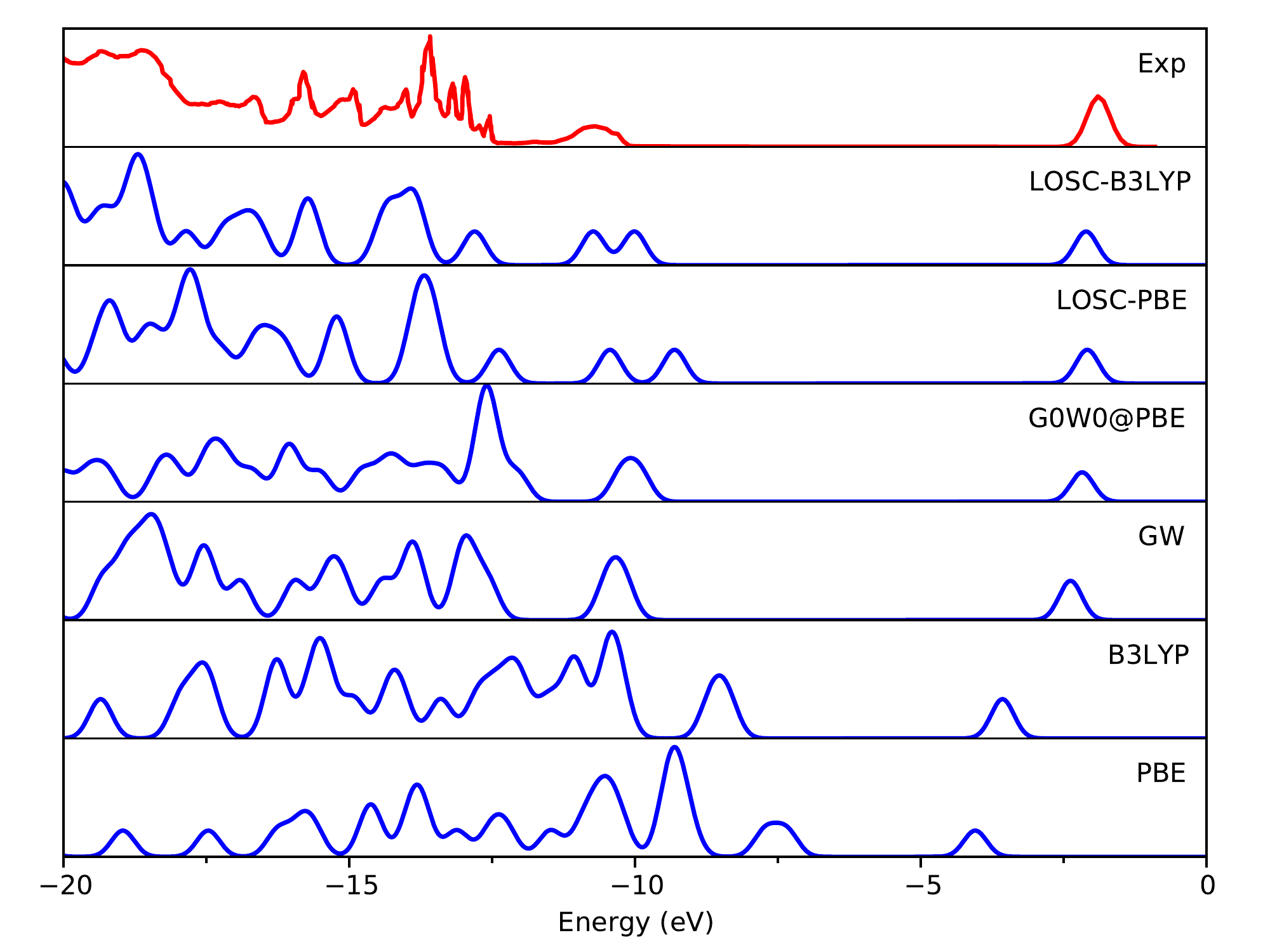}
    \caption{Photoemission spectrum of \ce{F4}-benzenedicarbonitrile.
             Experimental spectrum was obtained from Ref \citenum{neijzen1978photoelectron}}
\end{figure}

\begin{figure}
    \includegraphics[width=0.7\textwidth]{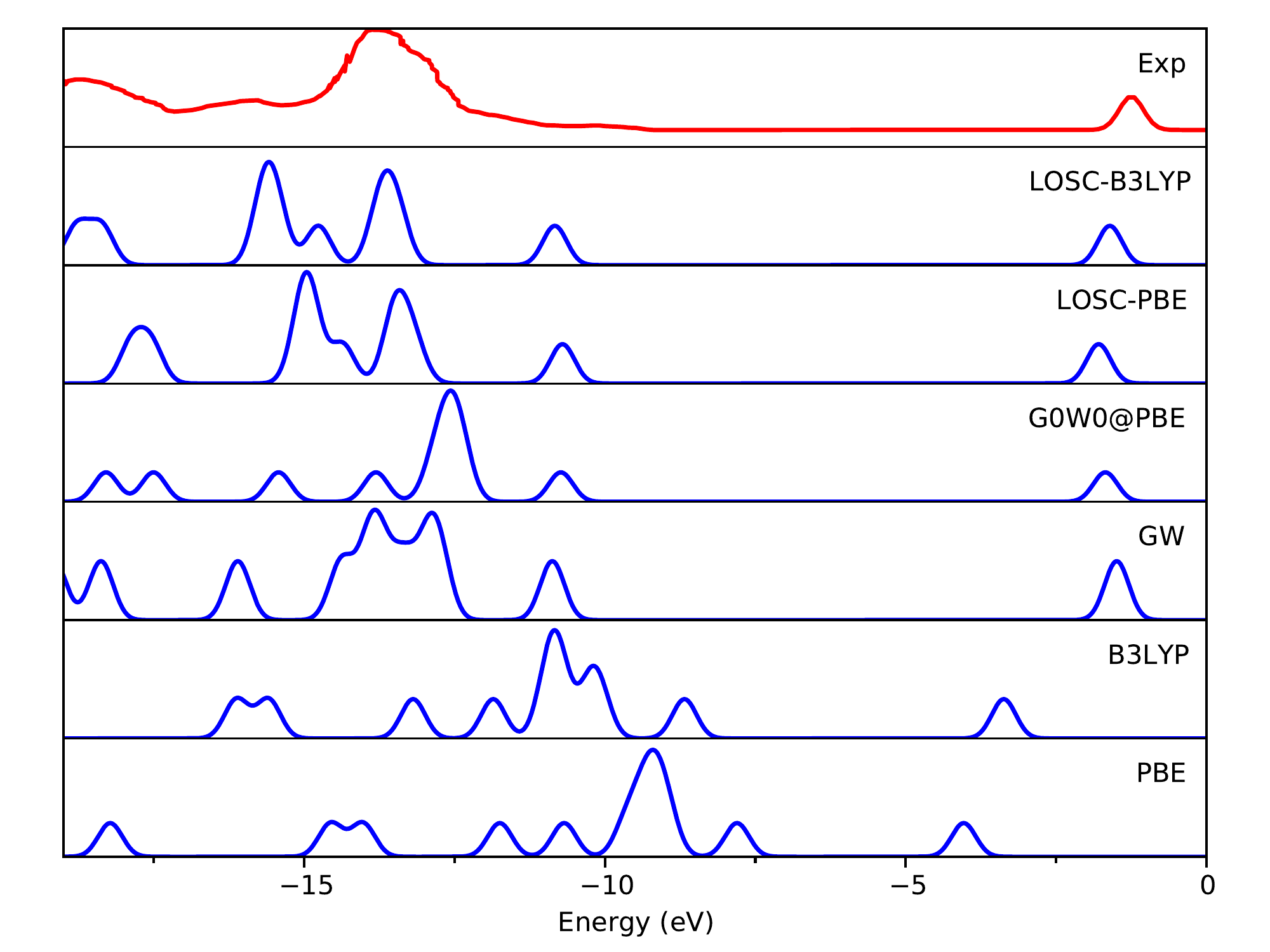}
    \caption{Photoemission spectrum of fumaronitrile.
             Experimental spectrum was obtained from Ref \citenum{fujikawa1976x}}
\end{figure}

\begin{figure}
    \includegraphics[width=0.7\textwidth]{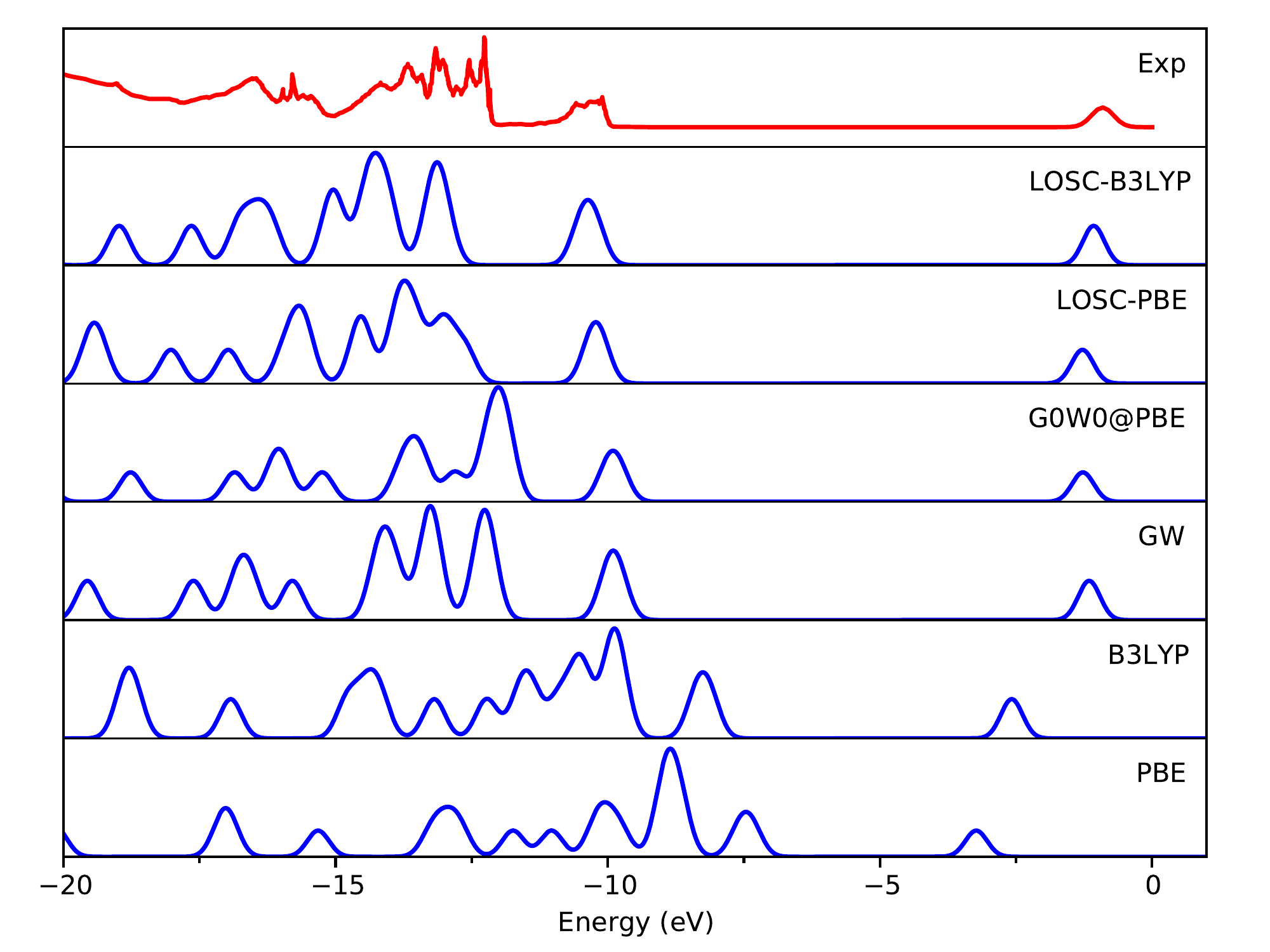}
    \caption{Photoemission spectrum of mDCNB.
             Experimental spectrum was obtained from Ref \citenum{neijzen1978photoelectron}}
\end{figure}

\begin{figure}
    \includegraphics[width=0.7\textwidth]{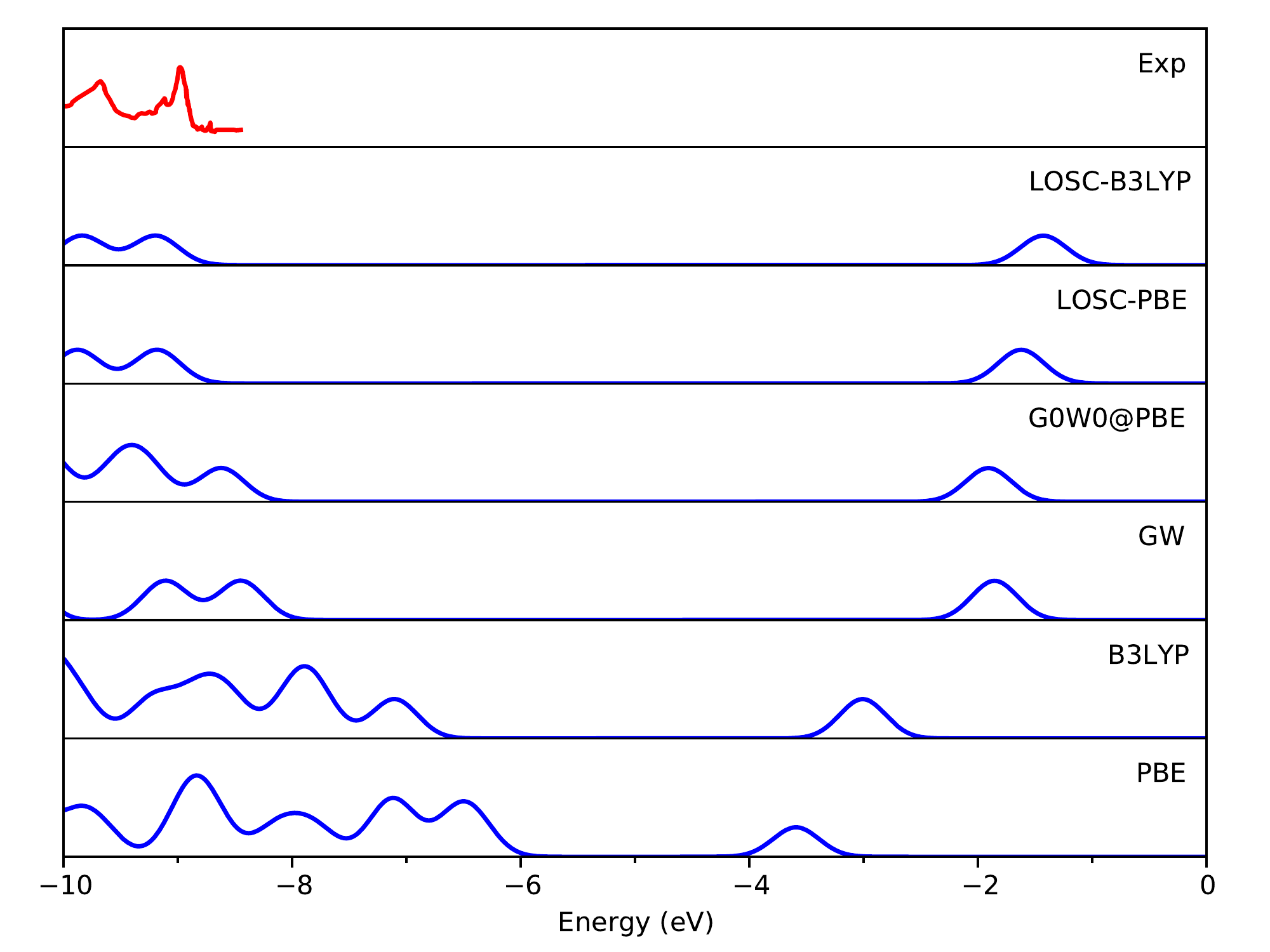}
    \caption{Photoemission spectrum of NDCA.
             Experimental spectrum was obtained from Ref \citenum{sauther2009gas}}
\end{figure}

\begin{figure}
    \includegraphics[width=0.7\textwidth]{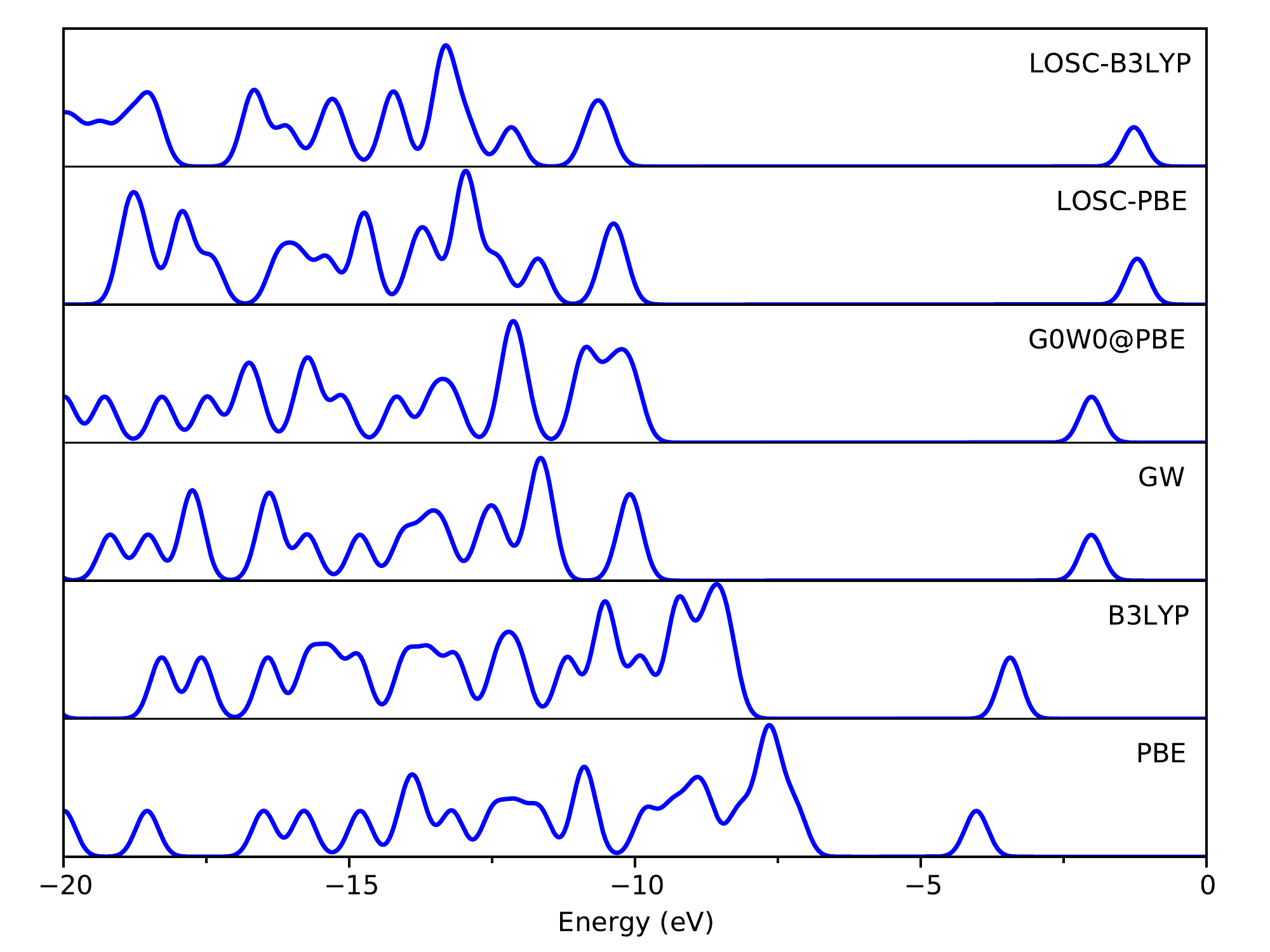}
    \caption{Photoemission spectrum of nitrobenzonitrile.}
\end{figure}

\begin{figure}
    \includegraphics[width=0.7\textwidth]{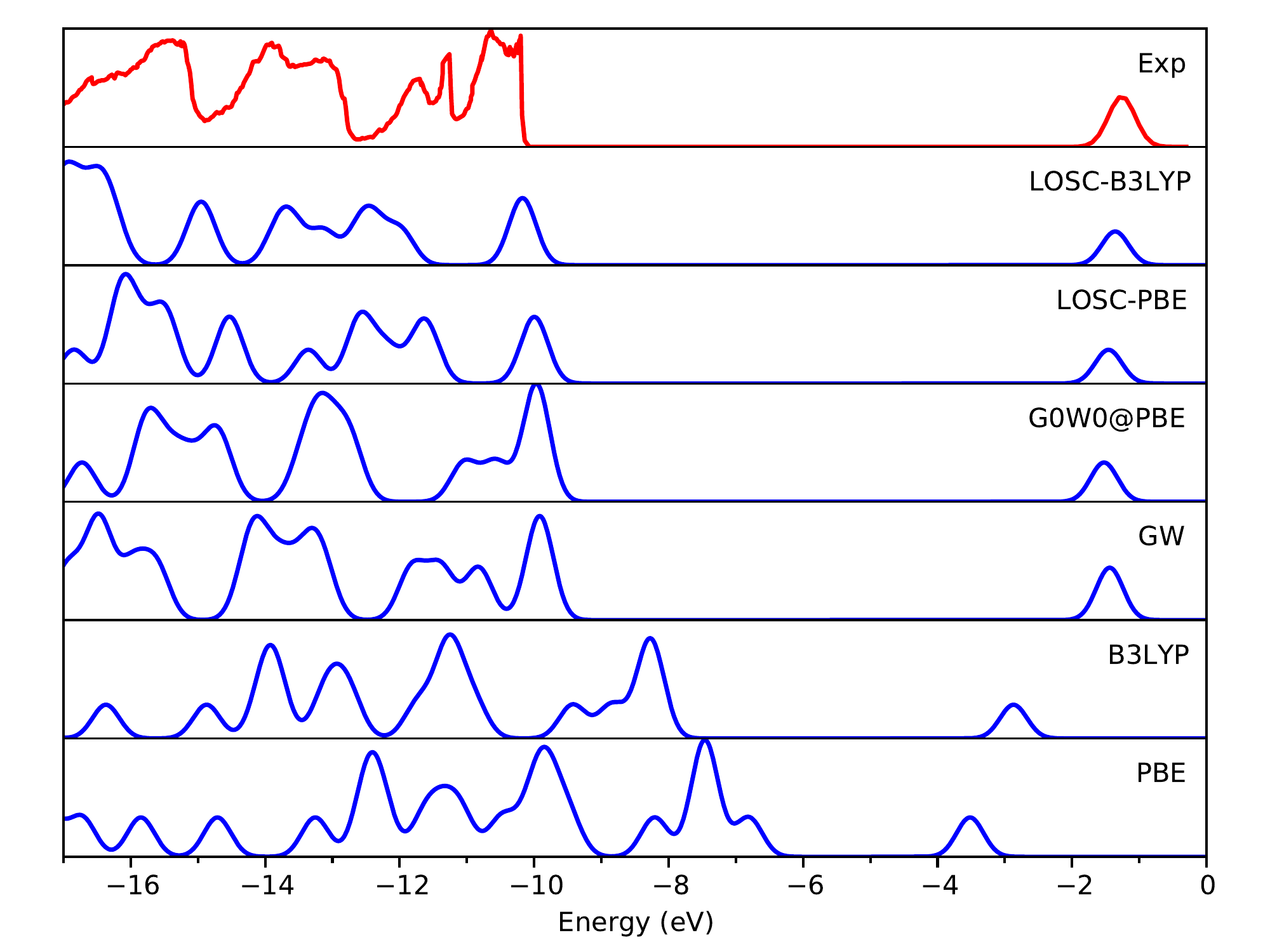}
    \caption{Photoemission spectrum of phthalic anhydride.
             Experimental spectrum was obtained from Ref \citenum{galasso1977photoelectron}}
\end{figure}

\begin{figure}
    \includegraphics[width=0.7\textwidth]{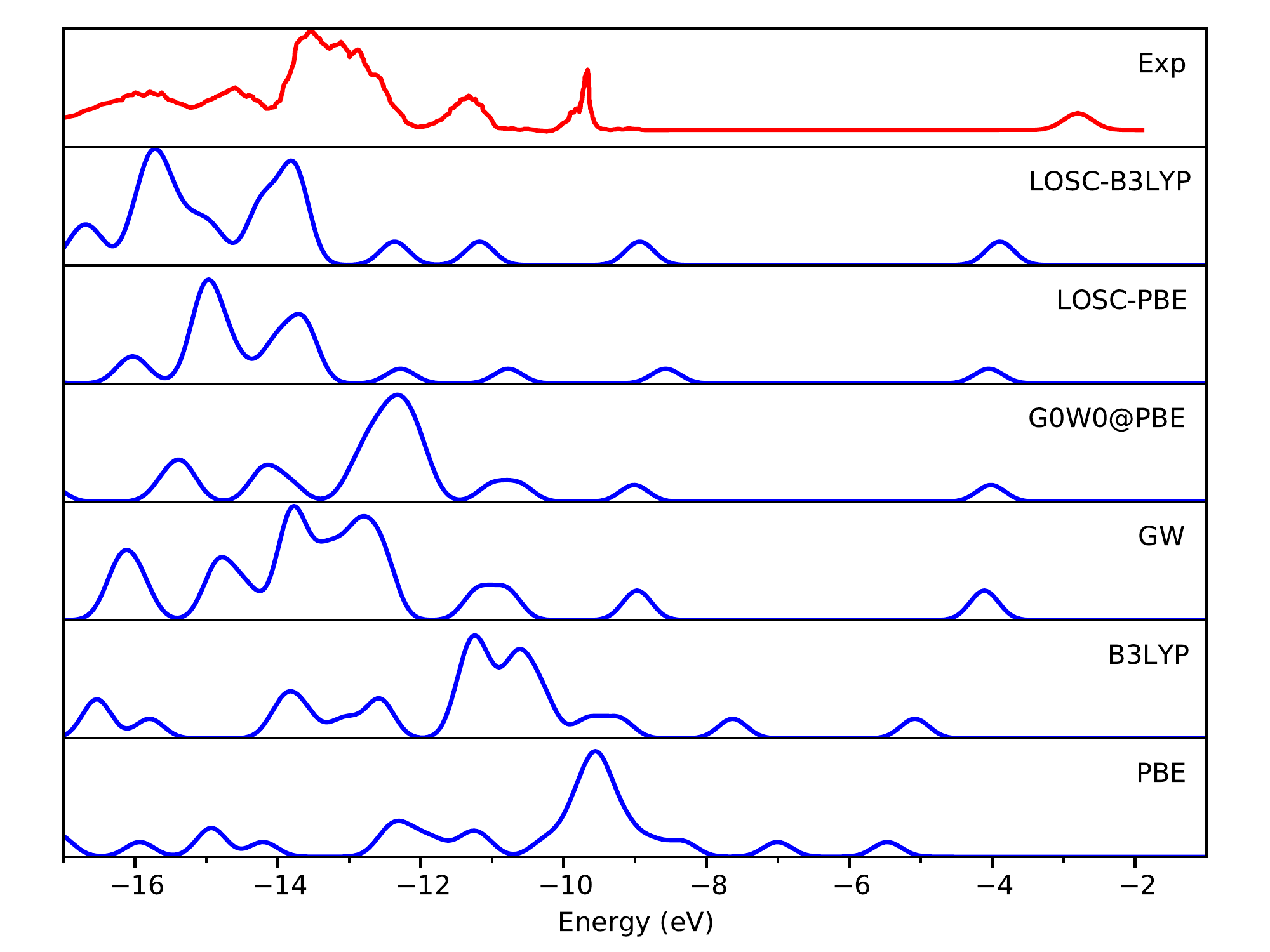}
    \caption{Photoemission spectrum of TCNQ.
             Experimental spectrum was obtained from Ref \citenum{ikemoto1974photoelectron}}
\end{figure}

\begin{figure}
    \includegraphics[width=0.7\textwidth]{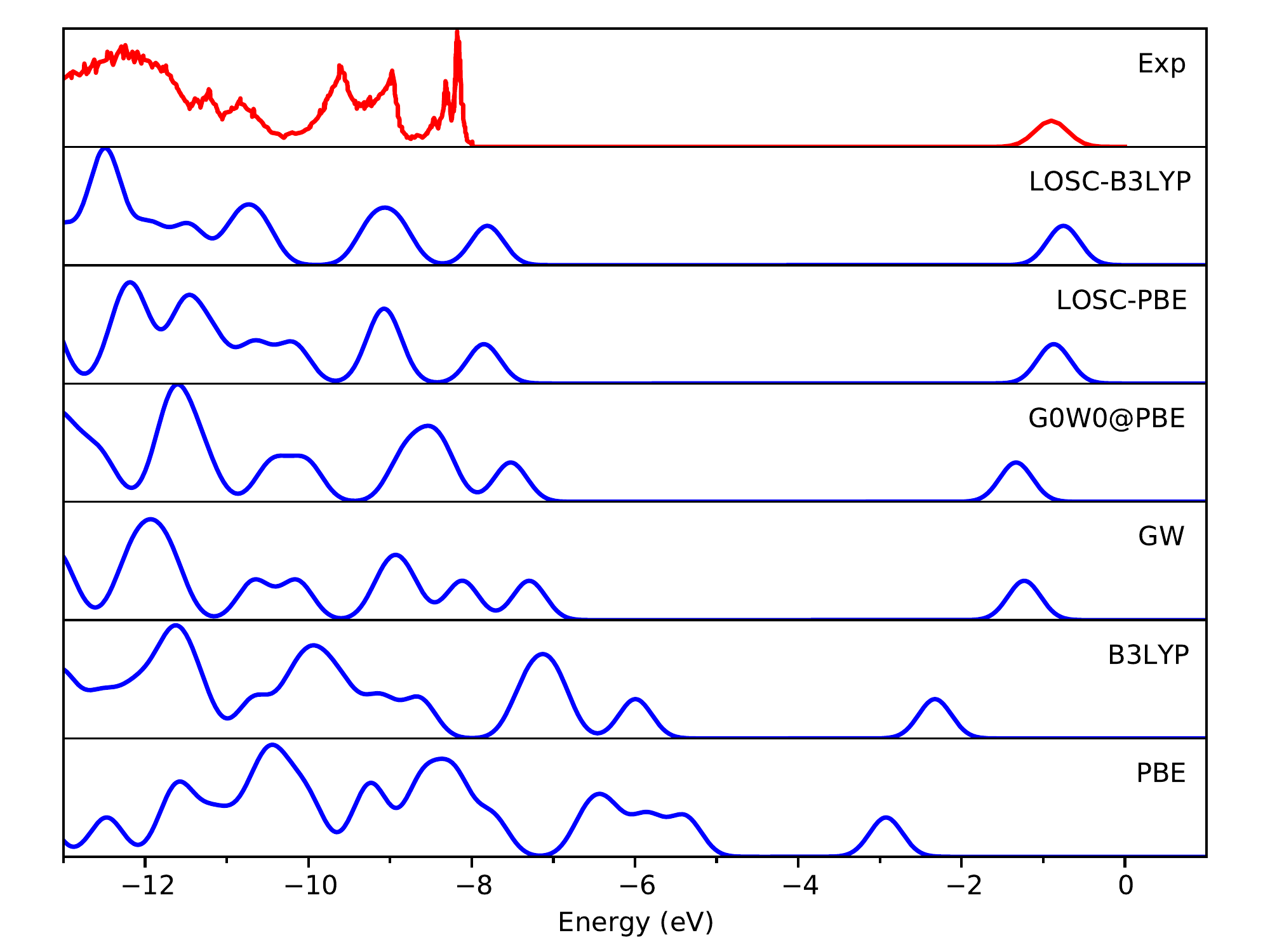}
    \caption{Photoemission spectrum of acridine.
             Experimental spectrum was obtained from Ref \citenum{maier1975ionisation}}
\end{figure}

\begin{figure}
    \includegraphics[width=0.7\textwidth]{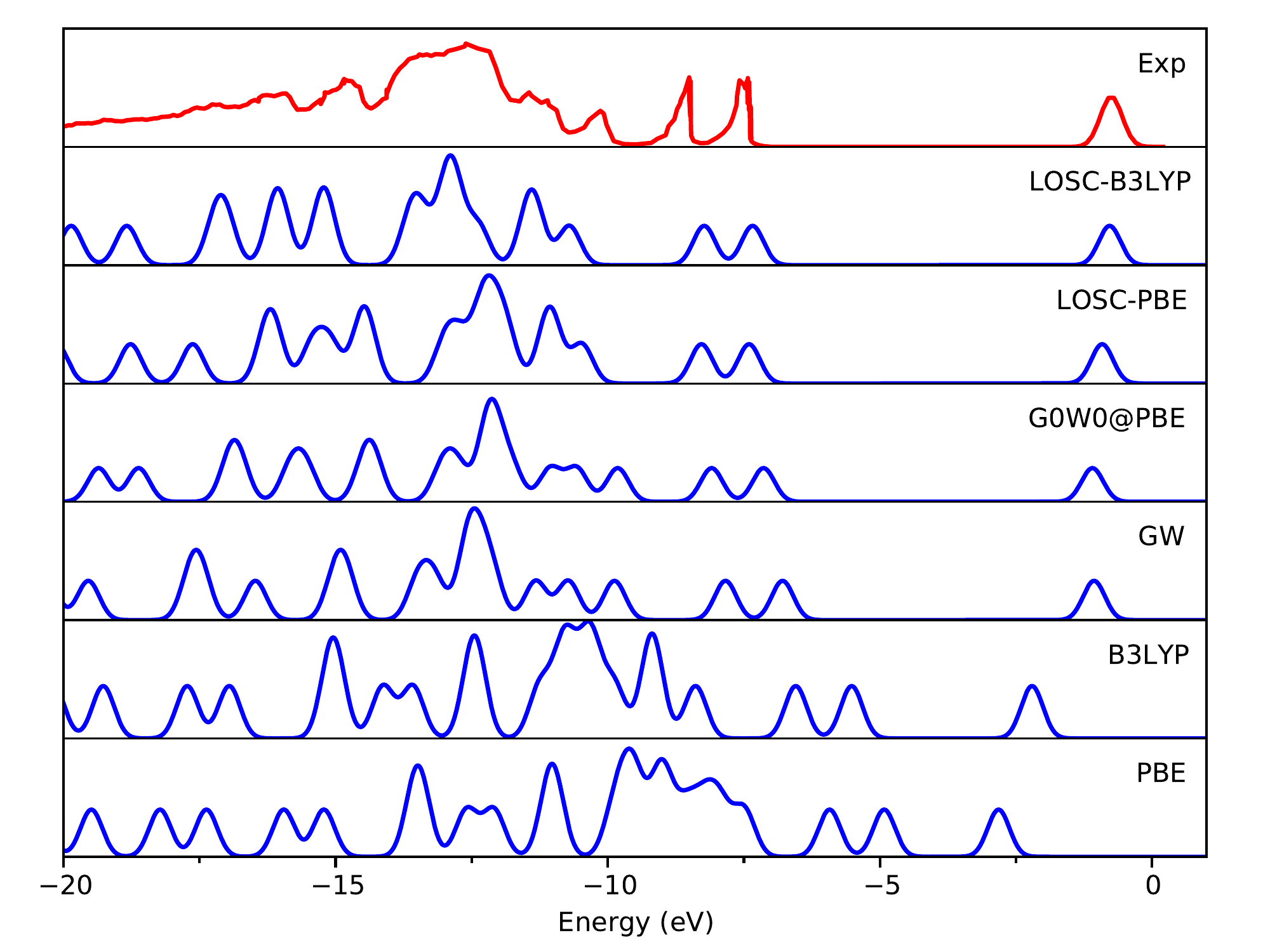}
    \caption{Photoemission spectrum of azulene.
             Experimental spectrum was obtained from Ref \citenum{dougherty1980photoelectron}}
\end{figure}

\begin{figure}
    \includegraphics[width=0.7\textwidth]{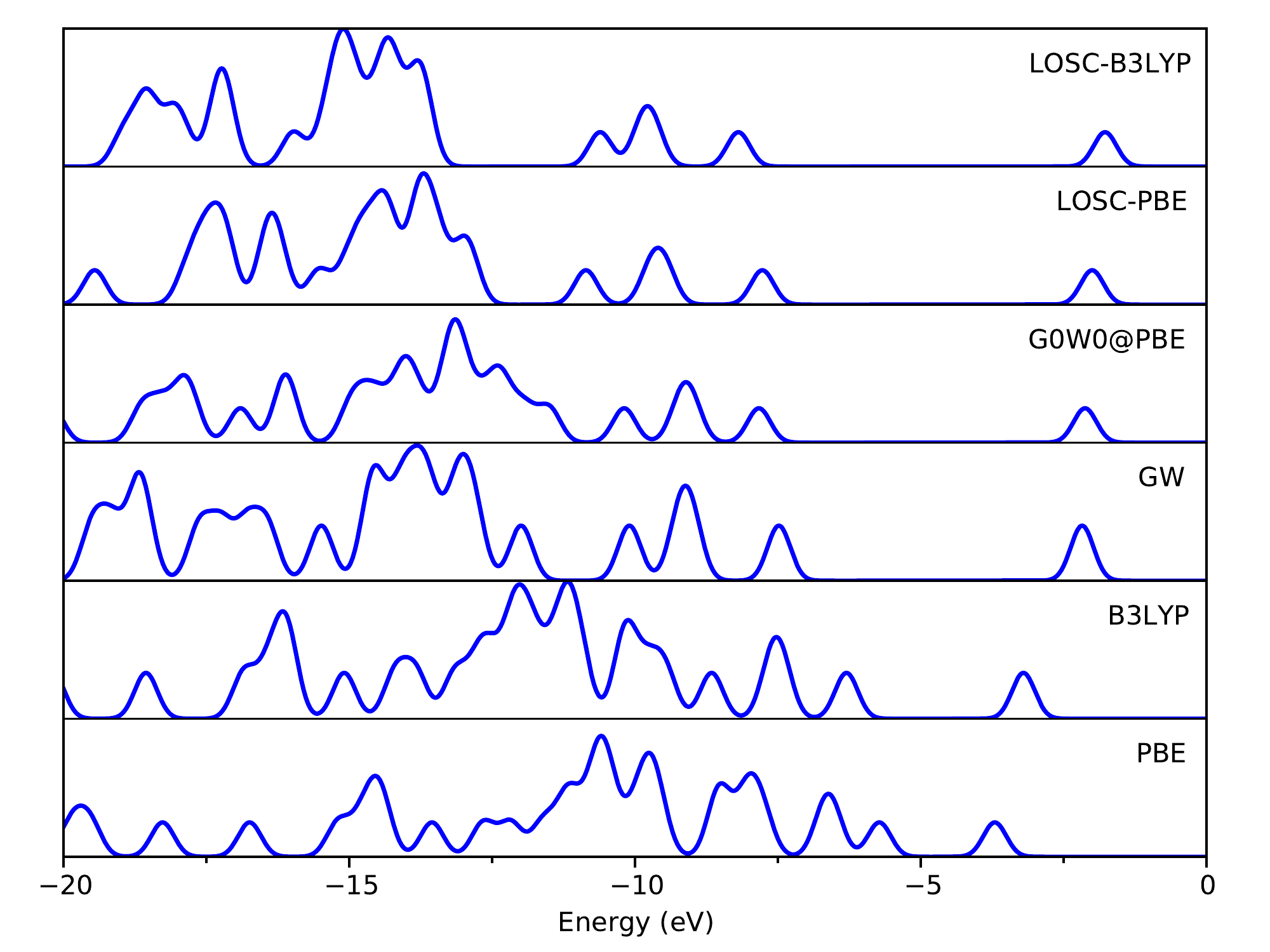}
    \caption{Photoemission spectrum of bodipy.}
\end{figure}

\begin{figure}
    \includegraphics[width=0.7\textwidth]{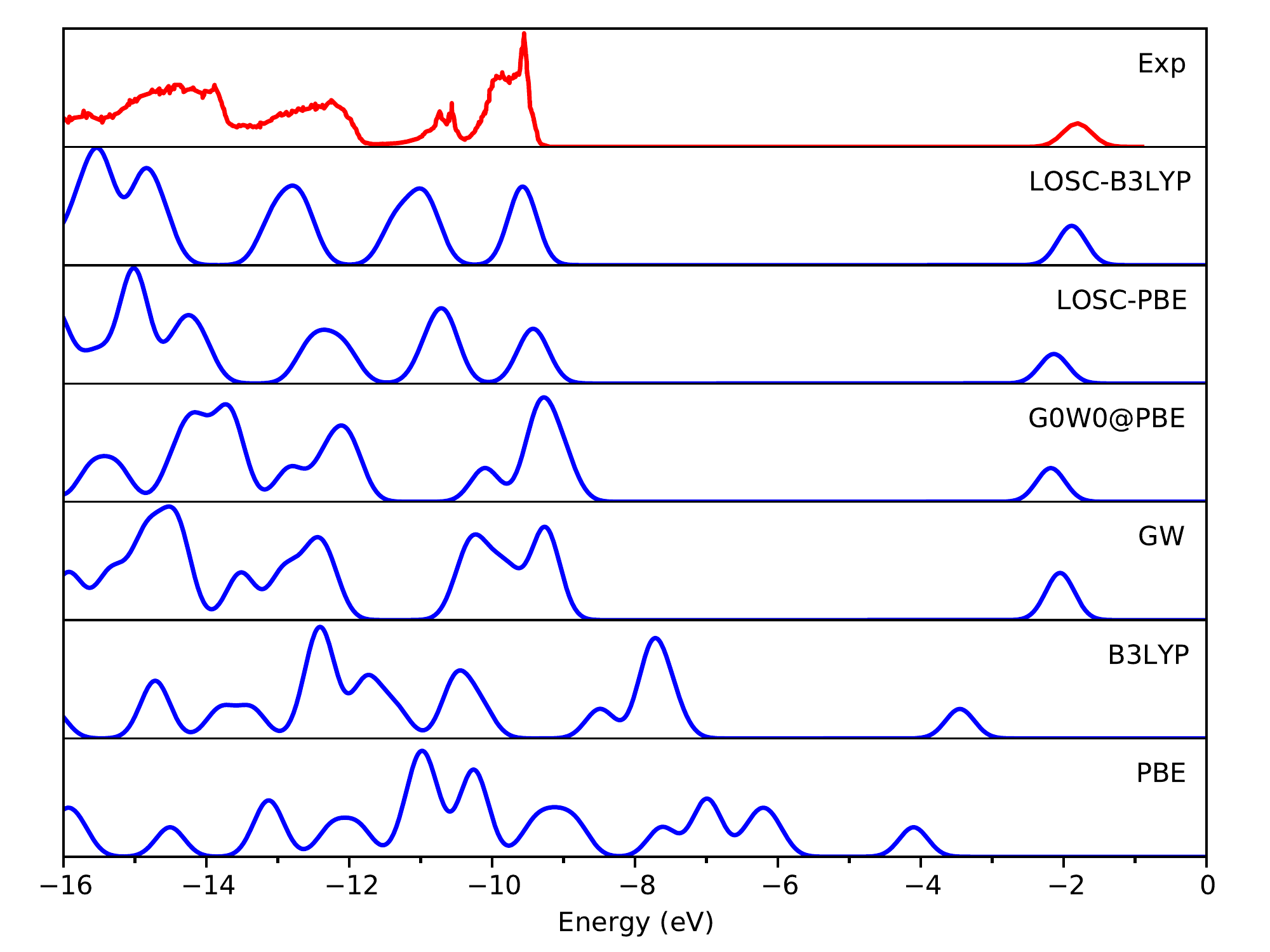}
    \caption{Photoemission spectrum of naphthalenedione.
             Experimental spectrum was obtained from Ref \citenum{millefiori1990uv}}
\end{figure}

\begin{figure}
    \includegraphics[width=0.7\textwidth]{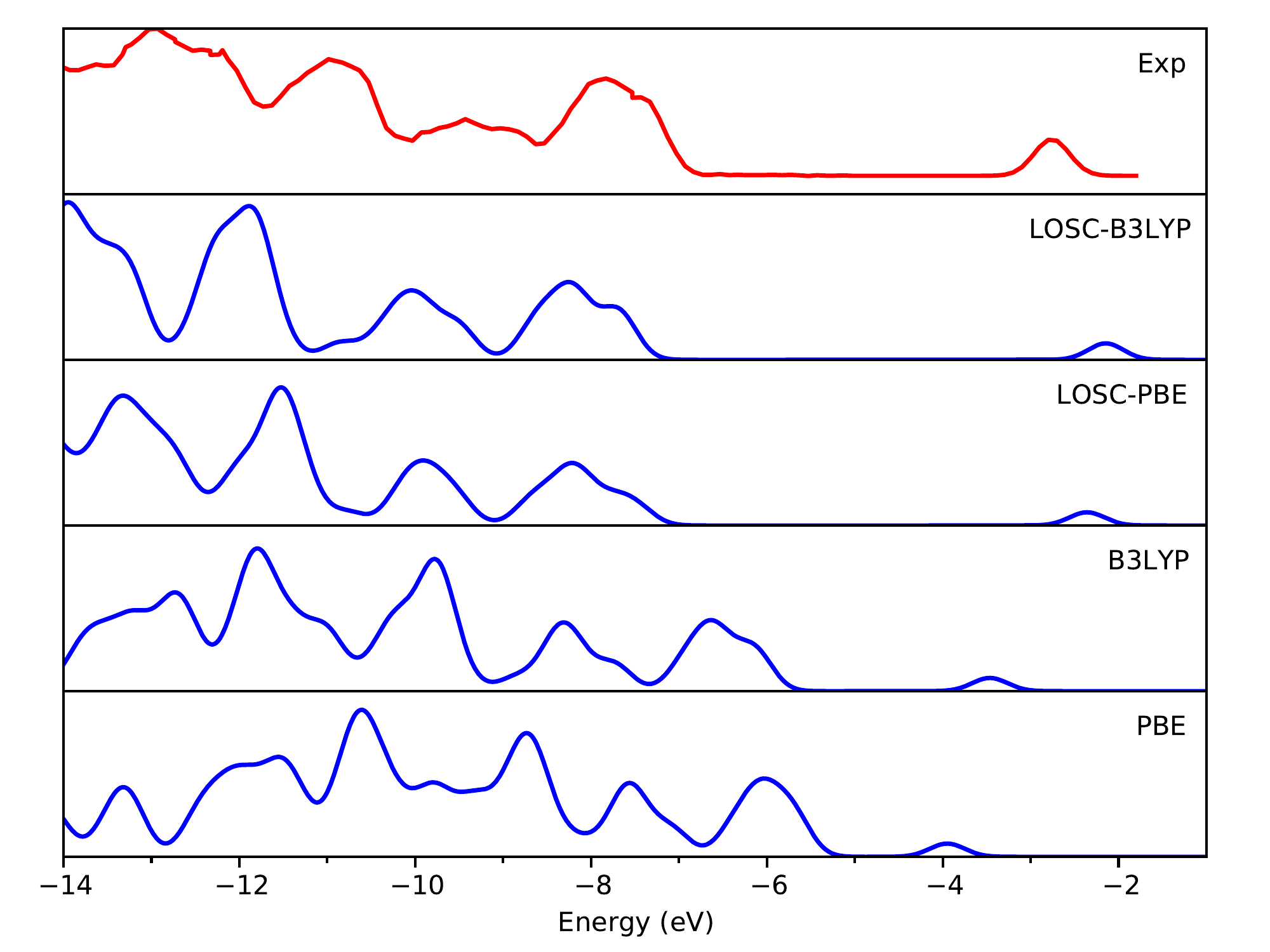}
    \caption{Photoemission spectrum of \ce{C70}.
             Experimental spectrum was obtained from Ref \citenum{hino1995photoelectron}}
\end{figure}

\begin{figure}
    \includegraphics[width=0.7\textwidth]{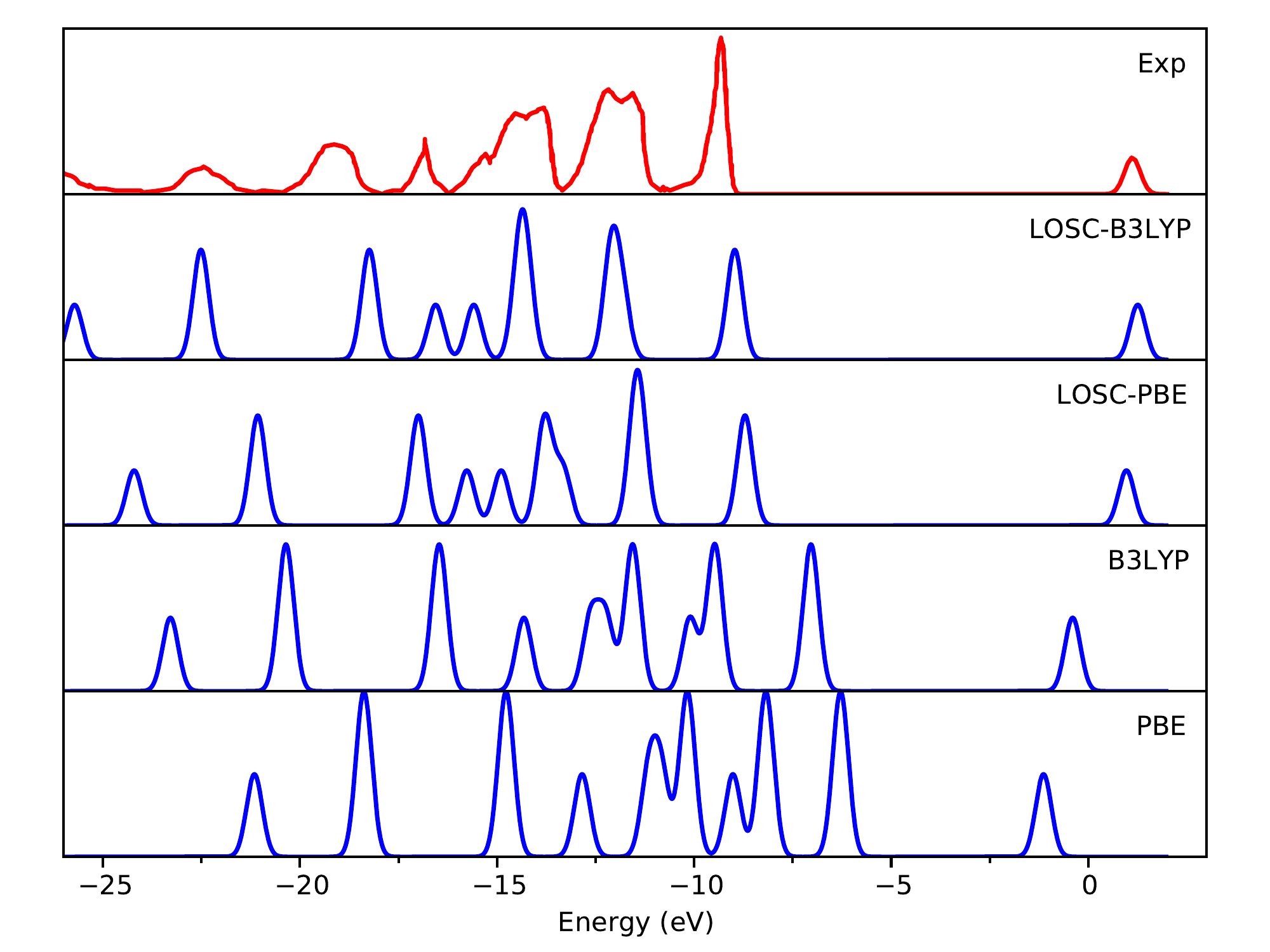}
    \caption{Photoemission spectrum of benzene.
             Experimental spectrum was obtained from Ref \citenum{potts1972photoelectron}}
\end{figure}

\begin{figure}
    \includegraphics[width=0.7\textwidth]{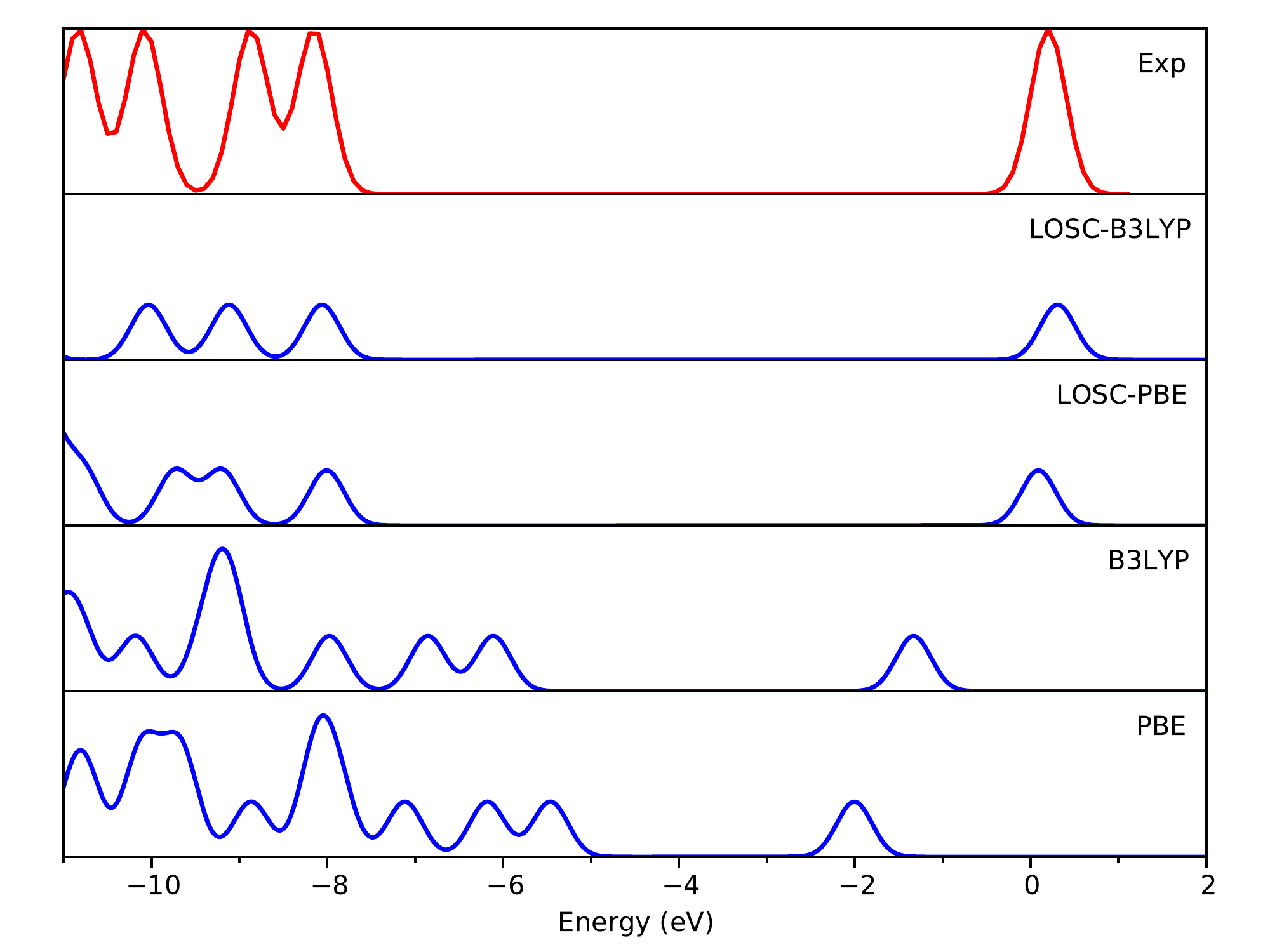}
    \caption{Photoemission spectrum of naphthalene. Experimental quasi-particle energies were
             obtained from Ref \citenum{schmidt1977photoelectron} and used to broaden the spectrum
             by Gaussian expansion with 0.2 eV.}
\end{figure}

\begin{figure}
    \includegraphics[width=0.7\textwidth]{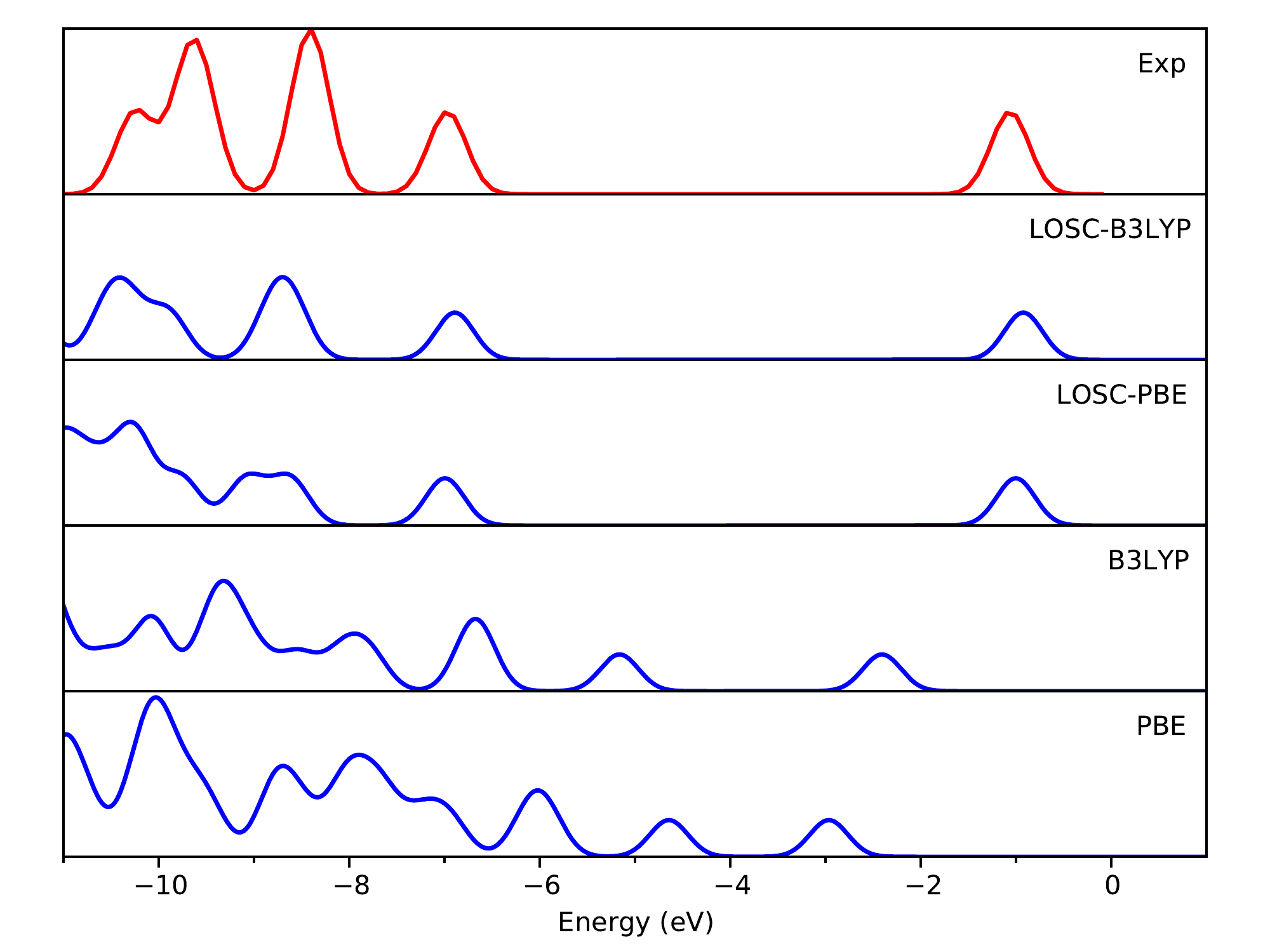}
    \caption{Photoemission spectrum of tetracene. Experimental quasi-particle energies were
             obtained from Ref \citenum{schmidt1977photoelectron} and used to broaden the spectrum
             by Gaussian expansion with 0.2 eV.}
\end{figure}

\begin{figure}
    \includegraphics[width=0.7\textwidth]{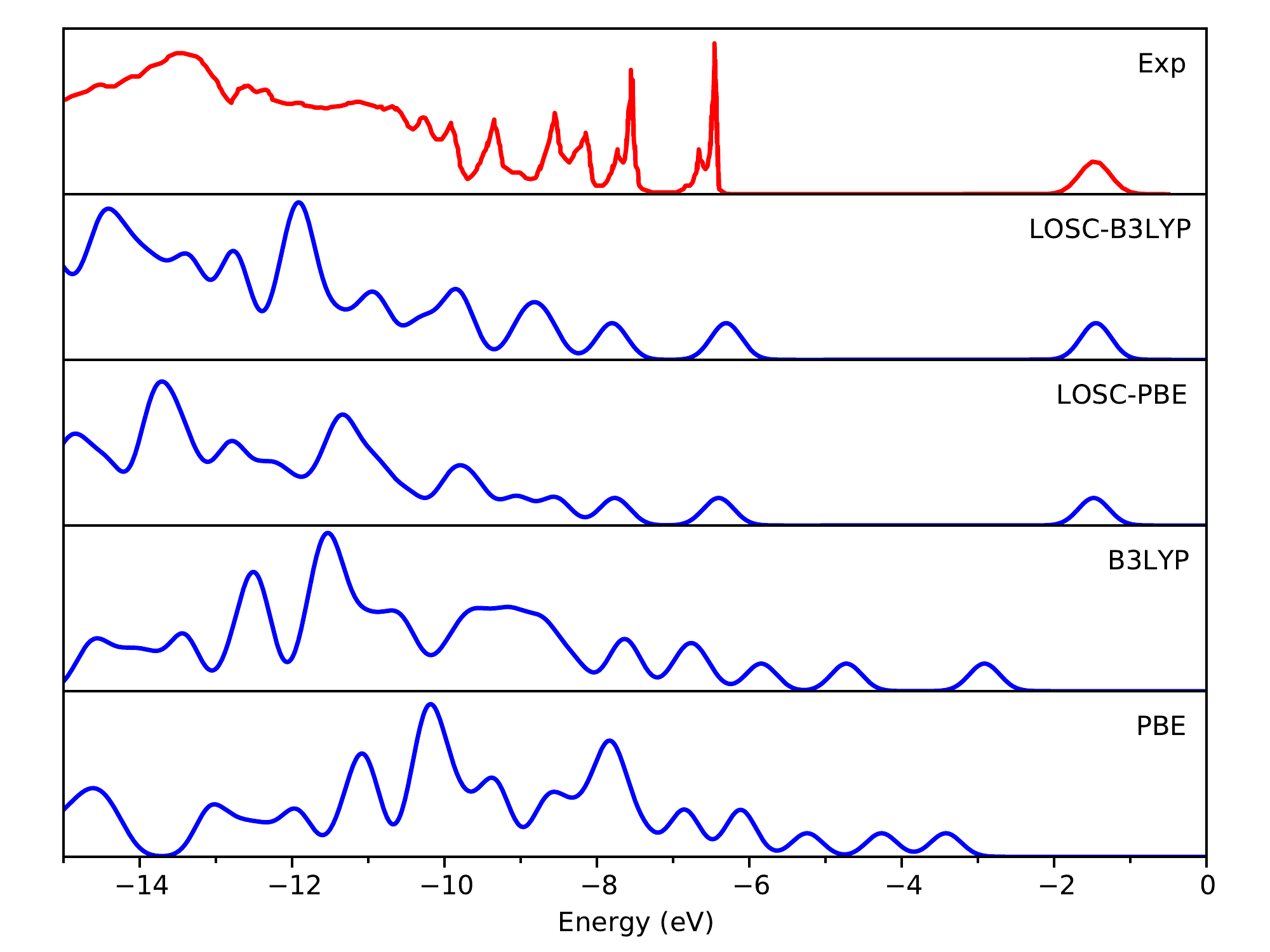}
    \caption{Photoemission spectrum of hexacene.
             Experimental spectrum was obtained from Ref \citenum{boschi1974photoelectron}}
\end{figure}

\clearpage
\section{Quasi-particle energy}
Table \ref{tab:HOMO_eig} and \ref{tab:LUMO_eig} show the detailed HOMO and LUMO energies
for 40 molecules from density functional approximations (DFAs), LOSC-DFAs, self-consistent $GW$ (sc$GW$),
$G_0W_0$@PBE and $\Delta$SCF methods. Most of the test molecules are selected from
Blase's \cite{blase2011first} and Marom's \cite{potentialselectron} test set, if not specified.
Geometries of Blase's test set are provided by Blase \cite{blase2011first}, and geometries of Marom's test set
can be found in Ref \citenum{potentialselectron}.

Table \ref{tab:valence_eig} shows the valence orbital energies from DFAs, LOSC-DFAs and $\Delta$SCF methods.
Test molecules includes polyacene (n = 1 - 6) and other three small molecules.
Geometries of polyacene (n = 1 - 6) were optimized from B3LYP/6-31g* with Gaussian 09 pacage \cite{Gaussian09},
and attached at the end of SI. Other three geometries are obtained from Ref \citenum{van2015gw}.
The basis set used for DFT calculation is cc-pVTZ, if not specified. The fitting basis for LOSC is aug-cc-pVTZ.

\begin{landscape}
\renewcommand\arraystretch{0.5}
\begin{ThreePartTable}
    \begin{TableNotes}
        \footnotesize
        \item [a] cc-pVDZ basis set used for DFT calculation.
        \item [b] 6-31G basis set was used for DFT calculation. Molecular
            geometry was obtained form Ref \citenum{zheng2005performance}.
        \item [c] Ionization energy was read out from the experimental spectrum shown in
            Ref \citenum{hino1995photoelectron}.
        \item [d] Geometry was optimized from B3LYP/6-31G* with Gaussian 09 package. \cite{Gaussian09}
        \item [e] Ref \citenum{li2017localized}
        \item [f] Ref \citenum{blase2011first} and \citenum{potentialselectron}, if not specified.
    \end{TableNotes}

\begin{longtable}{lccccccccccc}
    \caption{
    Comparison of (negative) HOMO energies (in eV) obtained from DFAs, LOSC-DFAs, self-consistent $GW$ (sc$GW$),
    $G_0W_0$@PBE, and $\Delta$SCF (PBE and LOSC-PBE) methods with the experimental ionization potential.}
    \label{tab:HOMO_eig}\\
    \toprule
        \multirow{2}{*}{Molecule} & \multirow{2}{*}{Exp\tnote{f}} & \multirow{2}{*}{sc$GW$\tnote{f}} & \multirow{2}{*}{$G_0W_0$\tnote{f}}
        & \multirow{2}{*}{$\Delta$PBE}  & \multirow{2}{*}{$\Delta$LOSC-} & \multirow{2}{*}{LDA} & \multirow{2}{*}{PBE}
        & \multirow{2}{*}{B3LYP} & \multirow{2}{*}{LOSC-} &\multirow{2}{*}{LOSC-} &\multirow{2}{*}{LOSC-} \\
        \\
        & & & & &PBE& & PBE& B3LYP& LDA& PBE&B3LYP\\
    \midrule
    \endfirsthead
    Table \ref{tab:HOMO_eig}. (Continued.) \\
    \toprule
        \multirow{2}{*}{Molecule} & \multirow{2}{*}{Exp\tnote{f}} & \multirow{2}{*}{sc$GW$\tnote{f}} & \multirow{2}{*}{$G_0W_0$\tnote{f}}
        & \multirow{2}{*}{$\Delta$PBE}  & \multirow{2}{*}{$\Delta$LOSC-} & \multirow{2}{*}{LDA} & \multirow{2}{*}{PBE}
        & \multirow{2}{*}{B3LYP} & \multirow{2}{*}{LOSC-} &\multirow{2}{*}{LOSC-} &\multirow{2}{*}{LOSC-} \\
        \\
        & & & & &PBE& & PBE& B3LYP& LDA& PBE&B3LYP\\
    \midrule
    \endhead
    \midrule
    \endfoot
    \midrule
    \insertTableNotes
    \endlastfoot

Anthracene\tnote{d}              & 7.40\tnote{e}  & 6.77  & 7.04  & 7.09  & 7.47  & 5.16  & 4.94  & 5.51  & 7.62  & 7.40  & 7.30  \\
Benzothiadiazole                 & 9.00           & NA    & NA    & 8.73  & 8.86  & 6.36  & 6.13  & 6.85  & 8.77  & 8.54  & 8.75  \\
Benzothiazole                    & 8.80           & NA    & NA    & 8.50  & 8.80  & 6.17  & 5.96  & 6.69  & 8.84  & 8.63  & 8.63  \\
\ce{C60}\tnote{a}                & 7.60           & NA    & NA    & NA    & NA    & 5.92  & 5.66  & 6.18  & 7.98  & 7.76  & 7.86  \\
Fluorene                         & 7.90           & NA    & NA    & 7.64  & 8.08  & 5.62  & 5.41  & 6.04  & 8.33  & 8.10  & 7.93  \\
\ce{H2P}                         & 6.90           & NA    & NA    & 6.75  & 6.78  & 5.22  & 4.98  & 5.48  & 7.37  & 7.23  & 7.05  \\
\ce{H2PC}\tnote{a}               & 6.40           & NA    & NA    & 6.27  & 6.11  & 5.07  & 4.84  & 5.11  & 6.32  & 6.10  & 6.13  \\
\ce{H2TPP}\tnote{a}              & 6.40           & NA    & NA    & 6.16  & 6.55  & 4.84  & 4.63  & 5.12  & 6.64  & 6.43  & 6.60  \\
Pentacene\tnote{d}               & 6.60           & NA    & NA    & 6.16  & 6.69  & 4.63  & 4.41  & 4.87  & 6.75  & 6.52  & 6.42  \\
PTCDA                            & 8.20           & NA    & NA    & 7.86  & 8.59  & 6.40  & 6.13  & 6.68  & 8.94  & 8.63  & 8.41  \\
Thiadiazole                      & 10.10          & NA    & NA    & 10.13 & 10.13 & 7.10  & 6.90  & 7.73  & 9.63  & 9.44  & 9.75  \\
thiophene                        & 8.85           & NA    & NA    & 8.82  & 8.77  & 5.98  & 5.78  & 6.58  & 8.46  & 8.27  & 8.56  \\
Benzoquinone                     & 10.03          & 10.22 & 9.41  & 9.21  & 10.75 & 6.43  & 6.25  & 7.67  & 11.00 & 10.80 & 11.01 \\
\ce{Cl4}-isobenzofuranedione     & 10.80          & 9.43  & 9.36  & 9.36  & 10.38 & 7.27  & 7.04  & 7.94  & 10.12 & 9.88  & 10.04 \\
Dichlone                         & 9.59           & 9.26  & 9.22  & 8.94  & 10.22 & 6.73  & 6.52  & 7.64  & 9.57  & 9.33  & 9.55  \\
\ce{F4}-benzoquinone             & 10.83          & 10.65 & 10.27 & 10.03 & 10.86 & 7.52  & 7.23  & 8.39  & 10.08 & 9.80  & 10.48 \\
Maleic anhydride                 & 11.09          & 11.26 & 10.46 & 10.36 & 11.82 & 7.25  & 7.02  & 8.45  & 11.85 & 11.61 & 11.94 \\
Nitrobenzene                     & 9.93           & 9.54  & 9.68  & 9.65  & 10.73 & 6.95  & 6.70  & 7.87  & 9.77  & 9.67  & 9.81  \\
Phenazine                        & 8.38           & 7.74  & 7.92  & 7.90  & 8.39  & 5.84  & 5.66  & 6.39  & 8.41  & 8.10  & 8.25  \\
Phthalimide                      & 9.84           & 9.45  & 9.37  & 9.24  & 10.76 & 6.48  & 6.25  & 7.66  & 9.68  & 9.51  & 9.69  \\
TCNE                             & 11.78          & 11.36 & 11.19 & 11.15 & 12.50 & 8.83  & 8.55  & 9.42  & 12.54 & 12.30 & 12.30 \\
Benzonitrile                     & 9.75           & 9.29  & 9.34  & 9.53  & 10.11 & 7.04  & 6.81  & 7.58  & 9.80  & 9.58  & 9.71  \\
\ce{Cl4}-benzoquinone            & 9.82           & 9.62  & 9.49  & 9.23  & 9.70  & 7.13  & 6.93  & 7.86  & 9.58  & 9.38  & 9.77  \\
Dinitrobenzonitrile              & N/A            & 10.55 & 10.52 & 11.21 & 11.76 & 7.89  & 7.61  & 8.90  & 11.52 & 11.21 & 11.26 \\
\ce{F4}-benzenedicarbonitrile    & 10.65          & 10.21 & 9.92  & 10.14 & 10.31 & 7.63  & 7.35  & 8.39  & 9.58  & 9.31  & 10.01 \\
Fumaronitrile                    & 11.23          & 10.88 & 10.73 & 10.80 & 11.06 & 8.05  & 7.80  & 8.68  & 10.94 & 10.71 & 10.83 \\
mDCBN                            & 10.40          & 9.80  & 9.80  & 9.86  & 10.46 & 7.60  & 7.35  & 8.14  & 10.38 & 10.13 & 10.24 \\
NDCA                             & 8.98           & 8.45  & 8.62  & 8.69  & 9.56  & 6.67  & 6.42  & 7.10  & 9.43  & 9.18  & 9.19  \\
Nitrobenzonitrile                & 10.40          & 10.02 & 10.02 & 9.94  & 11.60 & 7.45  & 7.19  & 8.37  & 10.51 & 10.28 & 10.53 \\
Phthalic anhydride               & 10.18          & 9.90  & 9.96  & 9.79  & 11.00 & 7.05  & 6.81  & 8.23  & 10.19 & 9.97  & 10.15 \\
TCNQ                             & 9.61           & 8.97  & 9.01  & 9.04  & 8.91  & 7.27  & 7.01  & 7.63  & 8.87  & 8.57  & 8.93  \\
Acridine                         & 7.99           & 7.30  & 7.52  & 7.52  & 8.06  & 5.62  & 5.38  & 5.99  & 8.05  & 7.85  & 7.81  \\
Azulene                          & 7.43           & 6.79  & 7.14  & 7.37  & 7.46  & 5.15  & 4.92  & 5.52  & 7.63  & 7.40  & 7.34  \\
Bodipy                           & N/A            & 7.48  & 7.83  & 8.04  & 7.76  & 5.97  & 5.72  & 6.30  & 8.01  & 7.77  & 8.19  \\
Naphthalenedione                 & 9.54           & 9.21  & 8.99  & 8.75  & 10.39 & 6.29  & 6.08  & 7.48  & 9.58  & 9.35  & 9.57  \\
\ce{C70}\tnote{b}                & 7.32\tnote{c}  & N/A   & N/A   & NA    & NA    & 5.91  & 5.69  & 6.13  & 7.87  & 7.65  & 7.63  \\
Benzene\tnote{d}                 & 9.24\tnote{e}  & N/A   & N/A   & 9.25  & 9.25  & 6.50  & 6.28  & 7.04  & 8.94  & 8.71  & 8.98  \\
Naphthalene\tnote{d}             & 8.11\tnote{e}  & N/A   & N/A   & 7.90  & 8.12  & 5.68  & 5.46  & 6.11  & 8.16  & 8.01  & 8.06  \\
Tetracene\tnote{d}               & 6.97\tnote{e}  & N/A   & N/A   & 6.55  & 7.03  & 4.87  & 4.64  & 5.16  & 7.20  & 7.00  & 6.89  \\
Hexacene\tnote{d}                & 6.36\tnote{e}  & N/A   & N/A   & 5.87  & 6.23  & 4.50  & 4.26  & 4.73  & 6.62  & 6.40  & 6.30  \\
\midrule
MAE                              &                & 0.47  & 0.51  & 0.43  & 0.34  & 2.58  & 2.81  & 2.00  & 0.34  & 0.37  & 0.26  \\
MSE                              &                & -0.43 & -0.51 & -0.42 & 0.20  & -2.58 & -2.81 & -2.00 & 0.04  & -0.18 & -0.05 \\
\end{longtable}
\end{ThreePartTable}
\end{landscape}

\begin{landscape}
\renewcommand\arraystretch{0.5}
\begin{ThreePartTable}
    \begin{TableNotes}
        \footnotesize
        \item[a] cc-pVDZ basise set was used for DFT calculation.
        \item[b] 6-31G basise set was used for DFT calculation. Molecular
                geometry was obtained form Ref \citenum{zheng2005performance}.
        \item[c] Ref \citenum{dabo2014piecewise}
        \item [d] Geometry was optimized from B3LYP/6-31G* with Gaussian 09 package. \cite{Gaussian09}
        \item [e] Ref \citenum{li2017localized}
        \item [f] Ref \citenum{blase2011first} and \citenum{potentialselectron}, if not specified.

    \end{TableNotes}

\begin{longtable}{lccccccccccc}
    \caption{
    Comparison of (negative) LUMO energies (in eV) obtained from DFAs, LOSC-DFAs, self-consistent $GW$ (sc$GW$),
    $G_0W_0$@PBE, and $\Delta$SCF (PBE and LOSC-PBE) methods with the experimental electron affinity.}
    \label{tab:LUMO_eig}\\
    \toprule
        \multirow{2}{*}{Molecule} & \multirow{2}{*}{Exp\tnote{f}} & \multirow{2}{*}{sc$GW$\tnote{f}} & \multirow{2}{*}{$G_0W_0$\tnote{f}}
        & \multirow{2}{*}{$\Delta$PBE}  & \multirow{2}{*}{$\Delta$LOSC-} & \multirow{2}{*}{LDA} & \multirow{2}{*}{PBE}
        & \multirow{2}{*}{B3LYP} & \multirow{2}{*}{LOSC-} &\multirow{2}{*}{LOSC-} &\multirow{2}{*}{LOSC-} \\
        \\
        & & & & &PBE& & PBE& B3LYP& LDA& PBE&B3LYP\\
    \midrule
    \endfirsthead
    Table \ref{tab:LUMO_eig}. (Continued.) \\
    \toprule
        \multirow{2}{*}{Molecule} & \multirow{2}{*}{Exp\tnote{f}} & \multirow{2}{*}{sc$GW$\tnote{f}} & \multirow{2}{*}{$G_0W_0$\tnote{f}}
        & \multirow{2}{*}{$\Delta$PBE}  & \multirow{2}{*}{$\Delta$LOSC-} & \multirow{2}{*}{LDA} & \multirow{2}{*}{PBE}
        & \multirow{2}{*}{B3LYP} & \multirow{2}{*}{LOSC-} &\multirow{2}{*}{LOSC-} &\multirow{2}{*}{LOSC-} \\
        \\
        & & & & &PBE& & PBE& B3LYP& LDA& PBE&B3LYP\\
    \midrule
    \endhead
    \midrule
    \endfoot
    \midrule
    \insertTableNotes
    \endlastfoot

Anthracene\tnote{d}           & 0.60\tnote{e}   & N/A  & N/A  & 0.50  & 0.29  & 2.92 & 2.68 & 2.09 & 0.87  & 0.62  & 0.48  \\
Benzothiadiazole              & N/A             & 0.86 & 0.98 & 0.72  & 0.65  & 3.48 & 3.25 & 2.70 & 1.31  & 1.08  & 1.03  \\
Benzothiazole                 & N/A             & N/A  & N/A  & -0.37 & -0.54 & 2.32 & 2.09 & 1.41 & -0.12 & -0.37 & -0.49 \\
\ce{C60}\tnote{a}             & 2.70            & N/A  & N/A  & NA    & NA    & 4.34 & 4.07 & 3.56 & 2.61  & 2.30  & 2.14  \\
Fluorene                      & N/A             & N/A  & N/A  & -0.30 & -0.45 & 2.07 & 1.83 & 1.19 & 0.02  & -0.29 & -0.43 \\
\ce{H2P}                      & N/A             & N/A  & N/A  & 1.31  & 1.74  & 3.29 & 3.04 & 2.60 & 1.89  & 1.71  & 1.37  \\
\ce{H2PC}\tnote{a}            & N/A             & N/A  & N/A  & 1.92  & 1.83  & 3.65 & 3.41 & 2.97 & 2.16  & 1.92  & 1.65  \\
\ce{H2TPP}\tnote{a}           & 1.72            & N/A  & N/A  & 1.33  & 1.17  & 3.03 & 2.82 & 2.37 & 1.58  & 1.32  & 1.16  \\
Pentacene\tnote{d}            & N/A             & N/A  & N/A  & 1.50  & 1.24  & 3.55 & 3.31 & 2.80 & 1.59  & 1.38  & 1.47  \\
PTCDA                         & N/A             & N/A  & N/A  & 2.90  & 2.63  & 4.90 & 4.60 & 4.18 & 3.02  & 2.69  & 2.71  \\
Thiadiazole                   & N/A             & N/A  & N/A  & -0.46 & -0.47 & 2.88 & 2.66 & 1.99 & 0.59  & 0.37  & 0.18  \\
thiophene                     & N/A             & N/A  & N/A  & -1.48 & -1.48 & 1.57 & 1.36 & 0.68 & -0.55 & -0.76 & -0.99 \\
Benzoquinone                  & 1.88            & 2.13 & 2.27 & 1.68  & 1.67  & 4.76 & 4.48 & 3.81 & 2.93  & 2.65  & 2.35  \\
\ce{Cl4}-isobenzofuranedione  & 1.96            & 2.32 & 2.25 & 1.70  & 1.54  & 4.28 & 3.99 & 3.43 & 2.14  & 1.87  & 1.74  \\
Dichlone                      & 2.21            & 2.55 & 2.58 & 2.00  & 1.94  & 4.59 & 4.32 & 3.73 & 2.68  & 2.41  & 2.23  \\
\ce{F4}benzoquinone           & 2.69            & 3.07 & 2.91 & 2.17  & 2.17  & 5.31 & 4.98 & 4.49 & 3.52  & 3.19  & 3.04  \\
Maleic anhydride              & 1.42            & 1.54 & 1.65 & 1.05  & 1.05  & 4.41 & 4.11 & 3.42 & 2.24  & 1.94  & 1.68  \\
Nitrobenzene                  & 1.01            & 1.18 & 1.24 & 0.55  & 0.28  & 3.65 & 3.35 & 2.71 & 0.64  & 0.35  & 0.34  \\
Phenazine                     & 1.31            & 1.70 & 1.77 & 1.16  & 1.11  & 3.60 & 3.34 & 2.75 & 1.54  & 1.29  & 1.20  \\
Phthalimide                   & 1.02            & 1.17 & 1.26 & 0.69  & 0.58  & 3.49 & 3.20 & 2.56 & 1.50  & 1.22  & 1.04  \\
TCNE                          & 3.17            & 3.80 & 3.78 & 3.13  & 3.22  & 6.03 & 5.74 & 5.22 & 4.08  & 3.78  & 3.70  \\
Benzonitrile                  & 0.26            & 0.26 & 0.40 & -0.15 & -0.15 & 2.71 & 2.45 & 1.77 & 0.58  & 0.28  & 0.28  \\
\ce{Cl4}-benzoquinone         & 2.77            & 3.15 & 3.11 & 2.50  & 2.53  & 5.21 & 4.94 & 4.39 & 3.21  & 2.96  & 3.01  \\
Dinitrobenzonitrile           & 2.16            & 2.53 & 2.46 & 2.12  & 1.36  & 4.79 & 4.47 & 3.88 & 1.86  & 1.54  & 1.60  \\
\ce{F4}-benzenedicarbonitrile & 1.89            & 2.38 & 2.18 & 1.59  & 1.46  & 4.36 & 4.05 & 3.57 & 2.39  & 2.09  & 2.11  \\
Fumaronitrile                 & 1.25            & 1.49 & 1.68 & 1.11  & 1.11  & 4.31 & 4.04 & 3.37 & 2.07  & 1.79  & 1.61  \\
mDCBN                         & 0.91            & 1.16 & 1.27 & 0.77  & 0.69  & 3.50 & 3.23 & 2.58 & 1.55  & 1.28  & 1.07  \\
NDCA                          & N/A             & 1.85 & 1.91 & 1.40  & 1.10  & 3.87 & 3.59 & 3.01 & 1.90  & 1.62  & 1.43  \\
Nitrobenzonitrile             & 1.71            & 2.02 & 2.01 & 1.49  & 1.14  & 4.33 & 4.03 & 3.44 & 1.54  & 1.21  & 1.27  \\
Phthalic anhydride            & 1.26            & 1.44 & 1.52 & 0.96  & 0.87  & 3.81 & 3.52 & 2.87 & 1.75  & 1.46  & 1.36  \\
TCNQ                          & 2.80            & 4.11 & 4.02 & 3.41  & 3.77  & 5.75 & 5.47 & 5.08 & 4.31  & 4.05  & 3.89  \\
Acridine                      & 0.90            & 1.23 & 1.33 & 0.80  & 0.72  & 3.18 & 2.93 & 2.33 & 1.13  & 0.87  & 0.75  \\
Azulene                       & 0.75            & 1.07 & 1.10 & 0.43  & 0.60  & 3.07 & 2.82 & 2.20 & 1.17  & 0.92  & 0.78  \\
Bodipy                        & N/A             & 2.18 & 2.13 & 1.43  & 1.87  & 3.96 & 3.70 & 3.20 & 2.25  & 2.00  & 1.77  \\
Naphthalenedione              & 1.81            & 2.05 & 2.18 & 1.62  & 1.56  & 4.38 & 4.10 & 3.45 & 2.43  & 2.14  & 1.89  \\
\ce{C70}\tnote{b}             & 2.76\tnote{c}   & N/A  & N/A  & 2.45  & 2.31  & 4.18 & 3.95 & 3.46 & 2.54  & 2.36  & 2.15  \\
Benzene\tnote{d}              & -1.12\tnote{e}  & N/A  & N/A  & -1.63 & -1.64 & 1.38 & 1.13 & 0.39 & -0.73 & -0.97 & -1.26 \\
Naphthalene\tnote{d}          & -0.20\tnote{e}  & N/A  & N/A  & -0.36 & -0.40 & 2.25 & 2.00 & 1.33 & 0.16  & -0.09 & -0.31 \\
Tetracene\tnote{d}            & 1.06\tnote{e}   & N/A  & N/A  & 1.08  & 0.85  & 3.22 & 2.96 & 2.41 & 1.26  & 1.00  & 0.92  \\
Hexacene\tnote{d}             & 1.47\tnote{e}   & N/A  & N/A  & 1.82  & 1.88  & 3.67 & 3.42 & 2.91 & 1.76  & 1.48  & 1.45  \\
\midrule
MAE                           &                 & 0.34 & 0.37 & 0.26  & 0.38  & 2.43 & 2.16 & 1.57 & 0.48  & 0.33  & 0.29  \\
MSE                           &                 & 0.34 & 0.37 & -0.19 & -0.28 & 2.43 & 2.16 & 1.57 & 0.39  & 0.11  & -0.02
\end{longtable}
\end{ThreePartTable}
\end{landscape}

\begin{landscape}
\renewcommand\arraystretch{0.5}
\begin{ThreePartTable}
    \begin{TableNotes}
    \footnotesize
    \item[a] The experimental data for polyacence (n = 1 - 6) were obtained from Ref \citenum{schmidt1977photoelectron},
            other experimental data were obtained from Ref \citenum{chong2002interpretation}.
    \end{TableNotes}

\begin{longtable}{lcccccccccccc}
    \caption{Occupied orbital energies (in eV) obtained from DFAs, LOSC-DFAs and $\Delta$SCF methods (B3LYP and LOSC-B3LYP)
            compared with the experimental reference.}
    \label{tab:valence_eig}\\
    \toprule
        \multirow{2}{*}{Molecule} & \multirow{2}{*}{State} & \multirow{2}{*}{Exp\tnote{a}} & \multirow{2}{*}{$\Delta$B3LYP}
        &\multirow{2}{*}{$\Delta$LOSC-} &\multirow{2}{*}{LDA} & \multirow{2}{*}{PBE}  & \multirow{2}{*}{BLYP} & \multirow{2}{*}{B3LYP} & \multirow{2}{*}{LOSC-}
        & \multirow{2}{*}{LOSC} & \multirow{2}{*}{LOSC-} &\multirow{2}{*}{LOSC-} \\
        \\
        & & & &B3LYP & & & & & LDA & PBE & BLYP & B3LYP\\
    \midrule
    \endfirsthead
    Table \ref{tab:valence_eig}. (Continued.) \\
    \toprule
        \multirow{2}{*}{Molecule} & \multirow{2}{*}{State} & \multirow{2}{*}{Exp\tnote{a}} & \multirow{2}{*}{$\Delta$B3LYP}
        &\multirow{2}{*}{$\Delta$LOSC-} &\multirow{2}{*}{LDA} & \multirow{2}{*}{PBE}  & \multirow{2}{*}{BLYP} & \multirow{2}{*}{B3LYP} & \multirow{2}{*}{LOSC-}
        & \multirow{2}{*}{LOSC} & \multirow{2}{*}{LOSC-} &\multirow{2}{*}{LOSC-} \\
        \\
        & & & &B3LYP & & & & & LDA & PBE & BLYP & B3LYP\\
    \midrule
    \endhead
    \midrule
    \endfoot
    \midrule
    \insertTableNotes
    \endlastfoot
Benzene       & $E_{1g}$   & 9.24   & 9.07       & 9.06            & 6.50  & 6.28  & 6.09  & 7.04  & 8.94     & 8.71     & 8.51      & 8.98       \\
              & $E_{1g}$   & 9.24   & 9.08       & 9.07            & 6.50  & 6.28  & 6.09  & 7.04  & 8.93     & 8.70     & 8.50      & 8.96       \\
              & $A_{2u}$   & 12.25  & NA         & NA              & 8.26  & 8.18  & 8.12  & 9.48  & 11.41    & 11.33    & 11.54     & 12.12      \\
              &  MAE       &        & 0.17       & 0.18            & 3.16  & 3.33  & 3.48  & 2.39  & 0.48     & 0.66     & 0.73      & 0.22       \\
\midrule
Naphthalene   & $A_u$      & 8.15   & 7.71       & 8.00            & 5.68  & 5.46  & 5.26  & 6.11  & 8.16     & 8.01     & 7.71      & 8.06       \\
              & $B_{1u}$   & 8.87   & 8.47       & 9.03            & 6.40  & 6.18  & 5.97  & 6.86  & 9.37     & 9.20     & 8.84      & 9.12       \\
              & $B_{2g}$   & 10.08  & 9.44       & 9.73            & 7.34  & 7.12  & 6.91  & 7.98  & 9.84     & 9.73     & 9.37      & 10.03      \\
              & $B_{3g}$   & 10.83  & 10.32      & 11.15           & 7.98  & 7.87  & 7.81  & 9.13  & 11.25    & 11.17    & 11.12     & 11.54      \\
              &  MAE       &        & 0.50       & 0.25            & 2.63  & 2.82  & 3.00  & 1.97  & 0.29     & 0.29     & 0.37      & 0.27       \\
\midrule
Anthracene    & $B_{2g}$   & 7.41   & 6.89       & 7.30            & 5.19  & 4.97  & 4.76  & 5.54  & 7.66     & 7.44     & 7.23      & 7.36       \\
              & $B_{3g}$   & 8.54   & 8.10       & 8.90            & 6.36  & 6.13  & 5.92  & 6.76  & 9.28     & 9.11     & 8.87      & 9.01       \\
              & $A_u $     & 9.19   & 8.38       & 9.24            & 6.68  & 6.45  & 6.24  & 7.22  & 9.52     & 9.36     & 9.15      & 9.43       \\
              & $B_{2g}$   & 10.18  & NA         & NA              & 7.68  & 7.45  & 7.24  & 8.35  & 10.36    & 10.20    & 9.99      & 10.52      \\
              & $B_{1u}$   & 10.28  & 9.52       & 10.52           & 7.83  & 7.60  & 7.38  & 8.50  & 10.84    & 10.67    & 10.44     & 10.86      \\
              &  MAE       &        & 0.63       & 0.19            & 2.37  & 2.60  & 2.81  & 1.84  & 0.41     & 0.24     & 0.18      & 0.33       \\
\midrule
Teracene      & $A_u$      & 6.97   & 6.35       & 6.86            & 4.87  & 4.64  & 4.44  & 5.16  & 7.20     & 7.00     & 6.86      & 6.89       \\
              & $AB_{2g}$  & 8.41   & 7.61       & 8.54            & 6.16  & 5.93  & 5.72  & 6.64  & 8.82     & 8.62     & 8.50      & 8.60       \\
              & $B_{1u}$   & 8.41   & 7.84       & 8.81            & 6.35  & 6.11  & 5.89  & 6.71  & 9.27     & 9.08     & 8.83      & 8.80       \\
              & $A_u$      & 9.56   & 8.66       & 9.70            & 7.19  & 6.95  & 6.74  & 7.80  & 9.93     & 9.76     & 9.61      & 9.90       \\
              & $B_{3g}$   & 9.70   & 8.92       & 10.08           & 7.48  & 7.25  & 7.02  & 8.08  & 10.50    & 10.30    & 10.06     & 10.32      \\
              & $B_{2g}$   & 10.25  & 9.46       & 9.93            & 7.85  & 7.62  & 7.40  & 8.54  & 10.41    & 10.22    & 10.04     & 10.54      \\
              &  MAE       &        & 0.74       & 0.25            & 2.23  & 2.47  & 2.68  & 1.73  & 0.47     & 0.29     & 0.21      & 0.32       \\
\midrule
Pentacene     & $B_{2g}$   & 6.61   & 5.96       & 6.47            & 4.66  & 4.42  & 4.22  & 4.90  & 6.78     & 6.59     & 6.42      & 6.46       \\
              & $A_u$      & 7.92   & 7.04       & 7.88            & 5.77  & 5.54  & 5.33  & 6.19  & 8.26     & 8.00     & 7.83      & 8.16       \\
              & $B_{3g}$   & 8.32   & 7.64       & 8.79            & 6.34  & 6.10  & 5.87  & 6.68  & 9.24     & 9.04     & 8.80      & 8.97       \\
              & $B_{2g}$   & 9.01   & 8.02       & 9.20            & 6.74  & 6.51  & 6.29  & 7.29  & 9.49     & 9.28     & 9.15      & 9.16       \\
              & $B_{1u}$   & 9.39   & 8.50       & 9.79            & 7.24  & 7.00  & 6.78  & 7.78  & 10.21    & 9.98     & 9.77      & 10.14      \\
              & $A_u$      & 9.80   & 8.79       & 10.04           & 7.48  & 7.24  & 7.02  & 8.12  & 10.11    & 9.85     & 9.70      & 10.11      \\
              & $B_{2g}$   & 10.23  & NA         & NA              & 7.88  & 7.72  & 7.50  & 8.65  & 11.48    & 10.46    & 10.28     & 10.67      \\
              & $MAE$      &        & 0.85       & 0.25            & 2.17  & 2.39  & 2.61  & 1.67  & 0.61     & 0.28     & 0.20      & 0.38       \\
\midrule
Hexacene      & $A_u$      & 6.36   & 5.67       & 5.97            & 4.50  & 4.26  & 4.06  & 4.73  & 6.62     & 6.40     & 6.18      & 6.30       \\
              & $B_{2g}$   & 7.35   & 6.60       & 7.41            & 5.48  & 5.24  & 5.03  & 5.84  & 7.95     & 7.77     & 7.56      & 7.80       \\
              & $B_{3u}$   & 8.12   & 7.49       & 8.36            & 6.33  & 6.09  & 5.86  & 6.66  & 9.26     & 9.06     & 8.78      & 8.96       \\
              & $A_u$      & 8.56   & 7.50       & 8.29            & 6.37  & 6.14  & 5.92  & 6.87  & 9.09     & 8.54     & 8.35      & 8.68       \\
              & $B_{1g}$   & 9.36   & 8.18       & 9.38            & 7.07  & 6.83  & 6.60  & 7.57  & 10.01    & 9.79     & 9.55      & 9.91       \\
              & $B_{2g}$   & 9.36   & 8.24       & 8.83            & 7.11  & 6.87  & 6.65  & 7.70  & 9.69     & 9.56     & 9.35      & 9.76       \\
              & $A_u$      & 9.95   & 8.83       & 9.43            & 7.66  & 7.43  & 7.20  & 8.32  & 10.42    & 9.97     & 9.82      & 10.32      \\
              & $B_{3u}$   & 9.95   & 9.14       & 10.62           & 7.88  & 7.75  & 7.55  & 8.66  & 11.51    & 11.36    & 10.57     & 11.02      \\
              & MAE        &        & 0.92       & 0.34            & 2.08  & 2.30  & 2.52  & 1.58  & 0.69     & 0.43     & 0.28      & 0.48       \\
\midrule
Thiophene     & 1$A_{2} $  & 8.87   & 8.65       & 8.56            & 5.98  & 5.78  & 5.59  & 6.58  & 8.46     & 8.27     & 8.08      & 8.56       \\
              & 3$B_{1} $  & 9.52   & 9.09       & 9.02            & 6.38  & 6.21  & 6.02  & 7.02  & 8.85     & 8.69     & 8.50      & 8.99       \\
              & 11$A_1$    & 12.10  & 11.46      & 11.52           & 8.49  & 8.37  & 8.22  & 9.46  & 11.58    & 11.45    & 11.34     & 11.96      \\
              & 2$B_{1} $  & 12.70  & 12.19      & 12.19           & 9.45  & 9.26  & 9.03  & 10.39 & 12.05    & 11.87    & 11.63     & 12.46      \\
              & 7$B_{2} $  & 13.30  & 12.36      & 12.30           & 9.47  & 9.38  & 9.24  & 10.55 & 12.10    & 12.02    & 11.93     & 12.65      \\
              & 10$A_1$    & 13.90  & 12.59      & 12.69           & 9.60  & 9.48  & 9.37  & 10.78 & 12.28    & 12.14    & 12.09     & 13.10      \\
              & 6$B_{2} $  & 14.30  & 12.96      & 13.01           & 9.95  & 9.87  & 9.78  & 11.20 & 12.59    & 12.57    & 12.50     & 13.29      \\
              & 9$A_{1} $  & 16.60  & 15.45      & 15.46           & 12.61 & 12.50 & 12.32 & 13.92 & 15.18    & 15.08    & 14.93     & 15.94      \\
              & 5$B_{2} $  & 17.60  & 16.46      & 16.24           & 13.50 & 13.47 & 13.29 & 14.99 & 15.93    & 15.90    & 15.78     & 17.00      \\
              & 8$A_{1} $  & 18.30  & 16.80      & 16.81           & 13.76 & 13.73 & 13.56 & 15.30 & 16.19    & 16.18    & 16.09     & 17.32      \\
              & MAE        &        & 0.92       & 0.94            & 3.80  & 3.92  & 4.08  & 2.70  & 1.20     & 1.30     & 1.43      & 0.59       \\
\midrule
Ethylene      & 1$B_{3u}$  & 10.68  & 10.53      & 10.53           & 6.92  & 6.73  & 6.55  & 7.62  & 10.63    & 10.45    & 10.26     & 10.59      \\
              & 1$B_{3g}$  & 12.80  & 12.39      & 12.39           & 8.48  & 8.51  & 8.47  & 9.82  & 11.33    & 11.35    & 11.31     & 12.10      \\
              & 3$A_{g }$  & 14.80  & 14.24      & 14.24           & 10.23 & 10.14 & 10.08 & 11.56 & 13.92    & 13.83    & 13.78     & 14.52      \\
              & 1$B_{2u}$  & 16.00  & 15.45      & 15.45           & 11.56 & 11.49 & 11.39 & 12.93 & 14.80    & 14.72    & 14.61     & 15.50      \\
              & 2$B_{1u}$  & 19.10  & 18.20      & 18.20           & 14.16 & 14.22 & 14.08 & 15.89 & 17.34    & 17.41    & 17.27     & 18.44      \\
              & 2$A_{g }$  & 23.60  & 23.25      & 23.25           & 18.72 & 18.79 & 18.60 & 20.82 & 23.18    & 23.25    & 23.06     & 24.40      \\
              & MAE        &        & 0.49       & 0.49            & 4.49  & 4.52  & 4.63  & 3.06  & 0.96     & 0.99     & 1.12      & 0.51       \\
\midrule
Water         & 1$B_{1 }$  & 12.62  & 12.60      & 12.60           & 6.99  & 6.83  & 6.77  & 8.49  & 13.11    & 12.98    & 12.92     & 13.42      \\
              & 3$A_{1 }$  & 14.74  & 14.71      & 14.71           & 9.02  & 8.94  & 8.85  & 10.55 & 15.13    & 15.03    & 14.94     & 15.41      \\
              & 1$B_{2 }$  & 18.55  & 18.74      & 18.74           & 12.98 & 12.90 & 12.81 & 14.44 & 18.46    & 18.38    & 18.28     & 18.83      \\
              & MAE        &        & 0.08       & 0.08            & 5.64  & 5.75  & 5.83  & 4.14  & 0.32     & 0.27     & 0.26      & 0.58       \\
\midrule
MAE Polyacene &            &        & 0.73       & 0.26            & 2.33  & 2.55  & 2.76  & 1.79  & 0.53     & 0.35     & 0.29      & 0.36       \\
MSE Polyacene &            &        & -0.73      & 0.04            & -2.33 & -2.55 & -2.76 & -1.79 & 0.42     & 0.19     & -0.02     & 0.29       \\
Total MAE     &            &        & 0.70       & 0.41            & 3.06  & 3.23  & 3.40  & 2.24  & 0.69     & 0.60     & 0.60      & 0.43       \\
Total MSE     &            &        & -0.69      & -0.22           & -3.06 & -3.23 & -3.40 & -2.24 & -0.06    & -0.23    & -0.41     & 0.08

\end{longtable}
\end{ThreePartTable}
\end{landscape}

\clearpage
\section{Excitation energy}
Table \ref{tab:exci_N-1} - \ref{tab:exci_error_N+1} show the detailed low-lying excitation energies
of 16 molecular test set from different methods.
The basis set used for DFT calculation was 6-311++G(3df, 3pd), if not specified.
Gaussian 09 package \cite{Gaussian09} was applied to perform TD-DFT calculation, and cc-pVTZ were used as basis set.
Geometries were obtained from Ref \citenum{yang_excitation_2014}.
Table \ref{tab:Rydberg_Li} - \ref{tab:Rydberg_Mg_Triplet}
show detailed excitation energies of 4 atoms (Li, Be, Mg and Na) in which Rydberg excited states were concerned.
To describe Rydberg states of these 4 atoms, greatly diffused basis set were used
to perform the calculation. Here even-tempered basis set was built, with its Gaussian orbital exponents
$\alpha$ satisfying $\alpha_i=2^{i-1}\alpha_1$.
Each of the even-tempered basis contains 17s, 15p and 11d functions with the
smallest exponents being $\alpha_1=0.000976525, 0.000976525,$ and $0.0039062500$,
respectively. For excitation calculation from DFT by orbital energies, unrestricted calculation were applied
for all test cases.

\begin{landscape}
\renewcommand\arraystretch{0.5}
\begin{ThreePartTable}
    \begin{TableNotes}
    \footnotesize
    \item[a] Reference data labeled with an asterisk are obtained from CC2/aug-cc-pVTZ calculation with TURBOMOLE package
            \cite{TURBOMOLE}, other reference values were obtained from Ref \citenum{schreiber2008benchmarks}.
    \end{TableNotes}

\begin{longtable}{lccccccccccc}
    \caption{Low-lying vertical excitation energies (in eV) obtained from N-1 system with Hartree Fock (HF), DFAs,
            LOSC-DFAs, TD-B3LYP and $\Delta$SCF-B3LYP.}
    \label{tab:exci_N-1}\\
    \toprule
    \multirow{2}{*}{Molecule} & \multirow{2}{*}{MO} & \multirow{2}{*}{Ref\tnote{a}} & \multirow{2}{*}{TD-}
    & \multirow{2}{*}{$\Delta$SCF-}  & \multirow{2}{*}{BLYP} & \multirow{2}{*}{B3LYP} & \multirow{2}{*}{LDA}
    & \multirow{2}{*}{HF} & \multirow{2}{*}{LOSC-} &\multirow{2}{*}{LOSC-} & \multirow{2}{*}{LOSC-}\\
    \\
    & & &B3LYP&B3LYP& & & & &BLYP&B3LYP&LDA\\
    \midrule
    \endfirsthead
    Table \ref{tab:exci_N-1}. (Continued.) \\
    \toprule
        \multirow{2}{*}{Molecule} & \multirow{2}{*}{MO} & \multirow{2}{*}{Ref\tnote{a}} & \multirow{2}{*}{TD-}
        & \multirow{2}{*}{$\Delta$SCF-}  & \multirow{2}{*}{BLYP} & \multirow{2}{*}{B3LYP} & \multirow{2}{*}{LDA}
        & \multirow{2}{*}{HF} & \multirow{2}{*}{LOSC-} &\multirow{2}{*}{LOSC-} & \multirow{2}{*}{LOSC-}\\
        \\
        & & &B3LYP&B3LYP& & & & &BLYP&B3LYP&LDA\\
    \midrule
    \endhead
    \midrule
    \endfoot
    \midrule
    \insertTableNotes
    \endlastfoot
    Ethene          & $^3B_{1u}$ & 4.50     & 4.05 & 4.38 & 4.61 & 4.43  & 4.90 & 3.09 & 4.15 & 4.05 & 4.40 \\
    Ethene          & $^1B_{1u}$ & 7.80     & 7.38 & 7.09 & 7.46 & 7.75  & 7.27 & 6.61 & 6.61 & 7.00 & 6.52 \\
    Furan           & $^3B_2$    & 4.17     & 3.70 & 3.98 & 4.05 & 3.95  & 4.31 & 2.95 & 3.74 & 3.71 & 3.88 \\
    Furan           & $^3A_2$    & 5.99$^*$ & 5.48 & 5.75 & 5.78 & 5.84  & 5.99 & 3.99 & 5.63 & 5.71 & 5.71 \\
    Furan           & $^1B_2$    & 6.32     & 5.94 & 5.78 & 5.73 & 6.11  & 5.60 & 5.07 & 5.76 & 6.08 & 5.43 \\
    Furan           & $^1A_2$    & 6.03$^*$ & 5.51 & 5.88 & 6.91 & 7.27  & 6.91 & 4.92 & 6.86 & 7.15 & 6.69 \\
    Benzoquinone    & $^3B_{1g}$ & 2.51     & 1.93 & 2.08 & 1.68 & 2.08  & 1.60 & 3.28 & 2.08 & 2.36 & 1.99 \\
    Benzoquinone    & $^3B_{3u}$ & 5.38$^*$ & 5.18 & 5.12 & 4.76 & 5.40  & 4.66 & 7.21 & 5.36 & 5.81 & 5.28 \\
    Benzoquinone    & $^1B_{1g}$ & 2.78     & 2.43 & 2.38 & 1.95 & 2.41  & 1.94 & 4.11 & 2.30 & 2.64 & 2.30 \\
    Benzoquinone    & $^1B_{3u}$ & 5.60     & 5.38 & 5.51 & 4.85 & 5.49  & 4.78 & 7.14 & 5.45 & 5.89 & 5.40 \\
    cyclopentadiene & $^3B_2$    & 3.25     & 2.74 & 3.11 & 3.21 & 3.12  & 3.46 & 2.30 & 2.82 & 2.82 & 3.06 \\
    cyclopentadiene & $^3A_2$    & 5.61$^*$ & 5.09 & 5.37 & 5.65 & 5.76  & 5.85 & 3.60 & 5.49 & 5.59 & 5.67 \\
    cyclopentadiene & $^1B_2$    & 5.55     & 4.95 & 4.82 & 4.74 & 5.06  & 4.66 & 4.93 & 4.29 & 4.70 & 4.22 \\
    cyclopentadiene & $^1A_2$    & 5.65$^*$ & 5.11 & 5.40 & 6.79 & 7.14  & 6.77 & 4.66 & 6.38 & 6.75 & 6.44 \\
    butadiene       & $^3B_u$    & 3.20     & 2.79 & 3.20 & 3.22 & 3.19  & 3.40 & 2.54 & 2.76 & 3.06 & 2.90 \\
    butadiene       & $^3B_g$    & 6.22$^*$ & 5.67 & 5.80 & 5.86 & 6.02  & 6.05 & 4.09 & 5.53 & 5.96 & 5.68 \\
    butadiene       & $^1B_u$    & 6.18     & 5.56 & 5.13 & 4.86 & 5.39  & 4.72 & 5.67 & 5.08 & 5.74 & 4.80 \\
    butadiene       & $^1B_g$    & 6.26$^*$ & 5.70 & 6.10 & 6.80 & 7.04  & 6.86 & 4.73 & 6.81 & 7.27 & 6.78 \\
    hexatriene      & $^3B_u$    & 2.40     & 2.12 & 2.52 & 2.49 & 2.51  & 2.62 & 1.92 & 2.76 & 2.53 & 2.94 \\
    hexatriene      & $^3A_u$    & 5.68$^*$ & 5.22 & 4.99 & 4.82 & 5.10  & 4.95 & 3.55 & 5.54 & 5.44 & 5.72 \\
    hexatriene      & $^1B_u$    & 5.10     & 4.60 & 4.11 & 3.66 & 4.22  & 3.54 & 5.20 & 4.02 & 4.70 & 3.97 \\
    hexatriene      & $^1A_u$    & 5.71$^*$ & 5.24 & 5.79 & 5.58 & 6.05  & 5.60 & 3.98 & 6.09 & 6.36 & 6.26 \\
    octetraene      & $^3B_u$    & 2.20     & 1.71 & 2.10 & 2.05 & 2.11  & 2.16 & 1.57 & 2.76 & 2.33 & 2.91 \\
    octetraene      & $^1B_u$    & 4.66     & 3.96 & 3.47 & 2.96 & 3.52  & 2.86 & 4.86 & 3.93 & 3.92 & 3.88 \\
    cyclopropene    & $^3B_2$    & 4.34     & 3.70 & 4.03 & 4.22 & 4.19  & 4.38 & 3.16 & 4.49 & 4.34 & 4.69 \\
    cyclopropene    & $^1B_2$    & 7.06     & 6.09 & 6.27 & 6.29 & 6.69  & 5.97 & 6.20 & 6.23 & 6.50 & 6.07 \\
    norbornadiene   & $^3A_2$    & 3.72     & 3.10 & 3.54 & 3.62 & 3.71  & 3.73 & 3.50 & 4.00 & 3.92 & 4.08 \\
    norbornadiene   & $^1A_2$    & 5.34     & 4.70 & 4.73 & 4.64 & 4.90  & 4.56 & 4.89 & 4.82 & 4.87 & 4.77 \\
    s-tetrazine     & $^3B_u$    & 1.89     & 1.47 & 1.62 & 1.32 & 1.63  & 1.21 & 2.48 & 1.28 & 1.72 & 1.22 \\
    s-tetrazine     & $^3A_u$    & 3.52     & 3.15 & 3.34 & 2.98 & 3.45  & 2.85 & 4.90 & 2.95 & 3.54 & 2.86 \\
    s-tetrazine     & $^1B_u$    & 2.24     & 2.27 & 2.12 & 1.77 & 2.17  & 1.74 & 3.57 & 1.73 & 2.25 & 1.74 \\
    s-tetrazine     & $^1A_u$    & 3.48     & 3.54 & N/A  & 3.13 & 3.59  & 3.06 & 4.88 & 3.08 & 3.66 & 3.06 \\
    formaldehyde    & $^3A_2$    & 3.50     & 3.10 & 3.19 & 3.37 & 3.17  & 3.38 & 1.17 & 2.94 & 2.73 & 3.08 \\
    formaldehyde    & $^1A_2$    & 3.88     & 3.83 & 3.48 & 3.63 & 3.41  & 3.89 & 1.20 & 3.09 & 2.87 & 3.47 \\
    acetone         & $^3B_1$    & 4.05     & 3.68 & 3.75 & 3.95 & 3.78  & 4.09 & 1.82 & 3.36 & 3.24 & 3.53 \\
    acetone         & $^3B_2$    & 5.87$^*$ & 5.71 & 6.20 & 7.95 & 7.47  & 8.38 & 2.81 & 6.23 & 5.99 & 6.78 \\
    acetone         & $^1B_1$    & 4.40     & 4.30 & 4.00 & 4.19 & 4.01  & 4.50 & 1.89 & 3.57 & 3.43 & 3.90 \\
    acetone         & $^1B_2$    & 5.92$^*$ & 5.77 & 6.24 & 8.02 & 7.54  & 8.46 & 2.85 & 6.30 & 6.01 & 6.65 \\
    pyridine        & $^3B_1$    & 4.25     & 4.05 & 4.42 & 4.37 & 4.42  & 4.43 & 2.82 & 3.22 & 4.21 & 2.93 \\
    pyridine        & $^3A_2$    & 5.28     & 4.96 & N/A  & 4.99 & 4.60  & 5.15 & 3.54 & 3.84 & 4.40 & 3.65 \\
    pyridine        & $^1B_1$    & 4.59     & 4.76 & 4.45 & 4.80 & 5.54  & 4.89 & 2.09 & 3.57 & 5.29 & 3.33 \\
    pyridine        & $^1A_2$    & 5.11     & 5.10 & N/A  & 5.36 & 6.40  & 5.32 & 3.58 & 4.13 & 6.07 & 3.77 \\
    pyridazine      & $^1B_1$    & 3.78     & 3.60 & 3.51 & 3.61 & 3.64  & 3.64 & 3.86 & 2.41 & 2.69 & 2.40 \\
    pyridazine      & $^1A_2$    & 4.32     & 4.19 & N/A  & 4.25 & 4.25  & 4.26 & 2.59 & 3.03 & 3.28 & 2.99 \\
    pyrizine        & $^1B_u$    & 3.95     & 3.93 & 3.81 & 3.53 & 3.90  & 3.52 & 5.12 & 3.57 & 3.55 & 3.55 \\
    pyrizine        & $^1A_u$    & 4.81     & 4.69 & N/A  & 4.39 & 4.76  & 4.34 & 5.56 & 4.35 & 4.35 & 4.30 \\
    pyrimidine      & $^1B_1$    & 4.55     & 4.25 & 4.11 & 3.97 & 4.17  & 4.01 & 3.40 & 3.72 & 3.77 & 3.75 \\
    pyrimidine      & $^1A_2$    & 4.91     & 4.60 & N/A  & 4.37 & 4.59  & 4.40 & 3.95 & 4.19 & 4.23 & 4.19 \\
    \midrule
    MAE             &            &          & 0.38 & 0.35 & 0.53 & 0.42  & 0.58 & 1.35 & 0.63 & 0.49 & 0.67 \\
    MSE             &            &          &-0.37 &-0.31 &-0.22 &-0.01  &-0.16 &-0.83 &-0.44 &-0.19 &-0.42 \\
\end{longtable}
\end{ThreePartTable}
\end{landscape}

\begin{landscape}
\renewcommand\arraystretch{0.5}
\begin{ThreePartTable}
    \begin{TableNotes}
    \footnotesize
    \item[a] Reference values were obtained from Ref \citenum{schreiber2008benchmarks}.
    \end{TableNotes}

\begin{longtable}{lccccccccccc}
    \caption{Low-lying vertical excitation energies (in eV) obtained from N+1 system with Hartree Fock (HF), DFAs,
            LOSC-DFAs, TD-B3LYP and $\Delta$SCF-B3LYP.}
    \label{tab:exci_N+1}\\
    \toprule
    \multirow{2}{*}{Molecule} & \multirow{2}{*}{MO} & \multirow{2}{*}{Ref\tnote{a}} & \multirow{2}{*}{TD-}
    & \multirow{2}{*}{$\Delta$SCF-}  & \multirow{2}{*}{BLYP} & \multirow{2}{*}{B3LYP} & \multirow{2}{*}{LDA}
    & \multirow{2}{*}{HF} & \multirow{2}{*}{LOSC-} &\multirow{2}{*}{LOSC-} & \multirow{2}{*}{LOSC-}\\
    \\
    & & &B3LYP&B3LYP& & & & &BLYP&B3LYP&LDA\\
    \midrule
    \endfirsthead
    Table \ref{tab:exci_N+1}. (Continued.)\\
    \toprule
    \multirow{2}{*}{Molecule} & \multirow{2}{*}{MO} & \multirow{2}{*}{Ref\tnote{a}} & \multirow{2}{*}{TD-}
    & \multirow{2}{*}{$\Delta$SCF-}  & \multirow{2}{*}{BLYP} & \multirow{2}{*}{B3LYP} & \multirow{2}{*}{LDA}
    & \multirow{2}{*}{HF} & \multirow{2}{*}{LOSC-} &\multirow{2}{*}{LOSC-} & \multirow{2}{*}{LOSC-}\\
    \\
    & & &B3LYP&B3LYP& & & & &BLYP&B3LYP&LDA\\
    \midrule
    \endhead
    \midrule
    \endfoot
    \midrule
    \insertTableNotes
    \endlastfoot
    Ethene          & $^1B_{1u}$ & 7.80 & 7.38     & 7.09      & 4.89 & 5.61  & 5.07 & 7.63  & 7.78      & 7.92       & 7.92     \\
    Ethene          & $^3B_{1u}$ & 4.50 & 4.05     & 4.38      & 4.86 & 5.59  & 5.02 & 7.65  & 7.75      & 7.88       & 7.89     \\
    Furan           & $^1B_2$    & 6.32 & 5.94     & 5.78      & 3.96 & 4.67  & 4.15 & 6.90  & 6.15      & 6.37       & 6.38     \\
    Furan           & $^3B_2$    & 4.17 & 3.70     & 3.98      & 3.95 & 4.65  & 4.12 & 6.90  & 6.22      & 6.33       & 6.34     \\
    Benzoquinone    & $^1B_{1g}$ & 2.78 & 2.43     & 2.38      & 1.81 & 2.39  & 1.79 & 4.71  & 4.67      & 4.55       & 4.51     \\
    Benzoquinone    & $^3B_{1g}$ & 2.51 & 1.93     & 2.08      & 1.54 & 2.06  & 1.45 & 2.71  & 4.34      & 4.20       & 4.15     \\
    cyclopentadiene & $^1B_2$    & 5.55 & 4.95     & 4.82      & 3.64 & 4.32  & 3.75 & 6.63  & 5.74      & 5.99       & 5.75     \\
    cyclopentadiene & $^3B_2$    & 3.25 & 2.74     & 3.11      & 3.63 & 4.31  & 3.66 & 6.63  & 5.73      & 5.98       & 5.67     \\
    butadiene       & $^1B_u$    & 6.18 & 5.56     & 5.13      & 3.92 & 4.71  & 3.95 & 6.68  & 6.36      & 5.74       & 6.14     \\
    butadiene       & $^3B_u$    & 3.20 & 2.79     & 3.20      & 3.52 & 3.27  & 3.50 & 6.67  & 5.81      & 3.88       & 5.56     \\
    hexatriene      & $^1B_u$    & 5.10 & 4.60     & 4.11      & 3.25 & 3.88  & 3.22 & 6.20  & 4.13      & 4.71       & 4.28     \\
    hexatriene      & $^3B_u$    & 2.40 & 2.12     & 2.52      & 2.45 & 2.55  & 2.56 & 6.17  & 3.24      & 2.92       & 3.40     \\
    octetraene      & $^1B_u$    & 4.66 & 3.96     & 3.47      & 2.72 & 3.30  & 2.68 & 5.87  & 4.12      & 3.83       & 3.99     \\
    octetraene      & $^3B_u$    & 2.20 & 1.71     & 2.10      & 2.04 & 2.13  & 2.12 & 1.82  & 3.17      & 2.61       & 3.26     \\
    cyclopropene    & $^1B_2$    & 7.06 & 6.09     & 6.27      & 4.42 & 5.18  & 4.49 & 7.83  & 6.74      & 7.04       & 6.74     \\
    cyclopropene    & $^3B_2$    & 4.34 & 3.70     & 4.03      & 4.40 & 5.16  & 4.45 & 7.81  & 6.72      & 7.02       & 6.71     \\
    norbornadiene   & $^1A_2$    & 5.34 & 4.70     & 4.73      & 3.76 & 4.51  & 3.77 & 6.99  & 6.56      & 6.72       & 6.42     \\
    norbornadiene   & $^3A_2$    & 3.72 & 3.10     & 3.54      & 3.75 & 4.50  & 3.66 & 6.98  & 6.55      & 6.72       & 6.33     \\
    s-tetrazine     & $^1B_u$    & 2.24 & 2.27     & 2.12      & 1.74 & 2.09  & 1.74 & 3.11  & 2.96      & 2.86       & 3.00     \\
    s-tetrazine     & $^3B_u$    & 1.89 & 1.47     & 1.62      & 1.37 & 1.61  & 1.30 & 1.90  & 2.64      & 2.60       & 2.57     \\
    formaldehyde    & $^1A_2$    & 3.88 & 3.83     & 3.48      & 3.34 & 3.67  & 3.28 & 5.65  & 6.18      & 4.78       & 5.84     \\
    formaldehyde    & $^3A_2$    & 3.50 & 3.10     & 3.19      & 3.19 & 3.29  & 3.02 & 4.82  & 5.99      & 4.35       & 5.55     \\
    acetone         & $^1B_1$    & 4.40 & 4.30     & 4.00      & 3.97 & 5.08  & N/A  & 8.79  & 7.51      & 7.77       & N/A      \\
    acetone         & $^3B_1$    & 4.05 & 3.68     & 3.75      & 3.95 & 5.07  & N/A  & 8.80  & 7.52      & 7.75       & N/A      \\
    pyridine        & $^1B_1$    & 4.59 & 4.76     & 4.45      & 4.12 & 5.36  & 4.00 & 7.70  & 7.67      & 8.16       & 7.29     \\
    pyridine        & $^3B_1$    & 4.25 & 4.05     & 4.42      & 4.04 & 5.29  & 3.85 & 7.69  & 7.61      & 8.09       & 7.16     \\
    pyridazine      & $^1B_1$    & 3.78 & 3.60     & 3.51      & 2.93 & 3.40  & 2.83 & 6.27  & 4.96      & 4.40       & 4.64     \\
    pyrizine        & $^1B_u$    & 3.95 & 3.93     & 3.81      & 3.28 & 3.74  & 3.27 & 5.27  & 6.40      & 5.58       & 5.57     \\
    pyrimidine      & $^1B_1$    & 4.55 & 4.25     & 4.11      & 3.64 & 4.07  & 3.58 & 7.81  & 6.82      & 5.75       & 6.57     \\
    \midrule
    MAE             &            &      & 0.41     & 0.40      & 0.91 & 0.78  & 0.98 & 2.05  & 1.72      & 1.51       & 1.47     \\
    MSE             &            &      &-0.40     &-0.38      &-0.83 &-0.23  &-0.87 & 2.01  &  1.58     &  1.39      &  1.33    \\
\end{longtable}
\end{ThreePartTable}
\end{landscape}

\begin{table}[htpb]
\centering
    \caption{Errors (in eV) of the vertical excitation energies from ($N+1$)-system with HF, DFAs,
            LOSC-DFAs, TD-B3LYP, and $\Delta$SCF-DFT. T1 stands for triplet HOMO to LUMO excitation,
            and S1 stands for singlet HOMO to LUMO excitation.}
    \label{tab:exci_error_N+1}
    \begin{tabular}{@{}lcccccc@{}}
    \toprule
    \multirow{2}{*}{Method} & \multicolumn{2}{c}{T1} & \multicolumn{2}{c}{S1} & \multicolumn{2}{c}{Total} \\ \cmidrule{2-7}
                        & MAE       & MSE        & MAE  & MSE   & MAE   & MSE   \\ \midrule
    HF                  & 2.56      &  2.51      & 1.64 &  1.62 & 2.05  &  2.01 \\
    BLYP                & 0.28      & -0.10      & 1.43 & -1.43 & 0.91  & -0.83 \\
    B3LYP               & 0.58      &  0.42      & 0.94 & -0.76 & 0.78  & -0.23 \\
    LDA                 & 0.35      & -0.10      & 1.48 & -1.48 & 0.98  & -0.87 \\
    LOSC-BLYP           & 2.25      &  2.25      & 1.29 &  1.04 & 1.72  &  1.58 \\
    LOSC-B3LYP          & 2.03      &  2.03      & 1.08 &  0.87 & 1.51  &  1.39 \\
    LOSC-LDA            & 2.05      &  2.05      & 1.00 &  0.75 & 1.47  &  1.33 \\
    TD-B3LYP            & 0.45      & -0.45      & 0.38 & -0.35 & 0.41  & -0.40 \\
    $\Delta$SCF-B3LYP   & 0.20      & -0.16      & 0.56 & -0.56 & 0.40  & -0.38 \\ \bottomrule
    \end{tabular}
\end{table}

\clearpage
\begin{table}[htpb]
    \centering
    \caption{Doublet excitation energy (in eV) of atom Li. Experimental data were obtained
            from Ref \citenum{NIST_ASD}}
    \label{tab:Rydberg_Li}
    \begin{tabular}{@{}lccccccc@{}} \toprule
    AO         & Exp  & LDA  & BLYP & B3LYP & LOSC-LDA & LOSC-BLYP & LOSC-B3LYP \\ \midrule
    2p         & 1.85 & 1.95 & 2.07 & 2.06  & 1.96     & 2.18      & 2.08       \\
    3s         & 3.37 & 4.30 & 4.80 & 4.56  & 2.78     & 3.53      & 3.46       \\
    3p         & 3.83 & 4.72 & 5.33 & 5.07  & 3.15     & 4.02      & 3.90       \\
    3d         & 3.88 & 4.98 & 5.56 & 5.27  & 3.43     & 4.29      & 4.15       \\
    4s         & 4.34 & 5.40 & 6.23 & 5.84  & 3.31     & 4.30      & 4.24       \\
    4p         & 4.52 & 5.56 & 6.42 & 6.03  & 3.43     & 4.45      & 4.43       \\
    4d         & 4.54 & 5.66 & 6.52 & 6.12  & 3.58     & 4.55      & 4.53       \\
    5s         & 4.75 & 5.85 & 6.81 & 6.19  & 3.54     & 4.66      & 4.39       \\
    5p         & 4.84 & 5.92 & 6.92 & 6.49  & 3.60     & 4.78      & 4.72       \\
    5d         & 4.85 & -    & 7.04 & 6.62  & -        & 5.00      & 4.98       \\
    6s         & 4.96 & 6.07 & 7.11 & 6.62  & 3.61     & 4.74      & 4.69       \\
    6p         & 5.01 & 6.12 & 7.12 & 6.68  & 3.67     & 4.71      & 4.76       \\ \midrule
    MAE        &      & 0.97 & 1.77 & 1.40  & 0.91     & 0.17      & 0.16       \\
    MSE        &      & 0.97 & 1.77 & 1.40  & -0.89    & 0.04      & -0.03      \\ \bottomrule
    \end{tabular}
\end{table}

\begin{table}[htpb]
    \centering
    \caption{Doublet excitation energy (in eV) of atom Na. Experimental data were obtained
            from Ref \citenum{NIST_ASD}}
    \label{tab:Rydberg_Na}
    \begin{tabular}{@{}lccccccc@{}} \toprule
    AO         & Exp  & LDA  & BLYP & B3LYP & LOSC-LDA & LOSC-BLYP & LOSC-B3LYP \\ \midrule
    3p         & 2.10 & 2.86 & 3.04 & 2.86  & 2.48     & 2.70      & 2.60       \\
    4s         & 3.19 & 4.52 & 5.34 & 4.83  & 3.29     & 4.04      & 3.82       \\
    3d         & 3.62 & 5.10 & 5.81 & 5.33  & 3.93     & 4.66      & 4.38       \\
    4p         & 3.75 & 5.15 & 5.89 & 5.39  & 3.81     & 4.49      & 4.28       \\
    5s         & 4.12 & 5.64 & 6.37 & 5.87  & 3.90     & 4.63      & 4.48       \\
    4d         & 4.28 & 5.85 & 6.53 & 6.04  & 4.14     & 4.93      & 4.73       \\
    5p         & 4.34 & 5.88 & 6.62 & 6.11  & 4.08     & 4.80      & 4.66       \\
    6s         & 4.51 & 6.10 & 6.82 & 6.32  & 4.15     & 4.89      & 4.77       \\
    6p         & 4.62 & 6.21 & 6.95 & 6.44  & 4.24     & 4.95      & 4.85       \\
    5d         & 4.59 & 6.34 & 7.00 & 6.52  & 4.60     & 5.29      & 5.16       \\
    7s         & 4.71 & 6.41 & 7.06 & 6.56  & 4.34     & 4.97      & 4.92       \\
    7p         & 4.78 & 6.78 & 7.13 & 6.63  & 4.35     & 5.08      & 4.99       \\ \midrule
    MAE        &      & 1.52 & 2.16 & 1.69  & 0.25     & 0.57      & 0.42       \\
    MSE        &      & 1.52 & 2.16 & 1.69  & -0.11    & 0.57      & 0.42       \\ \bottomrule
    \end{tabular}
\end{table}

\begin{table}[htpb]
    \centering
    \caption{Singlet excitation energy (in eV) of atom Be. Experimental data were obtained
            from Ref \citenum{NIST_ASD}}
    \label{tab:Rdberg_Be_Singlet}
    \begin{tabular}{@{}lccccccc@{}} \toprule
    AO         & Exp  & LDA   & BLYP  & B3LYP & LOSC-LDA & LOSC-BLYP & LOSC-B3LYP \\ \midrule
    2p         & 5.28 & 5.68  & 2.49  & 5.60  & 5.56     & 4.70      & 5.59       \\
    3s         & 6.78 & 9.02  & 6.14  & 8.98  & 7.27     & 7.54      & 7.42       \\
    3p         & 7.46 & 9.77  & 6.66  & 9.54  & 7.90     & 7.86      & 7.88       \\
    3d         & 7.99 & 10.29 & 7.30  & 10.08 & 8.44     & 8.32      & 8.34       \\
    4s         & 8.09 & 10.67 & 6.91  & 9.96  & 8.17     & 7.50      & 7.86       \\
    4p         & 8.31 & 10.90 & 7.11  & 10.19 & 8.37     & 7.79      & 8.14       \\
    4d         & 8.53 & 11.08 & 7.37  & 10.40 & 8.41     & 8.07      & 8.32       \\
    5s         & 8.59 & 11.25 & 7.48  & 10.53 & 8.44     & 7.91      & 8.22       \\
    5p         & 8.69 & 11.35 & 7.56  & 10.61 & 8.55     & 7.97      & 8.29       \\
    5d         & 8.80 & 11.55 & 7.87  & 10.87 & 8.84     & 8.47      & 8.27       \\
    6s         & 8.84 & 11.52 & 7.78  & 10.84 & 8.55     & 8.63      & 8.86       \\
    6p         & 8.90 & 11.57 & 7.82  & 10.88 & 8.57     & 8.04      & 8.41       \\ \midrule
    MAE        &      & 2.37  & 1.15  & 1.85  & 0.24     & 0.54      & 0.35       \\
    MSE        &      & 2.37  & -1.15 & 1.85  & 0.07     & -0.29     & -0.06      \\ \bottomrule
    \end{tabular}
\end{table}

\begin{table}[htpb]
    \centering
    \caption{Triplet excitation energy (in eV) of atom Be. Experimental data were obtained
            from Ref \citenum{NIST_ASD}}
    \label{tab:Rydberg_Be_Triplet}
    \begin{tabular}{@{}lccccccc@{}} \toprule
    AO         & Exp  & LDA   & BLYP  & B3LYP & LOSC-LDA & LOSC-BLYP & LOSC-B3LYP \\ \midrule
    2p         & 2.72 & 1.59  & 1.38  & 5.04  & 1.86     & 1.54      & 1.88       \\
    3s         & 6.46 & 8.37  & 7.87  & 7.74  & 6.90     & 6.22      & 6.45       \\
    3p         & 7.30 & 9.24  & 8.73  & 8.63  & 7.70     & 7.01      & 7.26       \\
    3d         & 7.69 & 9.62  & 9.21  & 9.10  & 8.18     & 7.55      & 7.77       \\
    4s         & 8.00 & 10.52 & 10.05 & 9.83  & 8.08     & 7.49      & 7.80       \\
    4p         & 8.28 & 10.77 & 10.32 & 10.07 & 8.29     & 7.69      & 7.97       \\
    4d         & 8.42 & 10.92 & 10.48 & 10.23 & 8.43     & 7.86      & 8.12       \\
    5s         & 8.56 & 11.19 & 10.73 & 10.47 & 8.39     & 7.81      & 8.21       \\
    5p         & 8.69 & 11.29 & 10.85 & 10.59 & 8.48     & 7.94      & 8.31       \\
    5d         & 8.75 & 11.42 & 10.97 & 10.74 & 8.71     & 8.14      & 8.55       \\
    6s         & 8.82 & 11.49 & 11.06 & 10.78 & 8.53     & 8.02      & 8.41       \\
    6p         & 8.89 & 11.54 & 11.15 & 10.86 & 8.54     & 8.07      & 8.45       \\ \midrule
    MAE        &      & 2.30  & 1.91  & 1.79  & 0.28     & 0.60      & 0.30       \\
    MSE        &      & 2.11  & 1.68  & 1.79  & -0.04    & -0.60     & -0.29      \\ \bottomrule
    \end{tabular}
\end{table}

\begin{table}[htpb]
    \centering
    \caption{Singlet excitation energy (in eV) of atom Mg. Experimental data were obtained
            from Ref \citenum{NIST_ASD}}
    \label{tab:Rydberg_Mg_Singlet}
    \begin{tabular}{@{}lccccccc@{}} \toprule
    AO         & Exp  & LDA  & BLYP & B3LYP & LOSC-LDA & LOSC-BLYP & LOSC-B3LYP \\ \midrule
    3p         & 4.35 & 5.00 & 5.61 & 5.13  & 4.62     & 5.28      & 4.86       \\
    4s         & 5.39 & 7.48 & 6.80 & 6.52  & 6.26     & 5.44      & 5.51       \\
    3d         & 5.75 & 8.33 & 8.21 & 7.63  & 7.02     & 6.79      & 6.48       \\
    4p         & 6.12 & 8.37 & 7.75 & 7.48  & 6.94     & 6.22      & 6.26       \\
    5s         & 6.52 & 9.00 & 8.52 & 8.19  & 6.96     & 6.41      & 6.57       \\
    4d         & 6.59 & 9.26 & 8.96 & 8.55  & 7.29     & 6.74      & 6.94       \\
    5p         & 6.78 & 9.29 & 8.83 & 8.50  & 7.25     & 6.70      & 6.89       \\
    6s         & 6.97 & 9.55 & 9.21 & 8.84  & 7.27     & 6.90      & 7.03       \\
    5d         & 6.98 & 9.70 & 9.48 & 9.06  & 7.36     & 7.25      & 7.13       \\
    6p         & 7.09 & 9.72 & 9.39 & 9.02  & 7.75     & 7.07      & 7.38       \\
    7s         & 7.19 & 9.81 & 9.53 & 9.13  & 7.40     & 7.05      & 7.21       \\
    7p         & 7.26 & 9.88 & 9.59 & 9.20  & 7.44     & 7.06      & 7.27       \\ \midrule
    MAE        &      & 2.37 & 2.07 & 1.69  & 0.55     & 0.26      & 0.21       \\
    MSE        &      & 2.37 & 2.07 & 1.69  & 0.55     & 0.16      & 0.21       \\ \bottomrule
    \end{tabular}
\end{table}

\begin{table}[htpb]
    \centering
    \caption{Triplet excitation energy (in eV) of atom Mg. Experimental data were obtained
            from Ref \citenum{NIST_ASD}}
    \label{tab:Rydberg_Mg_Triplet}
    \begin{tabular}{@{}lccccccc@{}} \toprule
    AO         & Exp  & LDA  & BLYP & B3LYP & LOSC-LDA & LOSC-BLYP & LOSC-B3LYP \\ \midrule
    3p         & 2.71 & 2.64 & 2.57 & 2.61  & 2.36     & 2.19      & 2.36       \\
    4s         & 5.11 & 7.03 & 6.67 & 6.45  & 5.69     & 5.21      & 5.39       \\
    3d         & 5.95 & 7.22 & 7.03 & 6.93  & 6.07     & 5.63      & 5.88       \\
    4p         & 5.93 & 8.05 & 7.62 & 7.37  & 6.66     & 6.09      & 6.22       \\
    5s         & 6.43 & 8.90 & 8.47 & 8.15  & 6.96     & 6.37      & 6.53       \\
    4d         & 6.72 & 9.08 & 8.65 & 8.35  & 7.23     & 6.65      & 6.79       \\
    5p         & 6.73 & 9.20 & 8.78 & 8.46  & 7.16     & 6.65      & 6.82       \\
    6s         & 6.93 & 9.50 & 9.17 & 8.81  & 7.29     & 6.87      & 7.03       \\
    5d         & 7.06 & 9.62 & 9.26 & 8.92  & 7.47     & 7.04      & 7.21       \\
    6p         & 7.07 & 9.64 & 9.35 & 8.98  & 7.42     & 6.99      & 7.16       \\
    7s         & 7.17 & 9.79 & 9.50 & 9.11  & 7.41     & 7.04      & 7.21       \\
    7p         & 7.25 & 9.86 & 9.57 & 9.19  & 7.44     & 7.04      & 7.23       \\ \midrule
    MAE        &      & 2.13 & 1.82 & 1.54  & 0.40     & 0.15      & 0.14       \\
    MSE        &      & 2.12 & 1.80 & 1.52  & 0.34     & -0.11     & 0.06       \\ \bottomrule
    \end{tabular}
\end{table}

\clearpage
\section{Supplementary polyacene geometry}
\begin{longtable}{@{}llll@{}}
Benzene     &           &           &           \\
12          &           &           &           \\
C           & 0.0000    & 1.3966    & 0.0000    \\
C           & 1.2095    & 0.6983    & 0.0000    \\
C           & 1.2095    & -0.6983   & 0.0000    \\
C           & 0.0000    & -1.3966   & 0.0000    \\
C           & -1.2095   & -0.6983   & 0.0000    \\
C           & -1.2095   & 0.6983    & 0.0000    \\
H           & 0.0000    & 2.4836    & 0.0000    \\
H           & 2.1509    & 1.2418    & 0.0000    \\
H           & 2.1509    & -1.2418   & 0.0000    \\
H           & 0.0000    & -2.4836   & 0.0000    \\
H           & -2.1509   & -1.2418   & 0.0000    \\
H           & -2.1509   & 1.2418    & 0.0000    \\
\end{longtable}

\begin{longtable}{@{}llll@{}}
Naphthalene &           &           &           \\
18          &           &           &           \\
C           & 0.0       & 0.714951  & 0.0       \\
C           & 0.0       & -0.714951 & 0.0       \\
C           & 2.426869  & 0.706561  & 0.0       \\
C           & -2.426869 & 0.706561  & 0.0       \\
C           & 2.426869  & -0.706561 & 0.0       \\
C           & -2.426869 & -0.706561 & 0.0       \\
C           & 1.242318  & 1.398661  & 0.0       \\
C           & -1.242318 & 1.398661  & 0.0       \\
C           & 1.242318  & -1.398661 & 0.0       \\
C           & -1.242318 & -1.398661 & 0.0       \\
H           & 3.369050  & 1.241625  & 0.0       \\
H           & -3.369050 & 1.241625  & 0.0       \\
H           & 3.369050  & -1.241625 & 0.0       \\
H           & -3.369050 & -1.241625 & 0.0       \\
H           & 1.241411  & 2.483152  & 0.0       \\
H           & -1.241411 & 2.483152  & 0.0       \\
H           & 1.241411  & -2.483152 & 0.0       \\
H           & -1.241411 & -2.483152 & 0.0       \\
\end{longtable}

\begin{longtable}{@{}llll@{}}
Anthracene  &           &           &           \\
24          &           &           &           \\
C           & 0.0       & 1.220369  & 0.720788  \\
C           & 0.0       & -1.220369 & 0.720788  \\
C           & 0.0       & 1.220369  & -0.720788 \\
C           & 0.0       & -1.220369 & -0.720788 \\
C           & 0.0       & 0.0       & 1.400039  \\
C           & 0.0       & 0.0       & -1.400039 \\
C           & 0.0       & 2.474059  & 1.403168  \\
C           & 0.0       & -2.474059 & 1.403168  \\
C           & 0.0       & 2.474059  & -1.403168 \\
C           & 0.0       & -2.474059 & -1.403168 \\
C           & 0.0       & 3.650786  & 0.711379  \\
C           & 0.0       & -3.650786 & 0.711379  \\
C           & 0.0       & 3.650786  & -0.711379 \\
C           & 0.0       & -3.650786 & -0.711379 \\
H           & 0.0       & 0.0       & 2.485281  \\
H           & 0.0       & 0.0       & -2.485281 \\
H           & 0.0       & 2.473747  & 2.487567  \\
H           & 0.0       & -2.473747 & 2.487567  \\
H           & 0.0       & 2.473747  & -2.487567 \\
H           & 0.0       & -2.473747 & -2.487567 \\
H           & 0.0       & 4.594838  & 1.243045  \\
H           & 0.0       & -4.594838 & 1.243045  \\
H           & 0.0       & 4.594838  & -1.243045 \\
H           & 0.0       & -4.594838 & -1.243045 \\
\end{longtable}

\begin{longtable}{@{}llll@{}}
Tetracene   &           &           &           \\
30          &           &           &           \\
C           & 0.0       & 0.724345  & 0.0       \\
C           & 0.0       & -0.724345 & 0.0       \\
C           & 1.232442  & 1.402849  & 0.0       \\
C           & -1.232442 & 1.402849  & 0.0       \\
C           & 1.232442  & -1.402849 & 0.0       \\
C           & -1.232442 & -1.402849 & 0.0       \\
C           & 2.443772  & 0.724216  & 0.0       \\
C           & -2.443772 & 0.724216  & 0.0       \\
C           & 2.443772  & -0.724216 & 0.0       \\
C           & -2.443772 & -0.724216 & 0.0       \\
C           & 3.702747  & 1.405473  & 0.0       \\
C           & -3.702747 & 1.405473  & 0.0       \\
C           & 3.702747  & -1.405473 & 0.0       \\
C           & -3.702747 & -1.405473 & 0.0       \\
C           & 4.875977  & 0.713742  & 0.0       \\
C           & -4.875977 & 0.713742  & 0.0       \\
C           & 4.875977  & -0.713742 & 0.0       \\
C           & -4.875977 & -0.713742 & 0.0       \\
H           & 1.232720  & 2.487982  & 0.0       \\
H           & -1.232720 & 2.487982  & 0.0       \\
H           & 1.232720  & -2.487982 & 0.0       \\
H           & -1.232720 & -2.487982 & 0.0       \\
H           & 3.702863  & 2.489850  & 0.0       \\
H           & -3.702863 & 2.489850  & 0.0       \\
H           & 3.702863  & -2.489850 & 0.0       \\
H           & -3.702863 & -2.489850 & 0.0       \\
H           & 5.820918  & 1.243786  & 0.0       \\
H           & -5.820918 & 1.243786  & 0.0       \\
H           & 5.820918  & -1.243786 & 0.0       \\
H           & -5.820918 & -1.243786 & 0.0       \\
\end{longtable}

\begin{longtable}{@{}llll@{}}
Pentacene   &           &           &           \\
36          &           &           &           \\
C           & 0.0       & 1.223172  & 0.726794  \\
C           & 0.0       & -1.223172 & 0.726794  \\
C           & 0.0       & 1.223172  & -0.726794 \\
C           & 0.0       & -1.223172 & -0.726794 \\
C           & 0.0       & 3.668639  & 0.726066  \\
C           & 0.0       & -3.668639 & 0.726066  \\
C           & 0.0       & 3.668639  & -0.726066 \\
C           & 0.0       & -3.668639 & -0.726066 \\
C           & 0.0       & 2.461665  & 1.404459  \\
C           & 0.0       & -2.461665 & 1.404459  \\
C           & 0.0       & 2.461665  & -1.404459 \\
C           & 0.0       & -2.461665 & -1.404459 \\
C           & 0.0       & 6.101733  & 0.714980  \\
C           & 0.0       & -6.101733 & 0.714980  \\
C           & 0.0       & 6.101733  & -0.714980 \\
C           & 0.0       & -6.101733 & -0.714980 \\
C           & 0.0       & 4.930253  & 1.406708  \\
C           & 0.0       & -4.930253 & 1.406708  \\
C           & 0.0       & 4.930253  & -1.406708 \\
C           & 0.0       & -4.930253 & -1.406708 \\
C           & 0.0       & 0.0       & 1.404933  \\
C           & 0.0       & 0.0       & -1.404933 \\
H           & 0.0       & 2.462192  & 2.489563  \\
H           & 0.0       & -2.462192 & 2.489563  \\
H           & 0.0       & 2.462192  & -2.489563 \\
H           & 0.0       & -2.462192 & -2.489563 \\
H           & 0.0       & 7.047138  & 1.244171  \\
H           & 0.0       & -7.047138 & 1.244171  \\
H           & 0.0       & 7.047138  & -1.244171 \\
H           & 0.0       & -7.047138 & -1.244171 \\
H           & 0.0       & 4.930599  & 2.491074  \\
H           & 0.0       & -4.930599 & 2.491074  \\
H           & 0.0       & 4.930599  & -2.491074 \\
H           & 0.0       & -4.930599 & -2.491074 \\
H           & 0.0       & 0.0       & 2.489940  \\
H           & 0.0       & 0.0       & -2.489940 \\
\end{longtable}

\begin{longtable}{llll}
Hexacene    &           &           &           \\
42          &           &           &           \\
C           & 0.728760  & 0.0       & 0.0       \\
C           & -0.728760 & 0.0       & 0.0       \\
C           & 1.406190  & 1.229341  & 0.0       \\
C           & -1.406190 & 1.229341  & 0.0       \\
C           & 1.406190  & -1.229341 & 0.0       \\
C           & -1.406190 & -1.229341 & 0.0       \\
C           & 0.728192  & 2.447851  & 0.0       \\
C           & -0.728192 & 2.447851  & 0.0       \\
C           & 0.728192  & -2.447851 & 0.0       \\
C           & -0.728192 & -2.447851 & 0.0       \\
C           & 1.405376  & 3.689513  & 0.0       \\
C           & -1.405376 & 3.689513  & 0.0       \\
C           & 1.405376  & -3.689513 & 0.0       \\
C           & -1.405376 & -3.689513 & 0.0       \\
C           & 0.727090  & 4.894222  & 0.0       \\
C           & -0.727090 & 4.894222  & 0.0       \\
C           & 0.727090  & -4.894222 & 0.0       \\
C           & -0.727090 & -4.894222 & 0.0       \\
C           & 1.407386  & 6.157229  & 0.0       \\
C           & -1.407386 & 6.157229  & 0.0       \\
C           & 1.407386  & -6.157229 & 0.0       \\
C           & -1.407386 & -6.157229 & 0.0       \\
C           & 0.715651  & 7.327785  & 0.0       \\
C           & -0.715651 & 7.327785  & 0.0       \\
C           & 0.715651  & -7.327785 & 0.0       \\
C           & -0.715651 & -7.327785 & 0.0       \\
H           & 2.491162  & 1.229591  & 0.0       \\
H           & -2.491162 & 1.229591  & 0.0       \\
H           & 2.491162  & -1.229591 & 0.0       \\
H           & -2.491162 & -1.229591 & 0.0       \\
H           & 2.490467  & 3.690188  & 0.0       \\
H           & -2.490467 & 3.690188  & 0.0       \\
H           & 2.490467  & -3.690188 & 0.0       \\
H           & -2.490467 & -3.690188 & 0.0       \\
H           & 2.491747  & 6.157716  & 0.0       \\
H           & -2.491747 & 6.157716  & 0.0       \\
H           & 2.491747  & -6.157716 & 0.0       \\
H           & -2.491747 & -6.157716 & 0.0       \\
H           & 1.244383  & 8.273439  & 0.0       \\
H           & -1.244383 & 8.273439  & 0.0       \\
H           & 1.244383  & -8.273439 & 0.0       \\
H           & -1.244383 & -8.273439 & 0.        \\
\end{longtable}

\bibliography{ref2}